\documentclass[11pt,twoside]{article}

\usepackage{amsmath,amsfonts,amssymb,amsthm}
\usepackage{epsfig}
\newcommand{\field}[1]{\mathbb{#1}}
\newcommand{\N}{\field{N}}
\newcommand{\R}{\field{R}}
\newcommand{\C}{\field{C}}

\newcommand{\F}{\mathcal{F}}
\renewcommand{\H}{\mathcal{H}}
\newcommand{\h}{\mathfrak{h}}
\renewcommand{\L}{\mathbf{B}}
\newcommand{\HxF}{\tilde{\mathcal{H}}}   

\newcommand{\Hgs}{\mathcal{H}_{\text{des}}} 
\newcommand{\Pgs}{P_{\text{des}}}           

\newcommand{\Hmod}{H_{\text{mod}}}         
\newcommand{\Hmodex}{\tilde{H}_{\text{mod}}}
\newcommand{\Emod}{E_{\text{mod}}}

\newcommand{\tf}{C_{0}^{\infty}}
\newcommand{\Hex}{\tilde{H}}
\newcommand{\uGamma}{\breve{\Gamma}}
\newcommand{\udGamma}{\mathrm{d}\breve{\Gamma}}
\newcommand{\dGamma}{\mathrm{d}\Gamma}

\newcommand{\eps}{\varepsilon}
\newcommand{\ph}{\varphi}
\newcommand{\const}{\mathrm{const}}
\newcommand{\ran}{\mathrm{Ran}}

\newcommand{\restricted}{|\grave{}\,}
\newcommand{\weak}{\rightharpoonup}

\newcommand{\sprod}[2]{\mbox{$\langle #1,#2 \rangle$}}       

\newcommand{\Ran}{\operatorname{Ran}}
\newcommand{\supp}{\operatorname{supp}}
\newcommand{\Ima}{\operatorname{Im}}
\newcommand{\Ree}{\operatorname{Re}}

\newtheorem{theorem}{Theorem}
\newtheorem{lemma}[theorem]{Lemma}
\newtheorem{corollary}[theorem]{Corollary}

\newtheorem{prop}[theorem]{Proposition}

\font\notefont=cmsl8 
\begin{document}
\title{\bf Asymptotic Completeness for Compton Scattering}
\footnotetext [2]{Work partially supported by U.S. National
Science Foundation grant DMS 01-00160.}
\author{\vspace{5pt} J. Fr\"ohlich$^1$\footnote{juerg@itp.phys.ethz.ch} ,
M. Griesemer$^2$\footnote{marcel@math.uab.edu}
and B. Schlein$^3$\footnote{schlein@cims.nyu.edu} \\
\vspace{-4pt}\small{$1.$ Theoretical Physics, ETH--H\"onggerberg,} \\
\small{CH--8093 Z\"urich, Switzerland}\\
\vspace{-4pt}\small{$2.$ Department of Mathematics, University of Alabama at
Birmingham,} \\
\small{Birmingham, AL 35294, USA}\\
\vspace{-4pt}\small{$3.$ Courant Institute, New York University,} \\
\small{New York, NY 10012, USA}\\
}
\date{\small 23 July, 2003}
\maketitle

\vspace{-0.5in}
\begin{center}
\em Dedicated to Freeman Dyson on the occasion of his 80$^{th}$
birthday.
\end{center}

\begin{abstract}
Scattering in a model of a massive quantum-mechanical particle, an
``electron'', interacting with massless, relativistic bosons,
``photons'', is studied. The interaction term in the Hamiltonian
of our model describes emission and absorption of ``photons'' by
the ``electron''; but ``electron-positron'' pair production is
suppressed. An ultraviolet cutoff and an (arbitrarily small, but
fixed) infrared cutoff are imposed on the interaction term. In a
range of energies where the propagation speed of the dressed
``electron'' is strictly smaller than the speed of light,
unitarity of the scattering matrix is proven, provided the
coupling constant is small enough; (asymptotic completeness of
Compton scattering). The proof combines a construction of dressed
one--electron states with propagation estimates for the
``electron'' and the ``photons''.
\end{abstract}

\noindent {\bf Contents}\\

\begin{tabular}{rl}
1 &Introduction\\ 2& Fock Space and Second Quantization\\
3 &The Model, Dressed One-Electron States, and Bounds on the Interaction\\
4 &Propagation Estimate for the Electron and Existence of the Wave Operator\\
5 &The Modified Hamiltonian\\ 6 &Propagation Estimates for Photons\\
7 &The Asymptotic Observable\\ 8 & The Inverse of the Wave
Operator\\ 9 & Putting It All Together: Asymptotic Completeness\\
10 &Outlook\\ A-G & Appendices
\end{tabular}

\section{Introduction}
The study of collisions between photons, the field quanta of the
electromagnetic field, and freely moving charged particles, in
particular electrons, at energies below the threshold for
electron-positron pair creation -- commonly called \emph{Compton
scattering} -- has played a significant r\^ ole in establishing
the reality of Einstein's photons, in the early days of quantum
theory. With the development of quantum electrodynamics (QED) it
became possible to calculate the cross section for Compton
scattering perturbatively, using the Feynman rules of relativistic
QED. The agreement between theoretical predictions and experiments
is astounding.

Yet, a careful theoretical analysis of Compton scattering uncovers
substantial difficulties mainly related to the so-called
\emph{infrared problem} in QED, \cite{BN, PF}: When, in the course
of a collision process, a charged particle, such as an electron,
undergoes an accelerated motion it emits \emph{infinitely} many
photons of \emph{finite} total energy. Unless treated carefully, a
perturbative calculation of scattering amplitudes is therefore
plagued by the infamous infrared divergencies.

Infrared divergencies can be eliminated by giving the photon a small mass, or,
alternatively, by introducing an infrared cutoff in the interaction term. Of
course, after having calculated suitable cross sections, one attempts to let the
photon mass or the infrared cutoff, respectively, tend to 0. This procedure,
carefully implemented, is known to work very well; see \cite{YFS}.

If the total energy of the incoming particles, photons and an
electron, is well below the threshold for electron-positron pair
creation it is a fairly good approximation to neglect all terms
in the Hamiltonian of relativistic QED describing pair creation-
and annihilation processes in a calculation of some cross section
for Compton scattering. The resulting model is a caricature of
QED without positrons in which the number of electrons is
conserved. It is this simplified model of QED, regularized in the
infrared region by an infrared cutoff, which has inspired the
analysis of Compton scattering presented in this paper.

To further simplify matters, we consider a toy model involving ``scalar
photons'' or ``pho\-nons'', and we also impose an ultraviolet cutoff in the
interaction Hamiltonian. But the methods developed in this paper can be applied
to the caricature of QED described above if one works in the Coulomb gauge and
introduces an ultraviolet and an infrared cutoff in the interaction Hamiltonian.

The main results of this paper can be described as follows: For the toy model
described above, we establish \emph{asymptotic completeness (AC) for Compton
scattering} below some threshold energy $\Sigma$, which depends on the
kinematics of the electron and on the coupling constant. The latter will have to
be chosen sufficiently small. This means that, on the subspace of physical state vectors containing one
electron and arbitrarily many ``scalar photons'' (massless bosons) of total
energy $\leq \Sigma$, the scattering operator of our toy model is
\emph{unitary}.

In a previous paper \cite{FGS2}, we have studied the scattering of
massless bosons at an electron bound to a static nucleus, below
the ionization threshold, in a similar toy model with an infrared-
and an ultraviolet cutoff. In other words, we have proven AC for
\emph{Rayleigh scattering} of ``photons'' at an atom, below the
ionization threshold, in the presence of an infrared- and
ultraviolet cutoff. By combining the methods in \cite{FGS2} with
those developed in this paper, we expect to be able to establish
AC in our toy model of an electron interacting with massless
bosons and with a static nucleus at energies below some threshold
energy $\Sigma$ (depending on the kinematics of the electron),
provided the coupling constant is small enough. Such a result
would apply to scattering processes encountered in the analysis of
the photoelectric effect (see \cite{BKZ}) and of Bremsstrahlung.
Further possible extensions of our results are described in Sect.
\ref{sec:out}.

As quite frequently the case in mathematical physics, our methods of proof are
considerably more interesting than the results we establish. We think that they
illustrate \emph{some} of the many subtleties of scattering theory in quantum
field theory in a fairly illuminating way. Before we are able to describe these
methods and give an outline of the strategy of our proof, we must define the
model studied in this paper more precisely.

To describe the dynamics of a conserved, unbound particle, here
called \emph{electron}, coupled to a quantized field of spin-0
massless bosons, we consider the Hamiltonian
\begin{equation*}
H_{g} = \Omega (p) + H_f + g\phi(G_x)
\end{equation*}
acting on the Hilbert space of state vectors \(\H = L^2(\R^3)\otimes \F\), where $\F$
is the bosonic Fock space
over $L^2(\R^3,dk)$, $k \in \R^3$ is the momentum of a boson,
$x\in \R^3$ the position of the electron, $p=-i\nabla_x$ the momentum of the
electron, and $\Omega (p)$ is the
energy of a non-interacting, free electron of momentum $p$.
The operator \( H_f=\int dk |k| a^{*}(k)a(k)\) is
the Hamiltonian of the free bosons; $a(k)$ and $a^{*}(k)$ being the usual
annihilation and creation operators obeying the canonical commutation relations
(CCR). The operator $\phi(G_x)$
describes the interaction between an electron at position $x$ and bosons. It is given
by
\begin{align}
\phi(G_x) &= \int dk \, ( \overline{G_x(k)} a(k) + G_x(k)a^{*}(k)), \quad
\text{with}\\
G_x(k) &= e^{-ik\cdot x}\kappa_{\sigma}(k) \label{eq:intro2}.
\end{align}
We impose an infrared cutoff by requiring that
\begin{equation*}
\kappa_{\sigma}(k) = 0\makebox[5em]{if}|k|<\sigma
\end{equation*}
where $\kappa_{\sigma} \in C_0^{\infty}(\R^3)$ is a form factor. The constant
$\sigma$ must be positive but can be arbitrarily small.
The smoothness and the decay assumptions on $\kappa_{\sigma}$ at
$|k|=\infty$ are technically convenient, but can be relaxed; see e.g.
\cite{Nel,Ammari}. The parameter $g \in \R$ is a coupling constant.

The Hamiltonian $H_g$ is invariant under translations in physical
space and thus admits a decomposition over the spectrum of the
total momentum \(P=p+P_f\), $P_f=\int dk\,k a^{*}(k)a(k)$, as a direct integral
\begin{align*}
H_g &\simeq \int_{\R^3}^{\oplus} H_g(P)\, dP \makebox[4em]{on}
\H \simeq \int^{\oplus}_{\R^3}
\F\, dP , \quad \text{with}\\
H_g(P) &= \Omega(P-P_f) + H_f + g\phi(\kappa_{\sigma}),
\end{align*}
where $\simeq$ indicates unitary equivalence.

On the dispersion law, $\Omega(p)$, of the free electron we only
impose minimal assumptions that are sufficient for our purpose and are
satisfied in examples of physical interest. We assume
that $\Omega\geq 0$, that $\Omega$ is twice continuously
differentiable, and that $\partial_i\partial_j\Omega$ and
$|\nabla\Omega|(\Omega+1)^{-1/2}$ are bounded functions. Most
importantly, we assume that, given an arbitrary $\beta>0$, there
exists a constant \(O_{\beta}>\inf\Omega\), such that
\begin{equation}\label{eq:O-beta}
|\nabla \Omega (p)| \leq \beta, \quad \text{for all $p$ with
$\Omega(p)\leq O_{\beta}$}.
\end{equation}
Note that these assumptions are satisfied in the examples where
\[\Omega (p) = \frac{p^2}{2M} \quad \text{(non-relativistic
kinematics)} \] and \[ \Omega (p) = \sqrt{p^2 + M^2} \quad
\text{(relativistic kinematics)}, \] for some positive mass $M$.
[We could also study an electron in a crystal interacting with
phonons.]

Our assumptions on $\Omega(p)$ and the presence of an infrared
cutoff $\sigma>0$ guarantee that the Hamiltonian $H_g(P)$ has a
unique one-particle eigenstate $\psi_P \in \F$ corresponding to the eigenvalue
(energy) $E_g (P) = \inf \sigma (H_g (P))$, for each $P$
with $\Omega(P)\leq O_{\beta}$, $\beta<1$, and for $|g|$
sufficiently small, depending on $\beta$. In fact, under these
assumptions, condition \eqref{eq:O-beta} allows us to show that
\begin{equation}
\inf_{|k| \geq \sigma} \left( E_{g}(P-k) + |k| - E_{g}(P)
\right)>0
\end{equation}
for all $\sigma >0$. This is the key ingredient for proving that
$H_{g}(P)$ has a one-particle eigenstate of energy $E_g(P)$ (c.f. \cite{Froe2}).
Uniqueness follows by standard Perron-Frobenius type arguments or
by a suitable positive commutator estimate.

Wave packets $\psi_f$, $f\in L^2(\R^3)$, of dressed one-particle
states $\psi_{P}$ are defined by
\begin{equation}\label{eq:wave-packet}
\psi_f(P) = f(P)\psi_P
\end{equation}
where $\supp f \subset \{P:\Omega(P)\leq O_{\beta}\}$. They
minimize the energy for a given distribution $|f|^2$ of the total
momentum, and they propagate according to
\[ e^{-iH_gt} \psi_f = \psi_{f_t},\qquad f_t(P)= e^{-iE_g(P)t}f(P).  \]

In nature, no excited one-electron states are observed, and,
correspondingly, one expects that every state $e^{-iH_g t}\psi$
eventually radiates off its excess energy and decays into a
dressed one-electron wave packet $\psi_f$. More precisely, for any
given $\psi$, $e^{-iH_g t}\psi$ should be well approximated, in
the distant future, by a linear combination of states of the form
\begin{equation}\label{eq:decay}
a^{*}(h_{1,t})\cdot \ldots \cdot a^{*}(h_{n,t}) e^{-iH_g t}\psi_f
\end{equation}
where \(h_{i,t} = e^{-i|k|t}h_i\), and $\psi_f$ is given by
\eqref{eq:wave-packet}. This is called {\em asymptotic
completeness (AC) for Compton scattering}. Mathematically more
convenient characterizations of AC may be given in terms of the
asymptotic field operators $a_{+}(h)$ and $a_{+}^{*}(h)$. Let
\(h\in L^2(\R^3,(1+|k|^{-1})dk)\) and let \(\ph\in
E_{\Sigma}(H_{g})\H\) for some $\Sigma<O_{\beta=1}$. Then the
limit
\begin{equation*}
  a_{+}^{\#}(h)\ph = \lim_{t\to\infty} e^{iH_g t} a^{\#}(h_{t})e^{-iH_g t}\ph
\end{equation*}
exists, and, moreover,
\begin{equation}\label{eq:field_prod}
a_{+}^*(h_1)\cdot\ldots\cdot a_{+}^{*}(h_n)\ph =
 \lim_{t\to\infty} e^{iH_g t} a^*(h_{1,t})\cdot\ldots\cdot a^{*} (h_{n,t})
 e^{-iH_g t}\ph
\end{equation}
if \(h_i\in L^2(\R^3,(1+|k|^{-1})dk)\), \(\ph\in
E_{\lambda}(H_{g})\H\), and \(\lambda+\sum_{i}M_i\leq \Sigma\),
where \(M_i:=\sup\{|k|:h_i(k)\neq 0\}\). Let $\H_{+}$ denote the
closure of the space spanned by vectors of the form
\(a_{+}^*(h_1)\cdot\ldots\cdot a_{+}^{*}(h_n)\psi_f \). From
\eqref{eq:decay} and \eqref{eq:field_prod} it is clear that AC means that $\H_{+}=\H$.
AC in this form asserts, on the one hand, that the asymptotic
dynamics of bosons which are not bound to the electron corresponds
to free motion, and, on the other hand, that $H_g(P)$ has no
eigenvalues above \(E_g (P) = \inf\sigma(H_g(P))\).

The main purpose of this paper is to show that
\begin{equation*}
  \H_{+}\supset E_{\Sigma}(H_g)\H,
\end{equation*}
for every \(\Sigma<O_{\beta=1/3}\) provided that $|g|$ is
sufficiently small depending on $\Sigma$ (Theorem~\ref{thm:ACph}).
Here $E_{\Sigma} (H_g)$ is
the spectral projection of $H_{g}$ onto vectors of energy $\leq
\Sigma$. We thus prove that all vectors in $\Ran E_{\Sigma}(H_g)$
decay into states of the form \eqref{eq:decay}. While the
assumption \(\Sigma<O_{\beta=1/3}\) may appear very restrictive,
it still allows for electrons with speeds as high as one third of
the speed of light ($\simeq 10^8$ m/s)!

Dressed one-electron states for the model discussed here with
relativistic and non-relativistic electrons were first constructed
in \cite{Froe1, Froe2}. For similar results on the related polaron
model, see \cite{Spohn2} and references therein.

First steps towards a scattering theory (construction of the M\o
ller operators) were previously made in \cite{Froe1}, \cite{Froe2}
and, for $\sigma = 0$, very recently in \cite{P}.

The scattering theory of a free electron in the
framework of non-relativistic QED in the dipole approximation has been studied
in \cite{Arai}. This model is
explicitly soluble and is not translation--invariant. Arai proves asymptotic
completeness after removing the infrared cutoff.

Asymptotic completeness in non-trivial models of quantum field
theory was previously established in \cite{Sp74, SZ76}, \cite{DG1,
DG2} and \cite{FGS2}. The papers \cite{DG1} and \cite{FGS2} are
devoted to an analysis of scattering in a system of a fixed number
of spatially confined particles interacting with massive
relativistic bosons. Confinement is enforced by a confining
(increasing) potential in \cite{DG1}  and by an energy cutoff in
\cite{FGS2}. In \cite{DG2} asymptotic completeness is proven for
spatially cutoff $P(\phi)_2$-Hamiltonians. For interesting results
in the scattering theory of systems of {\em
  massless} bosons and confined electrons {\em without infrared cutoff} see the
papers \cite{Spohn1, CG}. In none of these papers a translation
invariant model is studied. But, such models have been
analyzed in \cite{Froe1}, \cite{P}.\\

We now present an outline of our paper and explain the key ideas
underlying our proof of asymptotic completeness.

In \emph{Sect.} \ref{sec:Fock}, we introduce notation and recall
some well known facts about the formalism of second quantization
which will be used throughout our paper.

In \emph{Sect.} \ref{sec:sys}, we first give a mathematically
precise definition of our model and list all our hypotheses for
easy reference. We then summarize our key results on the
\emph{existence and uniqueness} of \emph{dressed one-electron
states}, $\psi_P$. All proofs concerning these matters are
deferred to Appendix \ref{app:DES}.

We also prove a fundamental \emph{positive--commutator estimate} and a
\emph{Virial Theorem}, which, by standard arguments of Mourre theory, show that
there are \emph{no excited} dressed one-electron states; i.e., there is \emph{no
binding} between a dressed electron and bosons. See Theorems \ref{thm:virial},
\ref{thm:pos_comm} and \ref{cor:gs-only}.

In the last part of Sect. \ref{sec:sys} we exhibit some simple
properties of the interaction Hamiltonian $g \phi (G_x)$. In
particular, we show that the strength of interaction between an
electron at position $x$ and a boson localized (in the sense of
Newton and Wigner) near a point $y \in \R^3$ tends rapidly to $0$,
as $|x-y| \to \infty$; (see Lemma \ref{lm:srdecay}). This property
is important in our proof of AC.

In \emph{Sect.} \ref{sec:waveop}, we construct \emph{M\o ller wave
operators} as a first step towards understanding scattering in our
model. Our construction is based on \cite{FGS1}. It involves the
following ideas.

(i) We prove a \emph{propagation estimate} saying that an electron
with dispersion law $\Omega (p)$ propagates with a group velocity
not exceeding $\beta$, for states with a finite total energy
$\Sigma$ if \(\| |\nabla\Omega|E_{\Sigma}(H_g)\|\leq \beta\), see
Proposition \ref{prop:pe-el} and \cite{FGS1}. A sufficient
condition for the latter assumption is that $\Sigma < O_{\beta}$
and that $|g|$ is sufficiently small.

(ii) This propagation estimate for the electron with $\beta<1$
combined with a stationary phase argument for the photon
guarantees that the interaction between a dressed electron and a
configuration of freely moving bosons tends to 0 at large times.
This can be used to establish \emph{existence of asymptotic
creation- and annihilation operators}, $a_{\pm}^*$, $a_{\pm}$:
\begin{equation*}
a_{\pm}^{\#} (h_1) \dots a_{\pm}^{\#} (h_n) \ph = \lim_{t\to \pm \infty}
e^{iH_g t} a^{\#} (h_{1,t}) \dots a^{\#} (h_{n,t})e^{-iH_g t} \ph ,
\end{equation*}
for an arbitrary $\ph \in E_{\lambda}(H_g)\H$, $h_j \in
L^2_{\omega}(\R^3) = L^2(\R^3,(1 + |k|^{-1}) d k)$,
$j=1,\dots,n$, $n\in\N$, and \(\lambda+\sum M_i\leq \Sigma\) where
\(\||\nabla\Omega|E_{\Sigma}(H_g)\|<1\). Here $h_t (k) :=
e^{-i|k|t} h(k)$, and $a^{\#} = a$ or $a^*$. See Theorem
\ref{thm:asy_fields}.

We then show that all dressed one-electron wave packets $\psi_f$,
with $\psi_f \in E_{\Sigma}(H_g)\H$, are ``\emph{vacua}'' for the
asymptotic creation-- and annihilation operators, in the sense
that \[ a_{\pm} (h) \psi_f =0 ,\] for arbitrary $h \in
L^2_{\omega} (\R^3)$; see Lemma \ref{lm:asy_vacua}.

(iii) We define the \emph{scattering identification map} $I$ by
\begin{equation*}
\begin{split}
I: \HxF \equiv \H \otimes \F &\longrightarrow \H \\
 \ph \otimes a^* (h_1) \dots a^* (h_n) \Omega &\longmapsto
 a^* (h_1) \dots a^* (h_n) \ph
\end{split}
\end{equation*}
and the \emph{extended Hamiltonian} $\Hex_g$ by \[ \Hex_g = H_g
\otimes 1 + 1 \otimes \dGamma (|k|). \] To say that asymptotic
creation operators exist - under the aforementioned assumptions -
is equivalent to saying that the operators
\begin{equation*}
\tilde{\Omega}_{\pm} \ph = \lim_{t \to \pm\infty} e^{iH_g t} I
e^{-i\Hex_g t } \ph
\end{equation*}
exist for $\ph$ in some dense subspace of $E_{\Sigma}(\Hex)\HxF$. The \emph{M\o ller wave
operators} are then defined by
\[ \Omega_{\pm} = \tilde{\Omega}_{\pm} (\Pgs \otimes 1) , \] where
$\Pgs$ is the orthogonal projection onto the subspace, $\Hgs$, of
$\H$ of dressed one-electron wave packets. Since vectors in $\Hgs$
are vacua for $a_{\pm}^{\#} (h)$, the operators $\Omega_{\pm}$
are \emph{isometric} on $\Hgs \otimes \F$; see Theorem
\ref{thm:Omega_+}.

\emph{Asymptotic completeness} of scattering on states of energy $\leq \Sigma$
can be formulated as the statement that
\begin{equation}\label{eq:asycompl1}
\ran \Omega_{\pm} \supset E_{\Sigma} (H_g) \H .
\end{equation}

In \emph{Sect.} \ref{sec:modham} we introduce a modified
Hamiltonian, $\Hmod$, which agrees with $H_g$, except that the
dispersion law, $|k|$, for soft bosons of momentum $k$ with $|k| <
\sigma$ is replaced by a new dispersion law $\omega (k)$, where
$\omega \in C^{\infty} (\R^3)$, $\omega (k)\geq |k|$, $ \omega(k)
= |k|$, for $|k| \geq \sigma$, and $\omega (k) \geq \sigma /2$,
for all $k$. Since bosons of momentum $k$ with $|k|\leq \sigma$ do
{\em not} interact with the electron, the Hamiltonians \(\Hmod\)
and $H_{g}$ have the same M{\o}ller operators. But since the boson
number operator is bounded by $\Hmod$, it is more convenient to
work with the Hamiltonian \(\Hmod\), instead of $H_g$. (Of course,
this trick does not survive the limit $\sigma \to 0 \,$!) \emph{In
the sections following Section~\ref{sec:modham} we work with $\Hmod$ exclusively and
$H\equiv \Hmod$!}


In \emph{Sect.} \ref{sec:propest} we establish the main propagation estimate for
the bosons. Denoting by $x$ the position of the electron and by $y$ the
Newton--Wigner position at time $t$ of an asymptotically free boson present in a state
of finite total energy, we show that
\begin{equation}\label{eq:improved_pe}
  \frac{1}{t}|J(y/t)\cdot (\nabla\omega(k)-y/t)+ h.c|\, F(|x|/t) \to 0 ,
  \qquad \text{as} \quad t \to \infty ,
\end{equation}
at an integrable rate, if \(J\in\tf(\R^3,\R^3),\, F\in \tf(\R)\)
and \(\supp(J)\subset \{|y|\geq \lambda\}\) while
\(\supp(F)\subset(-\infty,\beta]\) where $\beta<\lambda$. The
gradient, $\nabla\omega$, of $\omega$ is the group velocity of the
bosons. By Eq.~\eqref{eq:improved_pe} the asymptotic velocity of
bosons that escape the electron, is given by their group velocity
$\nabla\omega$.

\begin{figure}
\begin{center}
\epsfig{file=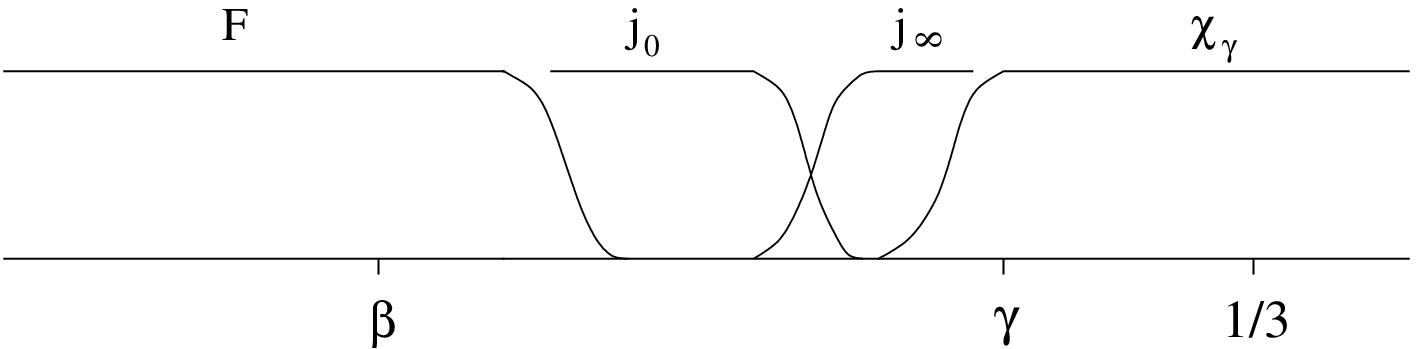,scale=.75}
\end{center}
\caption{Typical choice of the function $\chi_{\gamma}$, of the
electron space cutoff $F$ and of the partition in the photon space
$j_0, j_{\infty}$.\label{fig:beta0}}
\end{figure}

In \emph{Sect.} \ref{sec:observable}, we construct the asymptotic
observable $W$, which plays a crucial role in our proof of
asymptotic completeness. Given $\Sigma$ with
\(\sup_{p}|\nabla\Omega(p)|\chi(\Omega(p)\leq \Sigma)<\beta\) and
$g$ so small that \(\|\nabla\Omega E_{\Sigma}(H_g)\| \leq
\beta<1/3\), we choose $\gamma\in (\beta,1/3)$ and define
$\chi_{\gamma}$ as depicted in Figure~\ref{fig:beta0}. For every energy cutoff
$f$ with $\supp(f)\subset (-\infty,\Sigma]$ we define
\begin{equation}\label{eq:W}
  W= s-\lim_{t\to \infty}e^{iHt} f(H) \dGamma(\chi_{\gamma,t})f(H) e^{-iHt},
\end{equation}
where $\chi_{\gamma,t}$ denotes the operator of multiplication
with $\chi_{\gamma}(|y|/t)$. $W$ measures the number of bosons
that propagate into the region $\{|y|\geq \gamma t\}$ as
$t\to\infty$. They are asymptotically free since $\beta<\gamma$ is
an upper bound on the electron propagation speed by the electron
propagation estimate in Sect.~\ref{sec:waveop}. In fact, thanks to
this propagation estimate, we may add a suitable space cutoff
$F(|x|/t)$ (see Figure \ref{fig:beta0}) in Eq.~\eqref{eq:W} next to $\dGamma(\chi_{\gamma,t})$
without changing the limit, if it exists. For this reason the
propagation estimate \eqref{eq:improved_pe} is sufficient to prove
existence of $W$.

The \emph{key result} of Sect. \ref{sec:observable} is Theorem
\ref{thm:W_pos}, which says that $W$ is positive on the space of
states orthogonal to $\Hgs$, without soft bosons, and with
energies inside the support of the energy cutoff $f$. This
positivity is derived from an estimate of the form
\begin{equation}\label{eq:W-is-positive}
  \sprod{e^{-iH t_n} \psi}{\dGamma(|x-y|/t_n) e^{-iH t_n}\psi}
  \geq (1-\beta-\eps)\|f(H)\psi\|^2 + O(g),
\end{equation}
valid for all vectors $\psi$ with the properties specified above and for
smooth energy cutoffs $f$ with $\supp(f)\subset(-\infty,\Sigma]$
and with \(\||\nabla\Omega|E_{\Sigma}(H_g)\|\leq \beta\). Here
$\{t_n\}$ is a sequence of times, depending on $\eps$ and $\psi$,
with  $t_n\to\infty$ as $n\to \infty$. Some further explanations
are necessary at this point: (i) Soft bosons must be avoided
because, in $H=\Hmod$, their dispersion relation has been
modified. ``Without soft bosons'' means ``in the range of the
projector $\Gamma(\chi_i)$'', where \(\chi_i=\chi_{|k|\geq
\sigma}\). (ii) Inequality \eqref{eq:W-is-positive} would fail for
some \(\psi\in \Pgs^{\perp}\H\) were there \emph{excited}
one-electron states. But this has been excluded in Sect.
\ref{sec:sys}. (iii) The estimate \eqref{eq:W-is-positive} does
not easily translate into positivity of $W$ because in $W$ the
photon position is measured relative to the origin, rather than
relative to $x$. It is due to the assumptions $\beta<1/3$, and a
suitable choice of $\gamma\in(\beta,1/3)$ that
\eqref{eq:W-is-positive} implies positivity of $W$. See the
introduction to Section~\ref{sec:Wpos} for a detailed explanation.

The subject of \emph{Sect.} \ref{sec:inverse} is to show that, on states of
energy $\leq \Sigma$ and for sufficiently small coupling constant $g$ (depending
on the choice of $\Sigma$), the extended wave operator $\tilde{\Omega}_+$ can be
\emph{inverted}. Our proof is based on the construction of a \emph{Deift-Simon
wave operator} $W_+$ with the properties
\begin{equation*}
 W = \tilde{\Omega}_+ W_+ , \quad\text{and} \quad (1\otimes P_{\Omega})W_{+}=0.
\end{equation*}
In order to construct the operator
$W_+$, we have to split an arbitrary configuration of bosons into one staying
close to the electron and a configuration of bosons escaping ballistically from
the ``localization cone'' of the electron. This is accomplished by decomposing the space,
$\h = L^2 (\R^3 , dk)$, of one-boson wave functions into a direct sum of two
subspaces,
\begin{equation*}
  j_t : \h \ni h \longmapsto \left(j_{0,t} h , j_{\infty, t}h\right) \in \h\oplus \h ,
\end{equation*}
where $j_0$ and $j_{\infty}$ are $C^{\infty}$-functions on
$\R_{+}$ with $j_0+j_{\infty}\equiv 1$ and graphs as depicted in
Figure~\ref{fig:beta0}, and $j_{0,t}$, $j_{\infty ,t}$ are defined by
\begin{equation*}
  j_{\sharp, t} (y) := j_{\sharp} (y/t).
\end{equation*}
The operator $\uGamma (j_{t})$ is the second quantization
of the operator $j_{t}$. It maps the physical Hilbert space $\H = L^2 (\R^3,
d x) \otimes \F$ into the extended Hilbert space
\(\HxF = L^2 (\R^3, d x) \otimes \F \otimes \F\), and \[ I \uGamma (j_{t}) =1 .\]

The Deift-Simon wave operator $W_+$ is a linear operator from
$\H$ to $\HxF$ defined by
\begin{equation*}
 W_{+}  = s-\lim_{t\to\infty} e^{i\Hex t} f(\Hex)
 \uGamma(j_t) \dGamma(\chi_{\gamma,t}) f(H) e^{-iHt},
\end{equation*}
where \(\Hex = H \otimes 1 + 1 \otimes \dGamma (\omega)\) is the extended
modified Hamiltonian. The results of Section \ref{sec:inverse} are summarized in Theorem \ref{thm:W+exists}.

In \emph{Sect.} \ref{sec:AC}, our proof of \emph{asymptotic
completeness for Compton scattering} is completed. In order to
prove Eq.~\eqref{eq:asycompl1} we use an inductive argument, the
induction being in the number of bosons present in a scattering
state. Let $m := \sigma /2 >0$, where $\sigma$ is the infrared
cutoff, and let $n$ be an arbitrary positive integer. Our induction hypothesis is that
\begin{equation*}
  \ran \Omega_+ \supset E_{(-\infty,\Sigma-nm)} (H) \H ,
\end{equation*}
and our claim is that
\( \ran \Omega_+ \supset E_{(-\infty,\Sigma-(n-1)m)} (H) \H\). From
the definition of $\Omega_+$ it is clear that the dressed
one-electron wave packets are contained in $\ran \Omega_+$, and, thanks to the
infrared cutoff, so are all states which differ from a given vector in
\(\Ran \Omega_{+}\) only by soft bosons.
Since, moreover, $\ran \Omega_+$ is closed, it is enough to show
that
\begin{equation*}
  \ran \Omega_+ \supset \Pgs^{\perp} \Gamma
  (\chi_i) E_{\Delta} (H_g) \H,
\end{equation*}
where $\Delta $ is an arbitrary compact subinterval of $(-\infty, \Sigma -(n-1)m)$,
$\Pgs^{\perp}= 1 - \Pgs$, and $\Pgs$ is the orthogonal
projection onto the subspace, $\Hgs$, of $\H$ of dressed
one-electron wave packets. Let $\psi \in \Pgs^{\perp}\Gamma
(\chi_i) E_{\Delta}(H)\H$. Since our asymptotic observable $W$ is strictly positive on
this space, there exists a vector $\ph=\Pgs^{\perp}\Gamma
(\chi_i) E_{\Delta}(H)\ph$ with \(\psi = \Gamma(\chi_i) \Pgs^{\perp} W \ph\).
By Theorem~\ref{thm:W+exists},
\begin{equation*}
  W\ph = \tilde \Omega_{+} W_{+}\ph = \tilde \Omega_{+} (1\otimes
  P_{\Omega}^{\perp}) W_{+}\ph.
\end{equation*}
Next, by the intertwining property of $W_{+}$ and since \(\ph \in
E_{\Sigma-(n-1)m}(H)\),
\begin{equation*}
   W_{+}\ph \in E_{\Sigma-(n-1)m}(\Hex)\HxF.
\end{equation*}
Hence,
\begin{equation*}
  (1 \otimes P_{\Omega}^{\perp} ) W_+ \ph = (E_{\Sigma-nm} (H)
\otimes P_{\Omega}^{\perp}) W_+ \ph ,
\end{equation*}
because $\Hex = H \otimes 1 + 1 \otimes \dGamma (\omega)$, and
$\dGamma (\omega) \restricted{\ran P_{\Omega}^{\perp} \geq m}$. By
our induction hypothesis, $(E_{(-\infty,\Sigma - nm)} (H_g)
\otimes P_{\Omega}^{\perp}) W_+ \ph$ can be approximated, with a
norm error of less than $\eps$, by vectors of the form
\[ \sum_i (\Omega_+ \chi^{(i)} ) \otimes \ph^{(i)}, \] where
$\eps>0$ is arbitrarily small, $\ph^{(i)} \in \F$ is orthogonal to
$\Omega$, and $\chi^{(i)} \in \HxF$. Our results in Sect.
\ref{sec:waveop} (see Lemma \ref{lm:wo_prop}, and proof thereof)
then show that
\begin{equation*}
  \lim_{t \to \infty} e^{iH_g t} I e^{-i \Hex_g t} \left( \sum_i
  \Omega_+ \chi^{(i)} \otimes \ph^{(i)}\right)
\end{equation*}
exists and belongs to the range of $\Omega_+$. Some technical
details may be found in the proofs of Lemma \ref{lm:induction} and
Theorem \ref{thm:AC} of Sect. \ref{sec:AC}.

At present, we do not see how to remove the infrared cutoff
$\sigma$ in the proofs of our results of Sect. \ref{sec:propest},
\ref{sec:observable} and \ref{sec:inverse}. However, it is
possible to construct scattering states and wave operators in the
limit $\sigma \to 0$. Elaborating on a proposal in \cite{Froe1},
this has recently been shown by Pizzo in a remarkable paper
\cite{P}.

A more unpleasant assumptions in our work is the energy bound
$\Sigma < O_{\beta=1/3}$, forcing the electron speed to be less
than one third of the speed of light. One would expect asymptotic
completeness to hold true under the assumption
$\Sigma<O_{\beta=1}$, which is suffices for the existence of the
wave operator. The need for $\Sigma < O_{\beta=1/3}$ is due to a
lack of Lorentz invariance; the speed of light is \emph{not
independent} of the frame of reference (see
Section~\ref{sec:Wpos}). This problem can be avoided by defining
all observables relative to the electron position $x$, rather than
relative to the origin, but then one runs into serious technical
problems with non-$H$-bounded commutators.


In Sect. \ref{sec:out}, we conclude with an outlook.

Some technical details are discussed in several appendices.\\

\emph{Acknowledgements.} We thank V. Bach for his hospitality at
the University of Mainz, where part of this work was done, and we
are indebted to Gian Michele Graf for pointing out a serious gap
in an earlier version of this paper.

\section{Fock Space and Second Quantization}
\label{sec:Fock}

Let $\h$ be a complex Hilbert space, and let $\otimes_s^n\h$ denote the
$n$-fold symmetric tensor product of $\h$. Then the bosonic Fock space over $\h$
\[  \F=\F(\h)=\oplus_{n\geq0}\otimes_s^n\h \]
is the space of sequences \(\ph=(\ph_n)_{n\geq 0}\),
with $\ph_0\in \C$, \(\ph_n\in \otimes_s^n\h\), and with the scalar product
given by
\[ \sprod{\ph}{\psi} := \sum_{n\geq 0} ( \ph_n  , \psi_n ), \]
where \(( \ph_n,\psi_n)\) denotes the inner product in
\(\otimes^n_s \h\). The vector \(\Omega=(1,0,\ldots)\in\F\) is
called the vacuum. By $\F_0\subset\F$ we denote the dense
subspace of vectors $\ph$ for which $\ph_n=0$, for all but
finitely many $n$. The number operator $N$ is defined by
\((N\ph)_n=n\ph_n\).

\subsection{Creation- and Annihilation Operators}

The creation operator $a^*(h)$, $h\in\h$, on $\F$ is defined by
\[ a^*(h)\ph = \sqrt{n}\,S(h\otimes \ph),\hspace{3em}\mbox{for}\
\ph\in\otimes_s^{n-1}\h ,\] and extended by linearity to $\F_0$.
Here \(S\in \L(\otimes^n\h)\) denotes the orthogonal projection
onto the symmetric subspace \(\otimes_s^n\h\subset \otimes^n\h\). The
annihilation operator $a(h)$ is the adjoint of $a^*(h)$
restricted to $\F_0$. Creation- and annihilation operators
satisfy the canonical commutation relations (CCR)
\begin{equation*}
[a(g),a^{*}(h)] = (g,h),\hspace{3em}
[a^{\#}(g),a^{\#}(h)] =0.
\end{equation*}
In particular, \([a(h),a^{*}(h)] = \|h\|^2\), which implies that
the graph norms associated with the closable operators $a(h)$ and
$a^{*}(h)$ are equivalent. It follows that the closures of $a(h)$
and $a^{*}(h)$ have the same domain. On this common domain we
define the self-adjoint operator
\begin{equation}\label{eq:phi}
\phi(h) = \frac{1}{\sqrt{2}}(a(h)+a^{*}(h)).
\end{equation}
The creation- and annihilation operators, and thus $\phi(h)$, are
bounded relative to the square root of the number operator:
\begin{equation}\label{eq:aN_bound}
\|a^{\#}(h)(N+1)^{-1/2}\| \leq \|h\|
\end{equation}
More generally, for any $p\in\R$ and any integer $n$,
\begin{equation*}
\|(N+1)^pa^{\#}(h_1)\ldots a^{\#}(h_n)(N+1)^{-p-n/2}\| \leq
C_{n,p}\,\|h_1\|\cdot\ldots\cdot\|h_n\|.
\end{equation*}

\subsection{The Functor $\Gamma$}

Let $\h_1$ and $\h_2$ be two Hilbert spaces and let
\(b\in\L(\h_1,\h_2)\). We define $\Gamma(b)\ :\ \F(\h_1)\rightarrow\F(\h_2)$ by
\begin{equation*}
\Gamma(b)\restricted\otimes_s^n \h_{1} = b\otimes\ldots\otimes b.
\end{equation*}
In general $\Gamma(b)$ is unbounded; but if $\|b\|\leq 1$ then
\(\|\Gamma(b)\|\leq 1\). From the definition of $a^{*}(h)$ it
easily follows that
\begin{alignat}{2}
\Gamma(b)a^{*}(h) & = a^{*}(bh)\Gamma(b),\qquad  & h\in\h_1&
\label{geq1}\\ \Gamma(b)a(b^*h) &= a(h)\Gamma(b),   &
h\in\h_2&.\label{geq2}
\end{alignat}
If $b^{*}b=1$ on $\h_1$ then these equations imply
that
\begin{alignat}{2}
\Gamma(b)a(h) & = a(bh)\Gamma(b)\qquad & h\in\h_1 \label{geq3}&\\
\Gamma(b)\phi(h) & = \phi(bh)\Gamma(b) & h\in\h_1 \label{geq4}&.
\end{alignat}


\subsection{The Operator $\dGamma(b)$}

Let $b$ be an operator on $\h$. Then $\dGamma(b)\ :\ \F(\h)\rightarrow\F(\h)$ is
defined by
\begin{equation*}
\dGamma(b)\restricted\otimes_s^n\h =\sum_{i=1}^n(1\otimes\ldots b\otimes\ldots
1).
\end{equation*}
For example $N=\dGamma(1)$. From the definition of $a^{*}(h)$ we
get
\begin{equation*}
[\dGamma(b),a^*(h)] =  a^*(bh) \quad [\dGamma(b),a(h)] =  -a(b^*h),
\end{equation*}
and, if $b=b^{*}$,
\begin{equation}\label{eq:dgamma-phi}
i[\dGamma(b),\phi(h)]  = \phi(ibh).
\end{equation}
Note that \(\|\dGamma(b)(N+1)^{-1}\|\leq  \|b\|\).


\subsection{The Operator $\dGamma(a,b)$}\label{sec:dgamma(ab)}

Suppose $a,b\in \L(\h_1,\h_2)$. Then we define $\dGamma(a,b)\ :\
\F(\h_1)\rightarrow\F(\h_2)$ by
\begin{equation*}
\dGamma(a,b)\restricted\otimes_s^n\h =\sum_{j=1}^n
(\underbrace{a\otimes\ldots a}_{j-1}\otimes b\otimes
\underbrace{a\otimes\ldots a}_{n-j}).
\end{equation*}
For $a,b\in \L(\h)$ this definition is motivated by
\begin{equation*}
\Gamma(a)\dGamma(b) = \dGamma(a,ab),\makebox[5em]{and}
[\Gamma(a),\dGamma(b)] = \dGamma(a,[a,b]).
\end{equation*}
If $\|a\|\leq 1$ then \(\|\dGamma(a,b)(N+1)^{-1}\|\leq \|b\|\) and
\begin{equation}\label{eq:dG_bound}
\| N^{-1/2} \dGamma (a,b) \psi \| \leq \| \dGamma (b^* b)^{1/2} \psi \|.
\end{equation}

\begin{lemma}\label{lm:dGamma}
Suppose \(r_1:\h_1\to \h_2\), \(r_2^*:\h_2\to \h_3\) and \(q:\h_1\to \h_3\)
are linear operators and $\|q\|\leq 1$. Then
\begin{equation*}
   |\sprod{u}{\dGamma(q,r_2^* r_1)v}| \leq \sprod{u}{\dGamma(r_2^* r_2)u}^{1/2}\sprod{v}{\dGamma(r_1^*r_1)v}^{1/2}
\end{equation*}
for all $u\in \F(\h_3)$ and all $v\in \F(\h_1)$.
\end{lemma}

\begin{proof}
By definition of the inner product, of \(\dGamma(q,r_2^* r_1)\), and by assumption on $q$,
\begin{eqnarray*}
  |\sprod{u}{\dGamma(q,r_2^* r_1)v}|
  &=&  \Bigg| \sum_{n\geq 0} \sum_{j=1}^n
  \sprod{u_n}{(q\otimes\ldots \underbrace{r_2^* r_1}_{jth}\otimes \ldots q)v_n}\Bigg|\\
  &\leq & \sum_{n\geq 0} \sum_{j=1}^n\|(r_2)_ju_n \| \|(r_1)_j v_n \|
\end{eqnarray*}
where \((r_{\sharp})_j= 1\otimes \ldots r_{\sharp}\otimes\ldots 1\),
$r_{\sharp}$ in the $j$th factor.
The assertion now follows by the Schwarz inequality.
\end{proof}


\subsection{The Tensor Product of two Fock Spaces}
\label{sec:FxF}
Let $\h_1$ and $\h_2$ be two Hilbert spaces. We define a linear operator
\(U:\F(\h_1\oplus\h_2)\rightarrow\F(\h_1)\otimes\F(\h_2)\) by
\begin{equation}\label{ueq0}
\begin{split}
U\Omega &= \Omega\otimes\Omega\\
U a^*(h) &= [a^*(h_{(0)})\otimes 1+ 1\otimes
a^*(h_{(\infty)})]U\hspace{3em}\text{for
}h=(h_{(0)},h_{(\infty)})\in\h_1\oplus\h_2.
\end{split}
\end{equation}
This defines $U$ on finite linear combinations of vectors of the
form \(a^*(h_1)\ldots a^*(h_n)\Omega\). From the CCRs it follows
that $U$ is isometric. Its closure is isometric and onto, hence
unitary. It follows that
\begin{equation}\label{ueq1}
U a(h) = [a(h_{(0)})\otimes 1+ 1\otimes a(h_{(\infty)})]U.
\end{equation}
Furthermore we note that
\begin{equation}\label{ueq3}
U\dGamma(b) = [\dGamma(b_0)\otimes 1+ 1\otimes \dGamma(b_{\infty})]U \makebox[3em]{if}
b=\left(\begin{array}{cc} b_0 & 0 \\ 0 & b_{\infty} \end{array}\right) .
\end{equation}
For example \(UN= (N_0+N_{\infty})U\) where \( N_{0}=N\otimes 1\)
and \(N_{\infty}=1\otimes N\).

Let \(\F_n=\otimes_s^n\h\) and let $P_n$ be the projection from \(\F=\oplus_{n\geq
0}\F_n\) onto $\F_n$. Then the tensor product $\F\otimes\F$
is norm-isomorphic to \(\oplus_{n\geq 0}\oplus_{k=0}^n
\F_{n-k}\otimes\F_k\), the corresponding isomorphism being given
by \(\ph \mapsto (\ph_{n,k})_{n\geq 0,\; k=0..n}\) where $\ph_{n,k}=(P_{n-k}\otimes
P_k)\ph$. In this representation of
$\F\otimes\F$ and with $p_i(h_{(0)},h_{(\infty)})=h_{(i)}$, $U$ becomes
\begin{equation}\label{ueq2}
 U\restricted \otimes_s^n(\h\oplus\h)=
\sum_{k=0}^n\binom{n}{k}^{1/2}\underbrace{p_0\otimes\ldots\otimes
p_0}_{n-k\ \text{factors}}\otimes \underbrace{p_{\infty}\otimes
\ldots\otimes p_{\infty}}_{k\ \text{factors}}.
\end{equation}

\subsection{Factorizing Fock Space in a Tensor Product}
\label{sec:factfock}

Suppose $j_0$ and $j_{\infty}$ are linear operators on $\h$ and
\(j:\h\rightarrow\h\oplus\h\) is defined by
\(jh=(j_0h,j_{\infty}h),\ h\in\h\). Then
\(j^*(h_1,h_2)=j_0^*h_1+j_{\infty}^*h_2\) and consequently
\(j^*j=j_0^*j_0+j_{\infty}^*j_{\infty}\). On the level of Fock
spaces, \(\Gamma(j):\F(\h)\rightarrow\F(\h\oplus\h)\), and
we define
\[ \uGamma(j)=U\Gamma(j):\F\rightarrow\F\otimes\F.\]
It follows that \(\uGamma(j)^*\uGamma(j)=\Gamma(j^*j)\) which is
the identity if $j^*j=1$. Henceforth $j^*j=1$ is tacitly assumed
in this subsection. From \eqref{geq1} through \eqref{geq4},
\eqref{ueq0} and \eqref{ueq1} it follows that
\begin{align}\label{eq:ugamma-a}
\uGamma(j)a^{\#}(h) & = [a^{\#}(j_0h)\otimes 1+ 1\otimes
a^{\#}(j_{\infty}h)]\uGamma(j)\\
\label{eq:ugamma-phi}\uGamma(j)\phi(h) & = [\phi(j_0 h)\otimes 1+
1\otimes \phi(j_{\infty}h)]\uGamma(j).
\end{align}
Furthermore, if $\underline{\omega}=\omega\oplus\omega$ on
$\h\oplus\h$, then by \eqref{ueq3}
\begin{equation}\label{eq:ugamma-o}
\begin{split}
\uGamma(j)\dGamma(\omega) &= U\Gamma(j)\dGamma(\omega)
= U\dGamma(\underline{\omega})\Gamma(j)- U\dGamma(j,\underline{\omega}\, j-j\omega)\\
&= [\dGamma(\omega)\otimes 1+ 1\otimes\dGamma(\omega)]\uGamma(j) -
\udGamma(j,\underline{\omega}\, j-j\omega)
\end{split}
\end{equation}
where the notation \(\udGamma(a,b)=U\dGamma(a,b)\) is introduced.
In particular \(\uGamma(j)N = (N_0+N_{\infty})\uGamma(j)\).
We remark that, by \eqref{ueq2},
\begin{equation}\label{eq:ugamma|n}
\uGamma(j)\restricted \otimes_s^n \h =
\sum_{k=0}^n\binom{n}{k}^{1/2} \underbrace{j_0\otimes
\ldots\otimes j_0}_{n-k\ \text{factors}}
\otimes\underbrace{j_{\infty}\otimes\ldots\otimes j_{\infty}}_{k\
\text{factors}}.
\end{equation}

\begin{lemma}\label{lm:udGamma}
If \(j,k: \h\to \h\oplus \h\), $j^* j\leq 1$, and $k_0,\ k_{\infty}$ are self-adjoint, then
\begin{eqnarray*}
  |\sprod{u}{\udGamma(j,k)v}|
  &\leq &  \sprod{u}{\left(\dGamma(|k_{0}|)\otimes 1\right) u}^{1/2} \sprod{v}{\dGamma(|k_{0}|)v}^{1/2}\\
  & & + \sprod{u}{\left(1\otimes \dGamma(|k_{\infty}|)\right)u}^{1/2} \sprod{v}{\dGamma(|k_{\infty}|)v}^{1/2}
\end{eqnarray*}
for all $u\in \F\otimes \F$ and all $v\in \F$.
\end{lemma}

\begin{proof}
Write \(\sprod{u}{\udGamma(j,k)v} = \sprod{U^* u}{\dGamma(j,k^{(0)})v} +
\sprod{U^* u}{\dGamma(j,k^{(\infty)})v}\)
where \(k^{(0)}=(k_0,0)\) and \(k^{(\infty)}=(0,k_{\infty})\).
Then apply Lemma~\ref{lm:dGamma} to both terms. In the first term we
choose $r_2=(|k_0|^{1/2},0)$ and \(r_1=|k_0|^{1/2}\text{sgn}(k_0)\).
\end{proof}

\subsection{The "Scattering Identification"}

An important role will be played by the scattering identification
\( I:\F\otimes\F\to \F\) defined by
\begin{align*}
I (\ph\otimes\Omega) &= \ph\\ I \ph\otimes a^*(h_1)\cdots
a^*(h_n)\Omega &= a^*(h_1)\cdots a^*(h_n)\ph, \hspace{3em}\ph\in\F_0,
\end{align*}
and extended by linearity to $\F_0\otimes\F_0$. (Note that this
definition is symmetric with respect to the two factors in the
tensor product.) There is a second characterization of $I$ which
will often be used. Let $\iota:\h\otimes\h\to \h$ be defined by
$\iota(h_{(0)},h_{(\infty)})=h_{(0)}+h_{(\infty)}$. Then \(
I=\Gamma(\iota)U^*\), with $U$ as above.
Since $\|\iota\|=\sqrt{2}$, the operator $I$ is unbounded.

\begin{lemma}\label{lm:I_bound}
For each positive integer $k$, the operator \(I (N+1)^{-k}\otimes
\chi(N\leq k)\) is bounded.
\end{lemma}

Let \(j:\h\to \h\oplus\h\) be defined by $jh=(j_{0}h,j_{\infty}h)$
where $j_0,j_{\infty}\in \L(\h)$. If $j_0+j_{\infty}=1$, then
$\uGamma(j)$ is a right inverse of $I$, that is,
\begin{equation}\label{eq:Igamma}
I \uGamma(j) = 1.
\end{equation}
Indeed \(I\uGamma(j) = \Gamma(\iota)U^*U\Gamma(j) =\Gamma(\iota
j)= \Gamma(1)=1\).


\section{The Model, Dressed One-Electron States, and Bounds on the Interaction} \label{sec:sys}

In this section we describe our model in precise mathematical
terms and discuss its main properties. The main new result of this
section is Theorem~\ref{cor:gs-only}.

\subsection{The Model}\label{sec:model}

The Hamilton operator of the system described in the introduction
is defined by
\begin{equation}\label{eq:ham}
H_g = \Omega(p)\otimes 1 + 1 \otimes \dGamma(|k|) + g\phi(G_x)
\end{equation}
acting on the Hilbert space \(\H = L^2 (\R^3,dx) \otimes \F\),
where $\F$ is the bosonic Fock space over $L^2(\R^3 ,dk)$. Here
and henceforth $x\in \R^3$ denotes the position of the electron,
$k$ is the momentum of a boson and $p=-i\nabla_x$. In this paper
we are interested in both, relativistic electrons with
$\Omega(p)=\sqrt{p^2+M^2}$ and non-relativistic ones,
$\Omega(p)=p^2/2M$. Rather than treating these two cases
separately, we formulate a set of assumptions that are satisfied
in both cases.
\begin{quote}
{\bf Hypothesis 0.} $\Omega\in C^2(\R^3)$, $\Omega\geq 0$, and the
functions \(|\nabla\Omega|(\Omega+1)^{-1/2}\) and
\(\partial^2\Omega\) are bounded.
\end{quote}
The boundedness of \(|\nabla\Omega|(\Omega+1)^{-1/2}\) ensures
that $|\nabla\Omega|^2$ is $H_g$-bounded.

The coupling function $G_x(k)$ has the form
\[ G_x (k) = e^{-ik\cdot x} \kappa_{\sigma} (k)\]
with an infrared (IR) cutoff imposed on the form factor
$\kappa_{\sigma}$. Specifically, we assume that
\begin{quote}
{\bf Hypothesis 1.}  $\kappa_{\sigma} (k) = \kappa (k) \chi (|k|
/\sigma)$, for some $\sigma >0$. Here \(\kappa \in
\mathcal{C}_0^{\infty} (\R^3)\), $\kappa\geq 0$, and $\chi \in
C^{\infty} (\R,[0,1])$ with $\chi (s) = 0$ if $s \leq 1$ and
$\chi(s) =1 $ if $s \geq 2$.
\end{quote}
The fact that \(\int|\kappa_{\sigma}(k)|^2/|k|\, dk\leq
\int|\kappa(k)|^2/|k|\, dk<\infty\) for all $\sigma$ guarantees
that the smallness assumptions on $|g|$ in
Theorems~\ref{cor:gs-only} and \ref{thm:ACph} are independent of
$\sigma$. Incidentally, we put $\kappa_{\sigma =0} (k) = \kappa
(k)$ (this is used in Sect. \ref{sec:waveop} where most of the
results also hold without infrared cutoff). The assumption $\kappa
\geq 0$ in Hypothesis 1 is included for convenience. It allows us
to give a simple proof of Lemma~\ref{lm:unique} in Appendix
\ref{app:DES}, but it is otherwise not needed; (see the remark
after Theorem~\ref{cor:gs-only}).

By Lemma~\ref{lm:estim} below, the operator $\phi(G_x)$ is bounded
relative to \((\dGamma(|k|)+1)^{1/2}\) and thus also relative to
\((H_{g=0}+1)^{1/2}\). It follows that $\phi(G_x)$ is
infinitesimal w.r.~to $H_0=H_{g=0}$, and thus the operator $H_g$
is self-adjoint on $D(H_0)$ and bounded from below. Our main
results hold on spectral subspaces $E_{\Sigma}(H_g)\H$, where
$\||\nabla\Omega|E_{\Sigma}(H_g)\|\leq \beta$ for some
$\beta<1/3$. This bound can be derived from the following further
assumption on $\Omega$; (see Lemma~\ref{lm:grad-Omega2}).
\begin{quote}
{\bf Hypothesis 2.} For each $\beta>0$, there exists a constant
$O_{\beta}>\inf_{p}\Omega(p)$ such that
\[  |\nabla\Omega(p)| \leq \beta \qquad \text{for all}\ p\
\text{with}\ \Omega(p)\leq O_{\beta}.\]
\end{quote}
By lowering the values of $O_{\beta}$ we may achieve that
$\beta\mapsto O_{\beta}$ is non-decreasing and continuous from the
left. Under these assumptions, for each $\Sigma<O_{\beta}$, there
exists a $\beta'<\beta$ such that $\Sigma<O_{\beta'}<O_{\beta}$. A
function $O_{\beta}$ with these properties can also be defined by
\(O_{\beta}:=\sup\{\lambda:f(\lambda)<\beta\}\) where
\(f(\lambda):=\sup\{|\nabla\Omega(P)|: \Omega(P)<\lambda\}\).
Given Hypothesis 0, Hypothesis 2 is then equivalent to
$f(\lambda)\to 0$ as \(\lambda\to\inf\Omega(p)\).

An important consequence of Hypothesis~2 is that
\begin{equation}\label{eq:hyp2low}
\Omega(p-k) \geq \Omega(p)- \beta|k|,\qquad\text{if}\
\Omega(p)\leq O_{\beta},
\end{equation}
which is obvious from a sketch of the graph of a generic function
$\Omega$ satisfying Hypothesis 2.

As mentioned in the introduction, the number operator $N$ is not
bounded relative to $H_g$. However, by Hypothesis 1, an
interacting boson has a minimal energy $\sigma>0$ and thus the
number of interacting bosons is bounded w.r.~to the total energy,
while the number of soft bosons with energy below $\sigma$ is
conserved under the time evolution. To split the soft bosons from
the interacting ones, we use that $L^2(\R^3)= L^2 (\{k:|k| >
\sigma\}) \oplus L^2 (\{k:|k|\leq \sigma \})$ and thus that $\F$
is isomorphic to $\F_i\otimes\F_s$, where $\F_i$ and $\F_s$ are
the Fock spaces over $L^2(|k|> \sigma)$ and $L^2(|k|\leq\sigma)$,
respectively. Let $\chi_{i}$ denote the characteristic function of
the set \(\{k:|k|> \sigma\}\). Then the isomorphism
\(U:\F\to\F_i\otimes\F_s\) is given by
\begin{equation}\label{eq:U}
\begin{split}
U \, \Omega &= \Omega_i \otimes \Omega_s \\
U \, a^* (h) &= (a^* (\chi_i \, h) \otimes 1 + 1 \otimes a^* ((1-
\chi_i) h))\, U
\end{split}
\end{equation}
We also use the symbol $U$ to denote the operator $1_{L^2 (\R^3,
dx)}\otimes U: \H \to \H_i \otimes \F_s$, where \(\H_i=L^2 ( \R^3
, dx) \otimes \F_i\). On the Hilbert space $\H_i\otimes \F_s$ the
Hamiltonian is represented by
\begin{equation*}
\begin{split}
U H_g U^* &= H_i \otimes 1 + 1 \otimes \dGamma (|k|) \quad \text{with} \\
H_i &= \Omega(p) + \dGamma (|k|) + g \phi (G_x),
\end{split}
\end{equation*}
and the projector $\Gamma(\chi_i)$ onto the subspace of
interacting bosons becomes
\begin{equation*}
U \Gamma (\chi_i) U^* = 1 \otimes P_{\Omega_s},
\end{equation*}
where $P_{\Omega_s}$ is the orthogonal projection onto the vacuum
vector $\Omega_s \in \F_s$.

The Hamiltonian $H_g$ commutes with translations generated by the
total momentum \(P=p+\dGamma(k)\). It is therefore convenient to
describe $H_g$ in a representation of $\H$ in which the operator
$P$ is diagonal. To this end, we define the unitary map \(\Pi:\H
\to L^2(\R^3_P;\F)\), where $L^2(\R^3_P;\F) \equiv \int^{\oplus}
dP \, \F$ is the space of $L^2$-functions with values in $\F$. For
$\ph = \{ \ph_n (x,k_1 , \dots, k_n)\}_{n\geq 0} \in \H$ we define
$\Pi \ph \in L^2(\R^3_P;\F)$ by
\begin{equation*}
(\Pi\ph)_n(P,k_1,\ldots,k_n) = \hat{\ph}_n (P-\sum_{i=1}^n k_i ,
k_1, \ldots , k_n)
\end{equation*}
where
\begin{equation*}
\hat{\ph}_n(p,k_1,\ldots,k_n) = (2\pi)^{-3/2}\int e^{-ip\cdot
  x}\ph_n(x,k_1,\ldots,k_n) d^3x.
\end{equation*}
On \(L^2(\R^3_P;\F)\) the Hamiltonian $H_g$ is given by
\begin{equation*}
\begin{split}
(\Pi H_g \Pi^* \psi )(P) &= H_g (P) \psi (P), \quad \text{where} \\
H_g (P) &= \Omega (P- \dGamma (k)) + \dGamma (|k|) + g
\phi(\kappa_{\sigma}) .
\end{split}
\end{equation*}

\subsection{Dressed One-Electron States}\label{sec:DES}

Next we describe sufficient conditions for
$E_{g}(P)=\inf\sigma(H_g(P))$ to be an eigenvalue of $H_g(P)$.

If $g=0$ then clearly the vacuum vector is an eigenvector of
$H_{g=0}(P)$ and $\Omega(P)$ is its energy. Furthermore, if
$\Omega(P)\leq O_{\beta=1}$ then \(\Omega(P-k)+|k|\geq \Omega(P)\)
and hence
\[ \Omega(P)=\inf\sigma(H_{g=0}(P))=E_0(P). \]
At least for small $g$ and $\Omega(P)<O_{\beta=1}$, we expect that
\(\inf\sigma(H_{g}(P))\) remains an eigenvalue, and this is what
we prove below.

If $|\nabla\Omega(P)|>1$, however, then $\Omega(P)> E_0(P)$, and
the eigenvalue $\Omega(P)$ of $H_{g=0}(P)$ is expected to
disappear when the interaction is turned on.

\begin{theorem}\label{thm:groundstate}
Assume Hypotheses 0--2 are satisfied. Let $H_g (P)$ be defined as
above and let $E_g(P):=\inf\sigma(H_g(P))$. For every
$\Sigma<O_{\beta=1}$ there exists a constant $g_{\Sigma}>0$ such
that, for $|g|<g_{\Sigma}$ and $E_{g}(P)\leq \Sigma$,
\begin{itemize}
\item[(i)] $E_g (P)$ is a simple
eigenvalue of $H_g(P)$.
\item[(ii)] The (unique) ground state of $H_g(P)$ belongs
to $\ran \Gamma (\chi_i)$.
\end{itemize}
\end{theorem}

\begin{proof} It suffices to combine results proven
in Appendix~\ref{app:DES} to conclude Theorem
\ref{thm:groundstate}. (i) By Hypothesis 2 and the remarks
thereafter, there exists a $\beta<1$ such that $\Sigma<
O_{\beta}<O_1$. By Theorem~\ref{thm:gs}~(i), $E_{g}(P)$ is an
eigenvalue of $H_g(P)$ if $\Omega(P)\leq O_{\beta}$ and
$|g|<g_{\beta}$. By Lemma~\ref{lm:Omega-E}, the former assumption
is satisfied if $E_{g}(P)\leq \Sigma$ and $|g|\leq
(O_{\beta}-\Sigma)/(O_{\beta}+C)$. Hence (i) holds for
\(g_{\Sigma}:=\min(g_{\beta}, (O_{\beta}-\Sigma)/(O_{\beta}+C))\).

The uniqueness follows from Lemma~\ref{lm:unique}, and part (ii)
of Theorem \ref{thm:groundstate} from Theorem~\ref{thm:gs}, part
(ii).
\end{proof}

\emph{Remark.} If $\Omega(p)=\sqrt{p^2+M^2}$ then $H_{g}(P)$ has a
unique ground state for all values of $g\in\R,\ \sigma>0$ and all
$P\in \R^3$. An analogous result for $\Omega(p)=p^2/(2M)$ holds at
least for all $P\in \R^3$ with $|P|\leq(\sqrt{3}-1)/M$,
\cite{Froe2}.

In the following we denote by $\psi_P \in \F$ the (up to a phase)
unique ground state vector of $H_g (P)$ provided by
Theorem~\ref{thm:groundstate}. The space of dressed one-electron
wave packets \(\Hgs \subset \H\) is defined by
\begin{equation*}
\Pi \Hgs = \{\psi\in L^2(\{P : E_g (P) \leq \Sigma \} ; \F)|
\psi(P)\in \langle \psi_P \rangle \}
\end{equation*}
where $\langle \psi_{P} \rangle$ is the one-dimensional space
spanned by the vector $\psi_P$; $\Hgs$ is a closed linear subspace
which reduces $H_g$ in the sense that $H_g$ commutes with the
projection $\Pgs$ onto $\Hgs$. The latter is obvious from
 \(( \Pi \Pgs \Pi^* \ph )(P) = P_{\psi_P} \ph (P)\).

\subsection{Positive Commutator and Absence of Excited States}
\label{sec:poscomm}

The purpose of this section is to prove the absence of excited
eigenvalues of $H_g (P)$ below a given threshold $\Sigma$ if $g$
is small enough, depending on $\Sigma$. As usual this is done by
combining a positive commutator estimate with a virial theorem. A
priori we only have a virial theorem on $\ran \Gamma (\chi_i)$,
and therefore we only get absence of excited eigenvalues for
$H_g(P)$ {\em restricted to} $\ran\Gamma(\chi_i)$ in a first step.
- Recall that $\chi_i(k)$ is the characteristic function of the
set $\{ k \in \R^3 : |k| > \sigma \}$, where $\sigma >0$ is the
infrared cutoff defined in Hypothesis 1, and hence that $\Gamma
(\chi_i)$ is the orthogonal projection onto the subspace of
interacting bosons. - Thanks to the IR cutoff, however, this fact
then allows us to show that \(H_g(P)\restricted
\Gamma(\chi_i)^{\perp}\) has no eigenvalues, at all, below
$\Sigma$, and the desired result follows.

The conjugate operator we use is $A = \dGamma (a)$ where \[ a =
\frac{1}{2} \left( \frac{k}{|k|} \cdot y + y \cdot
\frac{k}{|k|}\right). \] On a suitable dense subspace of $\F$
\begin{equation} \label{eq:iHA}
[iH_g (P) , A ] = N - \nabla \Omega (P-\dGamma (k))\cdot \dGamma
(k / |k|) - g \, \phi ( ia \kappa_{\sigma}) .
\end{equation}
We use this identity to \emph{define} the quadratic form
$\sprod{\ph}{[iH_g (P),A]\ph}$ on $D(H_{g}(P))\cap D(N)$.

\begin{theorem}[Virial Theorem] \label{thm:virial}
Let Hypotheses 0 and 1 (Sect. \ref{sec:model}) be satisfied. If
$\ph \in \F$ is an eigenvector of $H_g(P)$ with $\Gamma (\chi_i)
\ph = \ph$, then
\[ \sprod{\ph}{[iH_g (P) , A] \ph} =0 .\]
\end{theorem}

\begin{proof} The theorem follows directly from Lemma \ref{lm:modivir} in
Appendix \ref{app:modivir}, where we prove the Virial Theorem for
a modified Hamiltonian $\Hmod (P)$, which is identical to $H_g
(P)$ on states without soft bosons.
\end{proof}

\begin{theorem}\label{thm:pos_comm}
Assume Hypotheses 0 -- 2 (Sect. \ref{sec:model}) are satisfied.
For each $\Sigma<O_{\beta=1}$, there exist constants
\(\delta_{\Sigma}>0,\ g_{\Sigma}>0\) and  $C_{\Sigma}$,
independent of $\sigma$, such that
\begin{equation*}
\sprod{\ph}{[iH_g(P),A]\ph} \geq \delta_{\Sigma} \sprod{\ph}{N\ph}
- C_{\Sigma} |g|\|\ph\|^2,
\end{equation*}
for all \(P\in \R^3,\ |g|<g_{\Sigma}\), and \(\ph\in D(N)\cap\ran
E_{\Sigma}(H_g(P))\).
\end{theorem}

\begin{proof}
Choose $f\in\tf(\R;[0,1])$ with $f\equiv 1$ on \([\inf_P
E_0(P)-1,\Sigma]\) and $f(s)=0$ for \(s\geq \Sigma+\eps\) where
\(\Sigma+\eps<O_{\beta=1}\). Let \(f=f(H_g(P))\) and
\(E_{\Sigma}\) as above. Since $fE_{\Sigma}=E_{\Sigma}$ and since
$[f,N^{1/2}]$ and $[f,N^{1/2}](H_g(P)+i)^{1/2}$ are of order $g$,
uniformly in $\sigma$, by Lemma~\ref{lm:grad-Omega2},
\begin{eqnarray*}
E_{\Sigma}\nabla\Omega(P-\dGamma(k))\cdot\dGamma(\hat{k})
E_{\Sigma} &\leq & E_{\Sigma}|\nabla\Omega(P-\dGamma(k)|N E_{\Sigma}\\
&=& E_{\Sigma} N^{1/2} f|\nabla\Omega(P-\dGamma(k))|f N^{1/2} E_{\Sigma}+O(g)\\
&\leq &
\||\nabla\Omega(P-\dGamma(k))|E_{\Sigma+\eps}(H_{g}(P))\|\,
E_{\Sigma} N E_{\Sigma} + O(g)\\
&\leq & \||\nabla\Omega|E_{\Sigma+\eps}(H_{g})\|\,
E_{\Sigma} N E_{\Sigma} + O(g)\\
&\leq & (1-\delta_{\Sigma})E_{\Sigma}N E_{\Sigma} + O(g)
\end{eqnarray*}
for some $\delta_{\Sigma}>0$ and $|g|$ small enough. Here $O(g)$
is independent of $\sigma$. By Eq.~\eqref{eq:iHA} defining
$[iH_{g}(P),A]$, this estimate and the boundedness of
$\phi(ia\kappa_{\sigma})E_{\Sigma}$ prove the theorem.
\end{proof}

In the next theorem, Theorems \ref{thm:virial} and
\ref{thm:pos_comm} are combined to prove absence of excited
eigenvalues below $\Sigma$. This is first done for \(H_g (P)
\restricted\ran \Gamma (\chi_i)\) (see Eq. \eqref{eq:gs-only1})
and then for $H_g(P)$.

\begin{theorem}\label{cor:gs-only}
Assume Hypotheses 0 -- 2 are satisfied and that $\Sigma <
O_{\beta=1}$, with $O_{\beta}$ given by Hypotheses 2. Then there
exists a constant $g_{\Sigma}>0$ such that
\[ \sigma_{\text{pp}} (H_g(P)) \cap (-\infty , \Sigma] = \{ E_g (P) \}, \]
for all $P\in \R^3$ with $E_g (P)\leq \Sigma$, and all $g$ with
\(|g|<g_{\Sigma}\).
\end{theorem}

\emph{Remark:} For those $P$ with $E_g (P) \leq \Sigma$ and for
$|g|$ small enough depending on $\Sigma$, the proof of this
theorem shows again that $E_g(P)$ is a {\em non-degenerate}
eigenvalue (cf. Theorem~\ref{thm:groundstate}). Here no assumption
on the sign of $\kappa$ is needed.

The proof also shows that \(\|\psi_P-\Omega\|=O(|g|^{1/2})\),
$g\to 0$, uniformly in $P$ for $E_{g}(P)\leq \Sigma$.

In the case of \emph{relativistic} electrons the theorem shows
that \(\sigma_{\text{pp}} (H_g(P)) \cap (-\infty,\Sigma] = \{ E_g
(P) \}\) for \emph{all} $\Sigma\in\R$ and for $|g|$ small enough,
depending on $\Sigma$.

\begin{proof}
Let $\psi_g = \Gamma (\chi_i) \psi_g$ be a normalized eigenvector
of $H_g (P)$ with energy $\leq \Sigma$, and choose the phase of
$\psi_g$ so that $\sprod{\psi_g}{\Omega}\geq 0$. By the Virial
Theorem and by Theorem~\ref{thm:pos_comm}
\begin{equation*}
0 \geq \delta_{\Sigma} \sprod{\psi_g}{(1-P_\Omega)\psi_g} -
C_{\Sigma}|g|
\end{equation*}
where
\[   \sprod{\psi_g}{(1-P_\Omega)\psi_g} = 1-|\sprod{\Omega}{\psi_g}|^2 \geq
1-|\sprod{\Omega}{\psi_g}| = \frac{1}{2} \|\psi_g - \Omega\|^2.\]
In the last equation the choice of the phase of $\psi_g$ was used.
We conclude that \(\|\psi_g-\Omega\| \leq (2 |g|
C_{\Sigma}/\delta_{\Sigma})^{1/2}\). Since it is impossible to
have two orthonormal vectors $\psi_g^{(1)}$ and $\psi_g^{(2)}$
with \(\|\psi_g^{(i)}-\Omega\| < 1/\sqrt{2}\), for \(|g|<
\delta_{\Sigma}/4C_{\Sigma}\) there exists only one eigenvalue of
$H_g (P) \restricted{\ran \Gamma (\chi_i)}$ below or equal to
$\Sigma$, and it is {\em simple}. By Theorem
\ref{thm:groundstate}, this eigenvalue is $E_g (P)$. Hence, for
these values of $g$,
\begin{equation}\label{eq:gs-only1}
\sigma_{\text{pp}} \left( H_g(P)\restricted{\ran \Gamma
(\chi_i)}\right) \cap (-\infty , \Sigma] = \{ E_g (P) \},
\end{equation}
for all $P$ with $E_g (P) \leq \Sigma$. The theorem now follows if
we show that
\begin{equation}\label{eq:gs-only2}
 \sigma_{\text{pp}} \left(H_g(P)\restricted{\ran \Gamma
(\chi_i)^{\perp}}\right) \cap (-\infty,\Sigma] = \emptyset .
\end{equation}
To prove \eqref{eq:gs-only2}, we use that $\F \cong \F_i \otimes
\F_s$, where $\F_i$ and $\F_s$ are the bosonic Fock spaces over
$L^2 (\{ k : |k|>\sigma \})$ and over $L^2(\{ k : |k|\leq\sigma
\})$, respectively, where $\sigma > 0$ is the infrared cutoff
defined in Hypothesis 1; ($\F_i$ and $\F_s$ are the spaces of
interacting and of soft, non-interacting bosons, respectively).
Consider the restriction of $H_g (P)$ to the subspace of $\F_i
\otimes \F_s$ of all vectors with exactly $n$ soft bosons. This
subspace is isomorphic to $\F_{s,n} = L^2_s (\R^{3n}, dk_1 \dots
dk_n ; \F_i)$, the space of all square integrable functions on
$\R^{3n}$, with values in $\F_i$ which are symmetric with respect
to permutations of the $n$ variables. The action of $H_g (P)$ on a
vector $\psi \in \F_{s, n}$ is given by
\begin{equation*}
\begin{split}
(H_g (P) \psi ) (k_1, \dots k_n ) &= H_P (k_1 , \dots, k_n)
\psi (k_1 , \dots k_n) \quad \text{with} \\
H_P (k_1 , \dots, k_n) &= H_g (P-k_1 - \dots - k_n ) + |k_1| +
\dots + |k_n| .
\end{split}
\end{equation*}
The operator $H_P (k_1 , \dots, k_n)$ acts on $\F_i$ and, by
\eqref{eq:gs-only1}, its only eigenvalue in the interval
$(-\infty,\Sigma]$ is given by $E_g(P - k_1 - \dots -k_n) + |k_1|
+ \dots +|k_n|$, as long as this number is smaller than $\Sigma$,
and if $|g| < \delta_{\Sigma}/(4C_{\Sigma})$. This implies that,
for $|g|<\delta_{\Sigma}/(4C_{\Sigma})$, a number $\lambda \in (-
\infty,\Sigma]$ is an eigenvalue of the restriction $H_g (P)
\restricted{\F_{s,n}}$ if and only if there exists a set
$M_{\lambda} \subset \R^{3n}$ of positive measure such that
\[ E_g (P - k_1 - \dots k_n) + |k_1| + \dots |k_n| = \lambda\]
\sloppy{for all $( k_1,\dots, k_n) \in M_{\lambda}$. Using that
\(|\nabla
E_g(P)|=|\sprod{\psi_P}{\nabla\Omega(P-\dGamma(k))\psi_{P}}|\leq
\sup_{P:E(P)\leq \Sigma} \|\nabla\Omega(P-P_f)\psi_P\|\leq \|
|\nabla\Omega| E_{\Sigma}(H_g)\|<1\), for $|g|$ small enough
(Lemma~\ref{lm:grad-Omega2}), it can easily be shown that such a
set $M_{\lambda}$ does not exist. This completes the proof of the
theorem.}
\end{proof}

\subsection{Bounds on the Interaction}

\begin{lemma}\label{lm:estim}
Let \(L^2_{\omega}(\R^3) \equiv L^2 (\R^3 , (1 + 1/|k|) dk) =
\left\{ h \in L^2 (\R^3) : \int dk (1 + 1/|k|) |h(k)|^2 < \infty
\right\}\) and let $h\in L^2_{\omega} (\R^3)$. Then
\begin{eqnarray*}
\| a(h) \ph \| &\leq &  \left( \int dk |h(k)|^2 /|k| \right)^{1/2} \, \| \dGamma (|k|)^{1/2} \ph \| \\
\| a^* (h) \ph \| &\leq & \| h \|_{\omega} \,\| (\dGamma (|k|)+1)^{1/2} \ph \| \\
\| \phi (h) \ph \| &\leq & \sqrt{2} \, \|h \|_{\omega}\,
\| (\dGamma (|k|)+1)^{1/2} \ph \|\\
\pm \phi(h) &\leq & \alpha \dGamma(|k|) +\frac{1}{\alpha}\int dk
\frac{|h(k)|^2}{|k|},\qquad \alpha>0 ,
\end{eqnarray*}
where \( \| h\|_{\omega}^2 = \int dk \, (1 + 1/|k|) |h(k)|^2 \).
\end{lemma}

\noindent For the easy proofs, see \cite{BFS1}, where similar
bounds are established.

In the analysis of electron-photon scattering it is important that
the interaction between bosons and electron decays sufficiently
fast with increasing distance. This decay is the subject of the
next lemma.

\begin{lemma}\label{lm:srdecay}
Assume Hypothesis 1 (Sect. \ref{sec:model}).
\begin{enumerate}
\item[i)] For arbitrary $n , \mu \in \N$ there is a constant
$C_{\mu , n} >0$ such that
\begin{equation*}
\sup_{x\in \R^3} \| \chi (|x-y|\geq R) |x-y|^n G_x \| \leq
C_{\mu,n} R^{-\mu}
\end{equation*}
for all $R>0$. In particular \( \| \phi \left(|x-y|^n G_x \right)
\, (N+1)^{-1/2} \|< \infty \), for all $n \in \N$.
\item[ii)] For every $\mu \in \N$ there is a constant $C_{\mu}>0$ such that
\begin{equation*}
\sup_{|x| \leq R} \| \chi (|y| \geq R') G_x  \| \leq C_{\mu} (R'
-R)^{-\mu}
\end{equation*}
for all $R' \geq R$.
\end{enumerate}
\end{lemma}

\begin{proof}
i) For all $x\in \R^3$
\begin{equation*}
\begin{split}
\|\chi (|x-y| \geq R) |x-y|^n G_x \|^2 &= \int_{|x-y|>R}  dy
|x-y|^{2n} | \hat{\kappa}_{\sigma} (x-y)|^2  = \int_{|y|>R} dy
|y|^{2n}
|\hat{\kappa}_{\sigma} (y)|^2 \\
&\leq R^{-2\mu} \int dy\,
|y|^{2(n+\mu)}|\hat{\kappa}_{\sigma}(y)|^2 = R^{-2\mu}C_{\mu ,n}
\end{split}
\end{equation*}
where, by Hypothesis~1, $C_{\mu ,n}$ is finite for all $\sigma\geq
0$ and all $n,\mu \in \N$.

Statement ii) follows from i), because if $|x|\leq R$ and $|y|
\geq R'$, then $|x-y| \geq R'-R$.
\end{proof}

The following lemma is used to apply Hypothesis 2, when we need to
control the velocity of the electron $|\nabla \Omega (p)|$ by
bounds on the total energy $H_g$.

\begin{lemma}\label{lm:grad-Omega2}
Assume Hypotheses 0 -- 2. For each $\beta>0$ and each
$\Sigma<O_{\beta}$, there exists  a constant $g_{\beta,\Sigma}>0$
independent of $\sigma$ such that
\[ \sup_{|g|\leq g_{\beta,\Sigma}}\| |\nabla\Omega|E_{\Sigma}(H_{g})\|\leq \beta \]
for all $\sigma>0$.
\end{lemma}

\emph{Remark.} This lemma holds equally for the modified
Hamiltonian $\Hmod$, introduced in Section \ref{sec:modham}.

\begin{proof}
Pick $\Sigma<O_{\beta}$ and pick $\eps>0$ such that
$\Sigma+\eps<O_{\beta}$. Choose $f\in\tf(\R,[0,1])$ with $f\equiv
1$ on \([\inf\sigma(H_{g=0})-1,\Sigma]\) and $f(s)=0$ for $s\geq
\Sigma+\eps$. Then
\begin{eqnarray*}
\||\nabla\Omega|E_{\Sigma}(H_g)\| & \leq &
\||\nabla\Omega|f(H_g)\|\\
& \leq & \||\nabla\Omega|f(H_{g=0})\| + O(g)\\
& \leq & \||\nabla\Omega|f(\Omega)\| + O(g) \leq \beta
\end{eqnarray*}
for $g$ small enough, because \(\||\nabla\Omega|f(\Omega)\|\leq
\sup\{|\nabla\Omega(p)|:\,\Omega(p)\leq \Sigma+\eps\} <\beta\) by
Hypothesis 2 and the remarks thereafter.
\end{proof}

For non-relativistic and relativistic electron kinematics the
constants $O_{\beta}$ and $g_{\Sigma, \beta}$ can be determined
explicitly:

\begin{lemma}\label{lm:grad-Omega}
Let $\Sigma\in \R$ and \(C:=\int |\kappa(k)|^2/|k|\, dk\) (which
is independent of the IR cutoff $\sigma$!)
\begin{itemize}
\item[(a)] If \(\Omega(p)= p^2 / 2M\) then
\begin{equation}
   \| |\nabla\Omega|E_{\Sigma}(H_g)\| \leq \left(\frac{2}{M}(\Sigma+g^2C)\right)^{1/2}.
\end{equation}
\item[(b)] If  \(\Omega(p)=\sqrt{p^2+M^2}\) then
\begin{equation}
   \| |\nabla\Omega|E_{\Sigma}(H_g) \| \leq \left(1-\frac{M^2}{(\Sigma+g^2C)^2}\right)^{1/2}.
\end{equation}
\end{itemize}
\end{lemma}

\begin{proof} From Lemma~\ref{lm:estim} with $\alpha=1/g$ and from
$|\kappa_{\sigma}|\leq |\kappa|$ it follows that
\begin{equation}\label{eq:Omega<H}
   \Omega \leq H_g + g^2 \int\frac{|\kappa(k)|^2}{|k|}\, dk
\end{equation}
in both cases.

Statement (a) follows from  \(|\nabla\Omega|^2 = 2\Omega/M\) and
\eqref{eq:Omega<H}. In case (b) we have \(|\nabla\Omega|^2 =
1-M^2/\Omega^2\) and we need an estimate on $\Omega^{-2}$ from
below. By  \eqref{eq:Omega<H}, \(\Omega^{-1}\geq (H+g^2 \int
|\kappa(k)|^2/|k|\, dk)^{-1}\) and hence
\begin{equation*}
   E_{\Sigma}(H_g) \Omega^{-2} E_{\Sigma}(H_g) \geq (E_{\Sigma}(H_g) \Omega^{-1}
   E_{\Sigma}(H_g) )^2
\geq (\Sigma+g^2C)^{-2}E_{\Sigma}(H_g).
\end{equation*}
This proves (b).
\end{proof}

\section{Propagation Estimate for the Electron and Existence of
the Wave Operator}\label{sec:waveop}

Wave operators map scattering states onto interacting states. In
our model the scattering states consist of dressed one-electron
(DES) wave packets and some asymptotically free outgoing bosons
described by asymptotic field operators, which act on the DES. The
DES were constructed in the previous section, and the existence of
asymptotic field operators in models such as the present one was
established in \cite{FGS1}. We recall that the key idea in
\cite{FGS1} was to utilize Huyghens' principle in conjunction with
the fact that massive relativistic particles propagate with a
speed strictly less than the speed of light. In the present
setting, where the electron dispersion law $\Omega(p)$ is more
general, we can limit the electron speed from above by imposing a
bound on the total energy. In fact, by the following propagation
estimate, the electron in a state from $\ran E_{\Sigma} (H_g)$
with \(\| |\nabla \Omega| E_{\Sigma} (H_g) \| \leq \beta\) will
stay out of the region $|x|>\beta t$ in the limit $t\to\infty$.
(see Proposition 6.3 in \cite{DG3} for a similar result in
$N$-body quantum scattering.)

\emph{No infrared cutoff is necessary in this section}. From
Hypothesis 1 we only need that \(\kappa_{\sigma}\in
C_0^{\infty}(\R^3)\) where $\sigma$ may be equal to zero.
Asymptotic completeness of the wave operator, stated at the end of
this section, of course does require that $\sigma$ is positive.

\begin{prop}[Propagation estimate for electron]\label{prop:pe-el}
Let Hypotheses 0 and 2 (Sect. \ref{sec:model}) be satisfied, and
assume that \(\kappa_{\sigma}\in C_0^{\infty}(\R^3)\) ($\sigma=0$
is allowed). Suppose $\beta$, $g$ and $\Sigma>\inf\sigma(H_g)$ are
real numbers for which $\| |\nabla \Omega| E_{\Sigma} (H_g) \|
\leq \beta$. Let $f \in \tf (\R)$ with $\supp f \subset (-\infty ,
\Sigma)$.
\begin{enumerate}
\item[i)] If \(\beta<\lambda<\lambda'<\infty\) then
there exists a constant \( C_{\lambda,\lambda'}\)  such that
\begin{equation*}
   \int_{1}^{\infty} \frac{dt}{t} \| \chi_{[\lambda,\lambda']}(|x| /t)
   f(H_g) e^{-iH_g t} \ph \|^2 \leq C_{\lambda,\lambda'} \| \ph \|^2 .
\end{equation*}
\item[ii)] Suppose $F\in C^{\infty} ( \R)$ with $F'\in \tf(\R)$
and $\supp(F)\subset (\beta,\infty]$. Then
\begin{equation*}
s-\lim_{t\to\infty} F(|x| /t) f(H_g) e^{-iH_g t} = 0
\end{equation*}
\end{enumerate}
\end{prop}

\emph{Remark.} This proposition equally holds on the extended
Hilbert space $\HxF = \H \otimes \F$ if $H_g$ is replaced by the
extended Hamiltonian $\Hex_g = H_g \otimes 1 + 1 \otimes \dGamma
(|k|)$ (see Eq.~\eqref{eq:Hexg} below).

Furthermore, the validity of the proposition does not depend on
the dispersion law of the bosons. Therefore we may replace $H_g$
(or $\Hex_g$) by the modified Hamiltonian $\Hmod$ (or $\Hmodex$)
to be introduced in Section \ref{sec:modham}, and the proposition
continues to hold.

\begin{proof}
i) Let $\eps >0$ be so small that \(\lambda-\eps>\beta\). Pick $h
\in \tf(\R)$ with $h=1$ on $[\lambda,\lambda']$ and
\(\supp(h)\subset[\lambda-\eps,\lambda'+1]\). Define
\(\tilde{h}(s)= \int_0^s d\tau h^2 (\tau) \), and set $h =h
(|x|/t)$ and $\tilde{h} = \tilde{h} (|x|/t)$. We work with the
\emph{propagation observable}
\begin{equation*}
\phi (t) = -f(H_g) \tilde{h} f(H_g).
\end{equation*}
Since $\phi (t)$ is a bounded operator, uniformly in $t$, it is
enough to prove the lower bound
\begin{equation}\label{eq:el-heisenberg}
D\phi (t) \equiv \frac{\partial \phi(t)}{\partial t} + [iH_g ,
\phi(t)]
 \geq \frac{C}{t} f h^2 f + O(t^{-2}),
\end{equation}
for a positive constant $C$. To prove \eqref{eq:el-heisenberg}, we
first note that
\begin{equation}\label{eq:el-partial}
\frac{\partial \phi(t)}{\partial t} = f(H_g) h^2 \frac{|x|}{t^2}
f(H_g) \geq \frac{(\lambda - \eps)}{t} f(H_g) h^2 f(H_g).
\end{equation}
Furthermore, by Lemma~\ref{lm:pdc1},
\begin{equation*}
\begin{split}
[iH_g, \phi(t)] =\; &- f(H_g) [i \Omega (p) , \tilde{h}] f(H_g) \\
=\; &-\frac{1}{2t} f(H_g) \left( \nabla \Omega \cdot \frac{x}{|x|}
h^2 + h^2 \frac{x}{|x|} \cdot \nabla
  \Omega \right)f(H_g) + O(t^{-2})\\
=\; &-\frac{1}{2t} f(H_g) h \left( \nabla \Omega \cdot
\frac{x}{|x|} +
  \frac{x}{|x|} \cdot \nabla \Omega \right) h f(H_g) + O(t^{-2}).
\end{split}
\end{equation*}
and thus
\begin{equation}\label{eq:el-comm1}
|\sprod{\ph_t}{[iH_g , \phi(t)] \ph_t}| \leq \frac{1}{t} \| |
\nabla \Omega | h
 f(H_g) \ph_t \| \| h f(H_g) \ph_t \| + O(t^{-2}).
\end{equation}
In order to estimate the factor $\| |\nabla \Omega| h
 f(H_g) \ph_t \|$, we choose $g \in \tf (\R)$ with $gf=f$ and with $\supp g
 \subset (-\infty , \Sigma)$, and we note that, since
 \([h,g(H_g)]=O(t^{-1})\),
\begin{equation}\label{eq:el-comm2}
f(H_g) h |\nabla \Omega |^2 h f(H_g) = f(H_g) h g(H_g) | \nabla
\Omega |^2 g(H_g) h f(H_g) +O(t^{-1}).
\end{equation}
By assumption on $|\nabla\Omega|$, \eqref{eq:el-comm2} combined
with \eqref{eq:el-comm1} shows that
\begin{equation*}
|\sprod{\ph_t}{[iH_g , \phi(t)] \ph_t}| \leq \frac{\beta}{t} \| h
f(H_g) \ph_t \|^2 + O(t^{-2})
\end{equation*}
where we commuted $g(H_g)$ with $h$ once again. This, together
with \eqref{eq:el-partial} and \(\lambda-\eps>\beta\), implies
\eqref{eq:el-heisenberg} and proves the first part of the
proposition.

ii) Clearly it is enough to prove that
\begin{equation}\label{eq:claim-b}
\lim_{t\to\infty} \phi(t) =0 \quad \text{where} \quad \phi(t) =
\sprod{\ph_t}{f(H_g) F(|x|/t) f(H_g) \ph_t},
\end{equation}
for $\ph \in \H$ and for an arbitrary $F$ satisfying the
assumptions of the proposition and such that $F(s) \geq 0$ for all
$s$. To this end we first note that the limit $\lim_{t\to\infty}
\phi(t)$ exists because $\int_1^{\infty} dt |\phi' (t)| < \infty$
by part i) of this proposition. Moreover, if $F$ has compact
support, then, by i), \( \int_1^{\infty} dt \, \phi (t)/t <
\infty\) and hence $\lim_{t\to\infty} \phi(t) =0$.

It remains to prove \eqref{eq:claim-b} if the support of $F$ is
not compact. Clearly it is enough to consider the case where
$F(s)=1$ for all $s$ sufficiently large and $F' \geq 0$. For such
functions $F$ we define
\begin{equation*}
\phi_{\lambda} (t) = \sprod{\ph_t}{f F(|x|/ \lambda t) f\ph_t},
\end{equation*}
for an arbitrary $\lambda\geq 1$. Computing the derivative of
$\phi_{\lambda}$ we find
\begin{equation*}
\begin{split}
\frac{d}{dt}\phi_{\lambda} (t) = \; &\sprod{\ph_t}{f\left(
-\frac{1}{t} F'
    \frac{|x|}{\lambda t} +
    \frac{1}{2\lambda t} (\nabla \Omega \cdot \frac{x}{|x|} F' + F'
\frac{x}{|x|} \cdot \nabla \Omega ) + O(\lambda^{-2}t^{-2})\right) f\ph_t} \\
\leq \; &O(\lambda^{-2} t^{-2})
\end{split}
\end{equation*}
for $\lambda$ large enough (because the sum of the terms
proportional to $t^{-1}$ is negative, if $\lambda$ is large
enough). Thus, for an arbitrary fixed $t_0$ (and for $\lambda$
large enough), we have that
\begin{equation*}
\phi_{\lambda} (t) = \phi_{\lambda} (t_0) + \int_{t_0}^t d\tau
\phi_{\lambda}^{\prime} (\tau) \leq \phi_{\lambda} (t_0)
+\frac{C}{\lambda^2 t_0} ,
\end{equation*}
for all $t >t_0$, and, in particular, for $t\to\infty$. Since
$\phi_{\lambda} (t_0)\to 0$ for $\lambda \to\infty$ it follows
that
\begin{equation}\label{eq:phi_R}
\lim_{\lambda\to\infty} \limsup_{t\to\infty} \phi_{\lambda} (t)
=0.
\end{equation}
Obviously
\begin{equation*}
\lim_{t\to\infty} \phi(t) = \lim_{t\to\infty} (\phi(t)
-\phi_{\lambda} (t)) + \lim_{t\to \infty} \phi_{\lambda} (t).
\end{equation*}
By \eqref{eq:phi_R} the second term can be made smaller than any
positive constant, by choosing $\lambda$ sufficiently large. After
having fixed $\lambda$, the first term on the r.h.s. of the last
equation is seen to vanish, because \[ \phi(t)-\phi_{\lambda} (t)
= \sprod{\ph_t}{f\left(F(|x|/t) -
    F(|x|/\lambda t)\right) f\ph_t} \] and because the function
$F(s) -F(s/\lambda)$ has compact support. Thus the l.h.s. of the
last equation is smaller than any positive constant. Since
$\phi(t) \geq 0$, for all $t$, Eq.~\eqref{eq:claim-b} follows.
\end{proof}

Using Proposition \ref{prop:pe-el} we can prove the existence of
asymptotic field operators, enabling us to construct states with
asymptotically free bosons. In order to prove the existence of the
asymptotic field operators we have to assume that $\| |\nabla
\Omega| E_{\Sigma}(H_g)\| < 1$; this will ensure that the photons
propagating along the light cone are far away from the electron
and hence move freely, as $t\to\infty$.

\begin{theorem}[Existence of asymptotic field operators]\label{thm:asy_fields}
Let Hypotheses 0 and 2 be satisfied and suppose
\(\kappa_{\sigma}\in C_0^{\infty}(\R^3)\) ($\sigma =0$ is
allowed). Let $g$ and $\Sigma$ be real numbers for which $\|
|\nabla\Omega| E_{\Sigma} (H_g) \| < 1$ (see Hypothesis 2 and
Lemma~\ref{lm:grad-Omega2}). Then the following statements hold
true.
\begin{enumerate}
\item[i)] Let $h \in L^2_{\omega} (\R^3 )$. Then the limit \[a_+^{\sharp} (h) \ph =
\lim_{t \to \infty} e^{iH_g t} a^{\sharp} (h_t) e^{-iH_g t} \ph \]
exists for all $\ph \in \ran E_{\Sigma} (H_g)$. Here $h_t (k) =
e^{-i|k|t} h(k)$.
\item[ii)] Let $h,g \in L^2_{\omega} (\R^3 )$. Then \[ [ a_+ (g) , a^*_+ (h)] = (g,h)
\quad \text{and} \quad [ a^{\sharp}_+ (g) , a^{\sharp}_+ (h)] = 0,
\] in the sense of quadratic forms on $\ran E_{\Sigma} (H_g)$.
\item[iii)] Let $h \in L^2_{\omega} (\R^3)$, and let $M:=\sup \{ |k| :h(k)\neq
0\}$ and $m := \inf \{ |k| :h(k)\neq 0\}$. Then
\begin{align*}
a_+^* (h) \ran \chi (H_g \leq E ) &\subset \ran \chi (H_g \leq E+M) \\
a_+ (h) \ran \chi (H_g \leq E ) &\subset \ran \chi (H_g \leq E-m),
\end{align*}
if $E\leq \Sigma$.
\item[iv)] Let $h_i \in L^2_{\omega} (\R^3)$ for $i=1,\dots n$. Put $M_i = \sup
\{ | k | :h_i (k)\neq 0\}$ and assume $\ph \in \ran E_{\lambda}
(H_g)$. Then, if $\lambda + \sum_{i=1}^n M_i \leq \Sigma$ we have
$\ph \in D(a_+^{\sharp} (h_1) \dots a_+^{\sharp} (h_n))$ and
\[ a_+^{\sharp} (h_1) \dots a_+^{\sharp} (h_n)\ph = \lim_{t \to \infty}
e^{iH_g t} a^{\sharp} (h_{1,t}) \dots a^{\sharp} (h_{n,t})e^{-iH_g
t} \ph \] and \[ \| a_+^{\sharp} (h_1) \dots a_+^{\sharp} (h_n)
(H_g +i)^{-n/2} \| \leq C \| h_1 \|_{\omega} \dots \| h_n
\|_{\omega} . \]
\end{enumerate}
\end{theorem}
\emph{Remark.} For \(\Omega(P)=\sqrt{P^2+m^2}\) the condition \(\|
|\nabla\Omega| E_{\Sigma}(H_g) \|<1\) is satisfied for all
$\Sigma\in\R$ and hence $a_+^{\sharp}(h)$ exists on
\(\cup_{\Sigma}E_{\Sigma}(H_g)\H\), and thus on $D(|H_g+i|^{1/2})$
by (iv). In this case part (iv) holds true for $h_i \in
L^2_{\omega} (\R^3)$, $i=1, \dots n$ and $\ph \in D(|H_g
+i|^{n/2})$, without any assumption on the support of the
functions $h_i$.

\begin{proof}
Similar results are proven in \cite{FGS1} for more involved
models. It is easy to make the necessary adaptations of the
arguments in \cite{FGS1} to the model at hand. The proof of i) in
\cite{FGS1} is based on a propagation estimate stronger than
Proposition \ref{prop:pe-el} (i), but Proposition \ref{prop:pe-el}
(i), enhanced by Proposition \ref{prop:pe-el} (ii), is actually
sufficient, as we now outline.

In order to prove i) it suffices to consider the case where
$h\in\tf (\R^3 \backslash \{ 0 \})$. This follows from the bound
$\| a^{\sharp} (h_t) (H_g +i)^{-1/2}\| \leq C \| h \|_{\omega}$,
which holds uniformly in $t$.

Choose $\eps>0$ so small that $\| |\nabla \Omega|E_{\Sigma}(H_g)
\|\leq  1-3\eps$ and pick $F \in \tf (\R)$, with $F(s)=1$, for
$s\leq 1 - 2\eps$, and $F(s) =0$, for $s>1-\eps$ . Then, by
Proposition \ref{prop:pe-el}, part ii), and since $[ (H_g +i)^{-1}
, F(|x|/t)] = O(t^{-1})$,
\begin{equation}\label{eq:asy_exists1}
e^{iH_g t} a^{\sharp} (h_t) e^{-iH_g t} \ph = \ph (t) + o(1) \quad
\quad (t \to \infty)
\end{equation}
where $\ph (t) = e^{iH_g t} a^{\sharp} (h_t) (H_g + i)^{-1}
F(|x|/t) e^{-iH_g t} (H_g + i) \ph$. By Cook's argument, the
existence of the limit $\lim_{t\to\infty} \ph (t)$ will follow if
we show that
\begin{equation}\label{eq:asy_int}
\int_1^{\infty} |\sprod{\psi}{\ph' (t)}| dt \leq C \| \psi \|,
\end{equation}
for all $\psi \in \H$ and some $C < \infty$. To this end we note
that
\begin{equation}\label{eq:asy_hei}
\begin{split}
\ph' (t) = \; &i g e^{iH_g t} [\phi (G_x) , a^{\sharp} (h_t)]
(H_g+i)^{-1} F(|x|/t) e^{-iH_g t} (H_g +i) \ph \\ &+ e^{iH_g t}
a^{\sharp} (h_t) (H_g +i)^{-1} D F  e^{-iH_g t} (H_g +i) \ph
\end{split}
\end{equation}
where $DF = [i\Omega (p) , F] + \partial F / \partial t$ is the
Heisenberg derivative of $F$. The first term gives an integrable
contribution to the integral in \eqref{eq:asy_int}, because $[\phi
(G_x), a^{\sharp} (h_t)] =\pm (G_x , h_t)$ and because
\begin{equation*}
\sup_{|x|<(1 -\eps) t} |(G_x , h_t)| \leq C_N /t^N
\end{equation*}
for any $N \in \N$; (here we use that $F(s) = 0$ if $s>1-\eps$ and
that $[(H_g+i)^{-1} , F] = O(t^{-1})$, by Lemma~\ref{lm:pdc1}).
The second term on the r.h.s. of \eqref{eq:asy_hei}, containing
the Heisenberg derivative of $F(|x|/t)$, gives an integrable
contribution too, by Proposition \ref{prop:pe-el}, part i) with
$\beta = 1-3\eps$, because
\begin{equation*}
DF = \frac{1}{t} \left( \nabla \Omega \cdot \frac{x}{|x|}
  -\frac{|x|}{t}\right) F'(|x|/t) + O(t^{-2})
\end{equation*}
where $\supp F' \subset [1-2\eps, 1-\eps]$. This proves
Eq.~\eqref{eq:asy_int}.
\end{proof}

Next we show, using Proposition \ref{prop:pe-el}, that the DES
wave packets $\ph \in \Hgs$ are vacua of these asymptotic fields.
It is known that \(E_g(P)=\inf\sigma(H_g(P))\) is an eigenvalue of
$H_g(P)$ if $\kappa_{\sigma}$ is sufficiently regular at the
origin (also if $\sigma =0$). Thus $\Hgs$ is non--empty. However,
we will not make any use of this, and no assertion about $\Hgs$ is
made in the following Lemma.

\begin{lemma}\label{lm:asy_vacua}
Suppose that Hypotheses 0 and 2 are satisfied and
\(\kappa_{\sigma}\in C_0^{\infty}(\R^3)\) ($\sigma =0$ is
allowed). Let $g$ and $\Sigma>\inf \sigma (H_g)$ be real numbers
for which $\| |\nabla\Omega| E_{\Sigma} (H_g) \| < 1$ (see
Hypothesis 2 and Lemma~\ref{lm:grad-Omega2}). Then, for all $\ph
\in E_{\Sigma} (H_g) \Hgs$ and $h \in L^2_{\omega} (\R^3)$,
\[ a_+ (h) \ph =0 .\]
\end{lemma}

\emph{Remark.} For \(\Omega(P)=\sqrt{P^2+m^2}\) one has the
stronger result that \(a_+ (h) \ph =0 \) for all $\ph\in \Hgs\cap
D(|H_g+i|^{1/2})$. This follows from the remark after
Theorem~\ref{thm:asy_fields}.

\begin{proof}
The intuition behind our proof is as follows: Because of the
assumption $\| | \nabla \Omega| E_{\Sigma} (H_g) \| <1$ the speed
of the electron is strictly less than one. Since, moreover, $\ph
\in \Hgs$, all bosons in $\ph_t$ are located near the electron,
and thus the overlap of the bosons in $\ph_t$ with a freely
propagating boson $h_t$ will vanish in the limit $t\to\infty$,
which implies that $a_{+}(h)\ph=0$.

This heuristic argument can be turned into a proof quite easily.
Since $\| a(h_t) (H_g +i)^{-1/2} \| \leq C \|h\|_{\omega}$
uniformly in $t$, we may assume that $h \in \tf (\R^3 / \{ 0 \})$.
Choose $\eps>0$ so small that $\| | \nabla \Omega| E_{\Sigma}(H_g)
\| \leq 1 -4\eps$ and pick $F\in \tf (\R)$, with $F(s)=1$ for
$s\leq 1-3\eps$ and $F(s)=0$ for $s \geq 1- 2\eps$. Then
\begin{equation}\label{eq:vac1}
  \ph_t = F(|x|/t)\ph_t + o(1), \qquad\text{as}\ t\to\infty
\end{equation}
by Proposition~\ref{prop:pe-el}, part ii) , with $\beta =1-3\eps$.
Given $\delta >0$, we next show that
\begin{equation}\label{eq:vac2}
  \ph_t = \Gamma(\chi_{[0,\delta]}(|x-y|/t))\ph_t + o(1), \qquad\text{as}\ t\to\infty.
\end{equation}
The operator on the right side, henceforth denoted by $Q_t$, is
translation invariant and hence leaves the fiber spaces $\H_P$
invariant. On the other hand, the time evolution of the component
of $\ph\in\Hgs$ in $\H_P$ is just a phase factor. Therefore
\(\|Q_t\ph_t\| = \|Q_t\ph\|\), which converges to $\|\ph\|$, as
$t\to\infty$. Since $Q_t$ is a projector this proves
\eqref{eq:vac2}. Combining \eqref{eq:vac1} with \eqref{eq:vac2}
for $\delta =\eps$ we get
\begin{equation}\label{eq:vac3}
   \ph_t = \Gamma(\chi_{\Delta}(|y|/t))\ph_t + o(1), \qquad\text{as}\  t\to\infty
\end{equation}
with \(\Delta = [0,1-\eps]\), because \(|x|/t\leq 1-2\eps\) and
\(|x-y|/t\leq \eps\) imply that \(|y|/t\leq 1-\eps\). Let
\(\psi\in D(H_g)\). By \eqref{eq:vac3}, and because
\(\|a^{*}(h_t)\psi_t\|\) is bounded uniformly in $t$,
\begin{eqnarray*}
   \sprod{\psi}{a_{+}(h)\ph} &=& \lim_{t\to\infty} \sprod{\psi_t}{a(h_t)\Gamma(\chi_{\Delta})\ph_t}\\
      &=&  \lim_{t\to\infty} \sprod{\psi_t}{\Gamma(\chi_{\Delta})a(\chi_{\Delta}h_t)\ph_t}.
\end{eqnarray*}
Using the Schwarz inequality and the bound \(\|
a(\chi_{\Delta}h_t)(H_g+i)^{-1/2}\|\leq \const
\|\chi_{\Delta}h_t\|_{\omega}\) we get
\begin{equation}\label{eq:vac4}
   \|a_{+}(h)\ph\| \leq C \limsup_{t\to \infty} \|\chi_{\Delta}h_t\|_{\omega},
\end{equation}
where
\begin{eqnarray*}
   \|\chi_{\Delta}h_t\|^2_{\omega}
     & = & \|\chi_{\Delta}h_t\|^2 + \sprod{\chi_{\Delta}h_t}{|k|^{-1}\chi_{\Delta}h_t}\\
     &\leq & 2 (1+t)\|\chi_{\Delta}h_t\|^2
\end{eqnarray*}
because \( \chi_{\Delta}|k|^{-1}\chi_{\Delta} \leq (\pi/2)\,
\chi_{\Delta}|y|\chi_{\Delta} \leq (\pi/2)t\chi_{\Delta}\), by
Kato's inequality (see \cite{Kato}, Section V.5). Since
\begin{equation*}
   \sup_{|y|/t \leq 1-\eps} |\hat{h}_t(y)| \leq C_N (1+t)^{-N}
\end{equation*}
for any integer $N$, and since the support of $y\mapsto
\chi_{\Delta}(|y|/t)$ has volume proportional to $t^3$, we
conclude that \( \|\chi_{\Delta}h_t \|_{\omega} \leq C_N
(1+t)^{2-N} \). For $N=3$ , this bound in conjunction with
\eqref{eq:vac4} completes the proof.
\end{proof}

Next, we define the M{\o}ller wave operator $\Omega_{+}$. We
introduce the extended Hilbert space $\HxF = \H \otimes \F$ and
the extended Hamilton operator
\begin{equation}\label{eq:Hexg}
\Hex_g = H_g \otimes 1 + 1 \otimes \dGamma (|k|).
\end{equation}
The wave operator $\Omega_{+}$ will be defined on a subspace of
$\HxF$.

\begin{theorem}[Existence of the wave operator]\label{thm:Omega_+}
Let Hypotheses 0 and 2 be satisfied and assume \(\kappa_{\sigma}
\in C_0^{\infty}(\R^3)\) ($\sigma=0$ is allowed). For every pair
of real numbers $g$ and $\Sigma$ with \(\| | \nabla \Omega|
E_{\Sigma} (H_g) \| < 1\), the limit
\begin{equation}\label{eq:Omega_+}
  \Omega_+ \ph := \lim_{t \to \infty} e^{iH_g t}I e^{-i\Hex_g t}
  (\Pgs \otimes  1)\ph
\end{equation}
exists, for $\ph$ in the dense subspace of $\ran
E_{\Sigma}(\Hex_{g})$ spanned by finite linear combinations of
vectors of the form \(\gamma \otimes a^* (h_1) \dots a^* (h_n)
\Omega\) with $\gamma=E_{\lambda}(H_{g})\gamma$, $h_i\in
L^2_{\omega}(\R^3)$, and \(\lambda+\sum_{i}\sup\{|k|:h_i(k)\neq 0
\}\leq \Sigma\). If \(\ph=\gamma \otimes a^* (h_1) \ldots
a^*(h_n)\Omega\) belongs to this space then
\begin{equation}\label{eq:Omega_+2}
  \Omega_+ \ph = a_+^* (h_1) \dots a_+^* (h_n) \Pgs \gamma.
\end{equation}
Furthermore $\| \Omega_+ \| =1$ and thus $\Omega_+$ has a unique
extension, also denoted by $\Omega_+$, to
$E_{\Sigma}(\Hex_g)\HxF$. On $(\Pgs \otimes
1)E_{\Sigma}(\Hex_g)\HxF$, the operator $\Omega_+$ is isometric,
and therefore $\ran \Omega_+$ is closed. For all $t\in \R$,
\[ e^{-iH_g t} \Omega_+ = \Omega_+ e^{-i\Hex_g t}.\]
\end{theorem}

{\em Remark.} i) In Section \ref{sec:modham} we will enlarge the
domain of the wave operator $\Omega_+$ to include arbitrarily many
soft, non-interacting bosons, regardless of their total energy.
ii) For \(\Omega(p)=\sqrt{p^2+m^2}\), the wave operator can be
defined as a partial isometry on the entire extended Hilbert space
$\HxF$. This follows from the remarks after
Theorem~\ref{thm:asy_fields} and Lemma~\ref{lm:asy_vacua}.

\begin{proof}
If \(\ph = \gamma \otimes a^* (h_1) \dots a^* (h_n) \Omega\), then
\begin{equation*}
e^{iH_gt}I e^{-\Hex_gt}\Pgs \gamma \otimes a^* (h_1) \dots
a^*(h_n) = e^{iH_gt}a^* (h_{1,t}) \dots
a^*(h_{n,t})e^{-iH_gt}\Pgs\gamma
\end{equation*}
and hence the existence of the limit \eqref{eq:Omega_+} and
equation \eqref{eq:Omega_+2} follow from Theorem
\ref{thm:asy_fields}, part iv). By Lemma~\ref{lm:dps}, the space
${\mathcal D}$ spanned by vectors of the form specified in the
theorem is dense in $\ran E_{\Sigma}(\Hex_{g})$. From
Eq.~\eqref{eq:Omega_+2}, in conjunction with
Theorem~\ref{thm:asy_fields}, part ii) and with Lemma
\ref{lm:asy_vacua}, it follows that $\Omega_+$ is a partial
isometry on \(\mathcal{D}\) and therefore $\|\Omega_+\|=1$. Hence
$\Omega_+$ has a unique extension to a partial isometry on
$E_{\Sigma}(\Hex_g)\HxF$. The remaining parts of the proof are
straightforward.
\end{proof}

The next result is a generalization of equation
\eqref{eq:Omega_+2} that will be needed for the proof of
asymptotic completeness.

\begin{lemma}\label{lm:wo_prop}
Suppose $\Omega_+$ is defined as in the preceding theorem. Assume
$\psi \in E_{\lambda} (\Hex_g) \HxF$ and $h_1, \dots h_n \in
L^2_{\omega} (\R^3)$, with $\lambda + \sum_{i=1}^n \sup \{ |k| :
h_i(k) \neq 0 \} \leq \Sigma$. Then
\begin{equation}\label{eq:wo_prop}
\Omega_+ (1 \otimes a^* (h_1) \dots a^* (h_n)) \psi = a_+^* (h_1)
\dots a_+^* (h_n) \Omega_+ \psi .
\end{equation}
\end{lemma}

\begin{proof}
If the vector $\psi$ is of the form
\begin{equation}\label{eq:psi}
  \psi = \gamma \otimes a^* (f_1) \dots a^* (f_m) \Omega ,
\end{equation}
where $\gamma \in E_{\eta} (H_g)\H$, $f_1 , \dots f_m \in \tf
(\R^3)$ with $\eta + \sum_i \sup \{ |k| : f_i (k) \neq 0 \} \leq
\lambda$, then
\begin{equation*}
\begin{split}
\Omega_+ (1 \otimes a^* (h_1) \dots a^* (h_n)) \psi &=  a^*_+
(h_{1}) \dots a^*_+ (h_{n})a^*_+ (f_{1})\dots
a^*_+ (f_{m}) \Pgs \gamma \\
&= a^*_+ (h_{1}) \dots a^*_+ (h_{n}) \Omega_+ \psi
\end{split}
\end{equation*}
by Eq.~\eqref{eq:Omega_+2}. This proves \eqref{eq:wo_prop} for all
$\psi$ which are finite linear combinations of vectors of the form
\eqref{eq:psi}. These vectors span a dense subspace of
$E_{\lambda} (\Hex) \HxF$ by Lemma~\ref{lm:dps} in Appendix
\ref{app:dps}. The lemma now follows by an approximation argument
using Theorem~\ref{thm:asy_fields} iv) and the intertwining
relation for $\Omega_{+}$.
\end{proof}

We are now prepared to formulate the \emph{main result} of this
paper.

\noindent\begin{theorem}[Asymptotic Completeness]\label{thm:ACph}
Assume that Hypotheses 0 -- 2 (Sect. \ref{sec:model}) are
satisfied, and let \(\Sigma\) be such that
\(\sup_p|\nabla\Omega(p)\chi (\Omega(p) \leq \Sigma )|<1/3\) (see
Hypotheses 2). Then, for $|g|$ small enough depending of $\Sigma$,
\[ \ran \Omega_+ \supset E_{\Sigma} (H_g) \H . \]
\end{theorem}

\emph{Remark.}  The assumption that \(\sup_p|\nabla\Omega(p)\chi
(\Omega(p) \leq \Sigma) | <1/3\) implies that \(\||\nabla\Omega|
E_{\Sigma}(H_{g=0})\|<1/3\), which, for small $|g|$, ensures that
\(\||\nabla\Omega| E_{\Sigma}(H_{g})\|<1/3\). This last inequality
is actually what we shall make use of. Since $|g|$ must be small
for reasons other than this one as well, we have chosen the above
formulation of the theorem.

This result follows from Theorem \ref{thm:AC} in Section
\ref{sec:AC}, where asymptotic completeness for a modified
Hamiltonian (with a modified dispersion law for the bosons) is
proved, and from Lemma \ref{lm:ran-Omega} in Section
\ref{sec:modham}, where the behavior of the soft bosons in the
scattering process is investigated.

In the most interesting cases of a relativistic dispersion
$\Omega(p) = \sqrt{p^2 +M^2}$ and of a non-relativistic dispersion
$\Omega (p) = p^2 /2M$ Theorem \ref{thm:ACph} implies the
following result.

\noindent\begin{corollary} Assume that Hypothesis 1 and one of the
following hypotheses hold.
\begin{itemize}
\item[1.] $\Omega(P)=P^2/2M$ and $0<\Sigma<M/18$,
\item[2.] $\Omega(P)=\sqrt{P^2+M^2}$ and
$M<\Sigma< 3M/\sqrt{8}$.
\end{itemize}
Then, for $|g|$ small enough,
\[\ran \Omega_+ \supset E_{\Sigma} (H_g) \H.\]
\end{corollary}

\begin{proof}
Hypotheses~0 and 2 are clearly satisfied in both cases and the
bounds on $\Sigma$ are chosen in such a way that
\(\sup_p|\nabla\Omega(p) \chi (\Omega(p) \leq \Sigma)| <1/3\).
Thus the corollary follows from Theorem~\ref{thm:ACph}.
\end{proof}

\section{The Modified Hamiltonian}\label{sec:modham}

Since the bosons in our model are massless, their number is not
bounded in terms of the total energy. This, however, is an
artefact, since the number of bosons with energy below $\sigma$
(the IR cutoff) is conserved. To avoid technical difficulties due
to the lack of a bound on the number operator, $N$, relative to
the Hamiltonian $H_g$, we work with a modified Hamiltonian $\Hmod$
whose photon-dispersion law, $\omega(k)$, is bounded from below by
a positive constant (in contrast to $|k|$).


We define
\begin{equation*}
\Hmod = \Omega (p) + \dGamma (\omega )+ g \phi (G_x),
\end{equation*}
and we assume that $\omega$ satisfies the following conditions.
\begin{quote}
{\bf Hypothesis 3.} $\omega \in \mathcal{C}^{\infty} (\R^3)$, with
$\omega(k)\geq |k|$, $\omega (k)=|k|$, for $|k| > \sigma$, $\omega
(k)\geq \sigma /2$, for all $k \in \R^3$,
\(\sup_{k}|\nabla\omega(k)|\leq 1\), and $\nabla\omega(k)\neq 0$
unless $k=0$. Furthermore, \(\omega(k_1+k_2)\leq
\omega(k_1)+\omega(k_2)\) for all $k_1,k_2\in\R^3$. Here $\sigma
>0$ is the infrared cutoff defined in Hypothesis 1.
\end{quote}
The Hamiltonian $\Hmod$ shares many of the properties derived for
$H_{g}$ in previous sections, such as Lemma~\ref{lm:grad-Omega2}
and Proposition~\ref{prop:pe-el} (see the remarks thereafter). We
now explore the similarities of $H_{g}$ and $\Hmod$ more
systematically.

The two Hamiltonians $H_g$ and $\Hmod$ act identically on states
of the system without soft bosons. Denoting by $\chi_i (k)$ the
characteristic function of the set $\{k:|k| > \sigma \}$, the
operator $\Gamma (\chi_i)$ is the orthogonal projection onto the
subspace of vectors describing states without soft bosons. Since
$\chi_i G_x = G_x$ it follows from Eqs.~\eqref{geq1} and
\eqref{geq2} that $H_g$ and $\Hmod$ commute with the projection
$\Gamma (\chi_i)$, and hence they leave the range of $\Gamma
(\chi_i)$ invariant. Moreover, since $\chi_i \omega (k) = \chi_i
\, |k|$,
\begin{equation}\label{eq:int_ham}
H_{g} \restricted{\ran \Gamma (\chi_i)} = \Hmod  \restricted{\ran
\Gamma (\chi_i)} .
\end{equation}
Let $U$ denote the unitary isomorphism \(U : \H \to \H_i \otimes
\F_s\) introduced in Section~\ref{sec:model}. Then, on the
factorized Hilbert space \(\H_i \otimes\F_s\), the Hamiltonians
$H_g$ and $\Hmod$ are given by
\begin{equation}\label{eq:fact_ham}
\begin{split}
U H_g U^* &= H_i \otimes 1 + 1 \otimes \dGamma (|k|) \\
U \Hmod U^* &= H_i \otimes 1 + 1 \otimes \dGamma (\omega) \quad \text{with} \\
H_i &= \Omega (p) + \dGamma (|k|) + g \phi (G_x).
\end{split}
\end{equation}
Again, we observe that the two Hamiltonians agree on states
without soft bosons.

The modified Hamiltonian $\Hmod$, like the physical Hamiltonian
$H_g$, commutes with spatial translations of the system, i.e.,
$[\Hmod , P] =0$, where $P=p + \dGamma (k)$ is the total momentum
of the system. In the representation of the system on the Hilbert
space $L^2 (\R^3_{P}; \F)$ the modified Hamiltonian $\Hmod$ is
given by
\begin{equation*}
\begin{split}
(\Pi \Hmod \Pi^* \psi )(P) &= \Hmod (P) \psi (P), \\
\Hmod (P) &= \Omega (P- \dGamma (k)) + \dGamma (\omega) + g
\phi(\kappa_{\sigma}),
\end{split}
\end{equation*}
where $\Pi: \H \to L^2 (\R^3 , dP;\F)$ has been defined in Section
\ref{sec:model}.

Like $H_g$ and $\Hmod$, the fiber Hamiltonians $H_g(P)$ and
$\Hmod(P)$ commute with the projection $\Gamma (\chi_i)$ and agree
on its range, that is
\begin{equation}\label{eq:restr}
H_g (P)\restricted{\ran \Gamma (\chi_i)} = \Hmod (P)
\restricted{\ran \Gamma (\chi_i)}.
\end{equation}
\begin{figure}\label{fig:omega}
\begin{center}
\epsfig{file=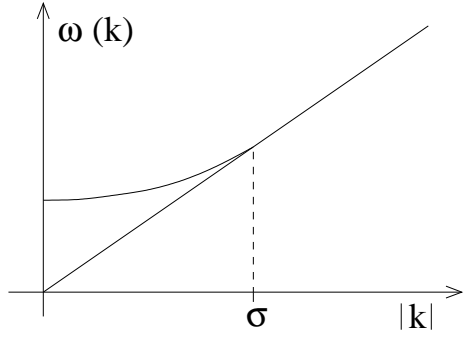,scale=.75}
\end{center}
\caption{Typical choice of the modified photon-dispersion law
$\omega (k)$.}
\end{figure}

In Appendix \ref{app:DES} (see Theorem~\ref{thm:gs}) it is shown
that, for \(\Omega(P)<O_{\beta=1}\) and $|g|$ small enough,
\begin{equation*}
\inf \sigma (\Hmod (P)) = \inf \sigma (H_g (P)) = E_g (P)
\end{equation*}
and that $E_{g}(P)$ is a simple eigenvalue of $H_g (P)$ and $\Hmod
(P)$. The corresponding dressed one-electron states coincide by
Theorem~\ref{thm:gs}, (ii). Since the subspace $\Hgs$ is defined
in terms of the dressed one-electron states $\psi_P$, it follows
that vectors in $\Hgs$ describe dressed one-electron wave packets
for the dynamics generated by the modified Hamiltonian $\Hmod$ as
well.

We remark that, in view of \eqref{eq:restr}, the proof of
Theorem~\ref{cor:gs-only} shows that
\[ \sigma_{\text{pp}} (\Hmod (P)) \cap (-\infty
, \Sigma) = \{E_g (P) \}, \] for all $P \in \R^3$ with $E_g (P)
\leq \Sigma$, and for $|g|$ sufficiently small.

Next, we consider the positive commutator discussed in Section
\ref{sec:poscomm}. Thanks to Eq. \eqref{eq:restr}, the inequality
established in Theorem~\ref{thm:pos_comm} continues to hold when
$H_g (P)$ is replaced by $\Hmod(P)$, provided we restrict it to
the range of the orthogonal projection $\Gamma (\chi_i)$. We need
to rewrite this commutator estimate in terms of $\Hmod$, rather
than $\Hmod(P)$, restricted to $\ran \Gamma(\chi_i)$. To this end
we define
\begin{equation*}
  a = \frac{1}{2} \left( \nabla \omega \cdot (y -x) + (y -x) \cdot
  \nabla \omega \right),
\end{equation*}
and we consider the conjugate operator $\dGamma (a)$. In the
representation of the system on the Hilbert space $L^2 (\R^3_P
;\F)$, the operator $\dGamma (a)$ is given by
\begin{align*}
&(\Pi \dGamma (a) \Pi^* \psi ) (P) = A \, \psi (P), \quad \text{where} \\
&A = \frac{1}{2} \, \dGamma  \left( \nabla \omega \cdot y  + y
\cdot \nabla \omega \right)
\end{align*}
is the conjugate operator used in Theorem \ref{thm:pos_comm} (if
restricted to states without soft bosons).

\begin{theorem}[Positive Commutator]\label{thm:pos_comm2}
Assume Hypotheses 0 -- 3 (see Sects. \ref{sec:model} and
\ref{sec:modham}) are satisfied. Let $\beta \leq 1$ and choose
$g_0$ and $\Sigma$ such that $\| | \nabla \Omega | E_{\Sigma}
(\Hmod) \| \leq \beta$, for all $g$ with $|g|\leq g_0$. Suppose
moreover that $f\in C_0^{\infty}(\R)$ and
\(\supp(f)\subset(-\infty,\Sigma)\). Then there exists a constant
$C$, independent of the infrared cutoff $\sigma$, such that, on
the range of the projector $\Gamma(\chi_i)$,
\begin{equation}
   f(\Hmod)[i\Hmod,\dGamma(a)]f(\Hmod) \geq (1-\beta)f(\Hmod) N f(\Hmod) - C g
   f(\Hmod)^2.
\end{equation}
for all $g$ with $|g|\leq g_0$.
\end{theorem}
\begin{proof}
Set $H\equiv \Hmod$. By definition
$$
  [iH,\dGamma(a)] = \dGamma(|\nabla\omega|^2) -
  \dGamma(\nabla\omega)\cdot\nabla\Omega - g \phi(iG_x)
$$
Since $\nabla\omega(k) = k/|k|$ on the range of $\chi_i$ and since
$\phi(iG_x)E_{\Sigma}(H)$ is bounded, it follows that
\begin{equation*}
\begin{split}
    f(H)\Gamma(\chi_i)[iH,\dGamma(a)]\Gamma(\chi_i)f(H)
    \geq \; &f(H)\Gamma(\chi_i) N \Gamma(\chi_i)f(H) - f(H)
    \Gamma(\chi_i)N|\nabla\Omega|\Gamma(\chi_i) f(H)
    \\ &- C g f(H)^2.
\end{split}
\end{equation*}
The assumption $\| |\nabla\Omega| E_{\Sigma} (H) \| \leq \beta$
implies \[ E_{\Sigma}(H) |\nabla\Omega| E_{\Sigma}(H) \leq \beta
E_{\Sigma}(H) .
\] Using this inequality and that $[f(H),N^{1/2}]$ and
$(H+i)^{1/2}[f(H),N^{1/2}]$ are of order $g$, uniformly in
$\sigma$, we conclude that
\begin{eqnarray*}
f(H) N |\nabla\Omega| f(H) &=& f(H) N^{1/2} |\nabla\Omega| N^{1/2} f(H)\\
& = & N^{1/2} f(H) |\nabla\Omega| f(H) N^{1/2} + O(g)\\
& \leq & \beta f(H) N f(H) + O(g),
\end{eqnarray*}
with $O(g)$ independent of $\sigma$. Since $\Gamma(\chi_i)$
commutes with $f(H)$, this proves the theorem.
\end{proof}

Next, we discuss the scattering theory for the modified
Hamiltonian. As in Theorem~\ref{thm:Omega_+} we assume that $g$
and $\Sigma>\inf\sigma(H_g)$ are real numbers for which $\| |
\nabla \Omega| E_{\Sigma} (H_g) \|< 1$. Hypothesis~2
(Sect.~\ref{sec:model}) and Lemma~\ref{lm:grad-Omega2} ensure the
existence of these numbers. Then, by the assumption that
$\omega(k)=|k|$ for wave vectors $k$ of interacting bosons (cf.
Hypotheses 1,3) we have that
\begin{equation}\label{eq:a_mod}
\begin{split}
e^{i\Hmod t} a^{\sharp} (e^{-i\omega t} h) e^{-i\Hmod t} &=
e^{iH_g t} e^{-i \dGamma (|k| - \omega) t} a^{\sharp} (e^{-i\omega
t} h) e^{i \dGamma (|k| - \omega) t}e^{-i H_g t} \\ &= e^{iH_g t}
a^{\sharp} (e^{-i|k|t} h) e^{-iH_g t}
\end{split}
\end{equation}
for all $t$. It follows that the limit
\begin{equation*}
a_{\text{mod},+}^{\sharp} (h) \ph =\lim_{t \to \infty} e^{i\Hmod
t} a^{\sharp} (e^{-i\omega t} h) e^{-i\Hmod t} \ph
\end{equation*}
exists and that $a^{\sharp}_{\text{mod},+} (h) \ph = a^{\sharp}_+
(h) \ph$, for all $\ph \in \ran E_{\Sigma} (\Hmod) \subset \ran
E_{\Sigma} (H_g)$ and for all $h \in L^2_{\omega} (\R^3)$. This
and the discussion of $\Hgs$, above, show that the asymptotic
states constructed with the help of the Hamiltonians $H_g$ and
$\Hmod$ coincide.

On the extended Hilbert space $\HxF =\H \otimes \F$, we define the
extended modified Hamiltonian \[ \Hmodex = \Hmod \otimes 1 + 1
\otimes \dGamma (\omega) . \] In terms of $\Hmod$ and $\Hmodex$ we
also define an extended (modified) version
$\tilde{\Omega}_+^{\text{mod}}$ of the wave operator $\Omega_+$
introduced in Section \ref{sec:waveop}.

\begin{lemma}\label{lm:tilde-Omega-mod}
Let Hypotheses 0, 2 and 3 be satisfied, and assume
\(\kappa_{\sigma} \in C_0^{\infty}(\R^3)\) ($\sigma =0$ is
allowed). For every pair of real numbers $g$ and $\Sigma$ with
\(\| | \nabla \Omega| E_{\Sigma} (H_g) \| < 1\), the limit
\begin{equation}\label{eq:tildeOmega}
  \tilde{\Omega}_+^{\text{mod}} \ph = \lim_{t \to \infty} e^{i\Hmod
  t} I e^{-i\Hmodex t} \ph
\end{equation}
exists for all \(\ph \in E_{\Sigma} (\Hmodex) \HxF\). The modified
wave operator $\Omega_+^{\text{mod}}$ defined by
$\Omega_+^{\text{mod}} = \tilde{\Omega}_+^{\text{mod}}(\Pgs
\otimes 1)$ agrees with $\Omega_{+}$ defined by
Theorem~\ref{thm:Omega_+}. More precisely
\begin{equation}\label{eq:Omega_+mod}
   \Omega_+^{\text{mod}} \ph = \Omega_+ \ph,
\end{equation}
for all $\ph \in \ran E_{\Sigma}(\Hmodex)\subset \ran
E_{\Sigma}(\tilde{H}_{g})$.
\end{lemma}

{\em Remark:} Recall from the discussion above that $\Pgs$ does
not depend on whether it is constructed using $H_g$ or $\Hmod$.

\begin{proof}
Since $IE_{\Sigma}(\Hmodex)$ is bounded, $e^{i\Hmod t}I
e^{-i\Hmodex t}E_{\Sigma}(\Hmodex)$ is bounded uniformly in
$t\in\R$ and hence it suffices to prove existence of
$\tilde{\Omega}_+^{\text{mod}}$ on a dense subspace of $\ran
E_{\Sigma}(\Hmodex)$. By Lemma~\ref{lm:dps}, finite linear
combinations of vectors of the form
\begin{equation*}
\ph = \gamma \otimes a^{*}(h_{1,t})\cdot\ldots\cdot
a^{*}(h_{n,t})\Omega
\end{equation*}
with \(\lambda+\sum_{i}M_i< \Sigma\), where
$\gamma=E_{\lambda}(\Hmod)\gamma$, and
\(M_i=\sup\{\omega(k):h_i(k)\neq 0\}\), form such a subspace.
Existence of $\tilde{\Omega}_+^{\text{mod}}$ on these vectors
follows from
\begin{equation*}
e^{i\Hmod t}I e^{-i\Hmodex t} = e^{iH_g t}I e^{-i\Hex_g t}
\end{equation*}
and from Theorem~\ref{thm:Omega_+}. This also proves
\eqref{eq:Omega_+mod}.
\end{proof}

We shall now extend the domain of $\Omega_{+}$ to include
arbitrarily many soft, non-interacting bosons. As a byproduct we
obtain a second proof of \eqref{eq:Omega_+mod}. To start with, we
recall the isomorphism $U: \F \to \F_i \otimes \F_s$ introduced in
Section \ref{sec:model} and define a unitary isomorphism $U
\otimes U: \HxF \to \H_i \otimes \F_i \otimes \F_s \otimes \F_s$
separating interacting from soft bosons in the extended Hilbert
space $\HxF$. With respect to this factorization the extended
Hamiltonian $\Hex_g$ becomes $\Hex_g = \Hex_i \otimes 1 \otimes 1
+ 1 \otimes 1 \otimes \dGamma (|k|) \otimes 1 + 1 \otimes 1
\otimes 1 \otimes \dGamma (|k|)$, where $\Hex_i = H_i \otimes 1 +
1 \otimes \dGamma (|k|)$. As an operator from $\H_i \otimes \F_i
\otimes \F_s \otimes \F_s$ to $\H_i \otimes \F_s$, the wave
operator $\Omega_+$ acts as
\begin{equation}\label{eq:wo_fact}
U \Omega_+ (U^* \otimes U^*) = \Omega_+^{\text{int}} \otimes
\Omega_+^{\text{soft}}
\end{equation}
where \(\Omega_+^{\text{int}}: \H_i \otimes \F_i \to \H_i \) is
given by
\begin{equation}\label{eq:wo_fact2}
\Omega_+^{\text{int}} = s-\lim_{t \to \infty} e^{iH_i t}I
e^{-i\Hex_i t} (\Pgs^{\text{int}} \otimes 1)
\end{equation}
while \( \Omega_+^{\text{soft}} : \F_s \otimes \F_s \to \F_s \) is
given by
\begin{equation}\label{eq:Omega_soft}
\Omega_+^{\text{soft}} = I (P_{\Omega} \otimes 1) ,
\end{equation}
where $P_{\Omega}$ is the orthogonal projection onto the vacuum
vector $\Omega \in \F_s$. In view of \eqref{eq:wo_fact} and
\eqref{eq:wo_fact2}, the domain of $\Omega_{+}$ can obviously be
extended to $\ran E_{\Sigma} (\Hex_i) \otimes \F_s \otimes
\F_s\supset \ran E_{\Sigma} (\Hex_g)$. For the modified wave
operator $\Omega_+^{\text{mod}} = \tilde{\Omega}_+^{\text{mod}}
(\Pgs\otimes 1)$, we have $\Omega_+^{\text{mod}}=\Omega_{+,
\text{mod}}^{\text{int}} \otimes \Omega_+^{\text{soft}}$, and from
$H_g \restricted{\ran \Gamma (\chi_i )} = \Hmod \restricted{\ran
\Gamma (\chi_i )}$ it follows that $\Omega_{+,
\text{mod}}^{\text{int}}=\Omega_{+}^{\text{int}}$. Consequently,
also $\Omega_+^{\text{mod}}$ is well defined on $\ran E_{\Sigma}
(\Hex_i) \otimes \F_s \otimes \F_s$ and $\Omega_+^{\text{mod}} =
\Omega_+$.

We summarize the main conclusions in a lemma.

\begin{lemma}\label{lm:ran-Omega}
Let the assumptions of Theorem~\ref{lm:tilde-Omega-mod} be
satisfied, and let \(\Omega_{+}\) be defined on \(\ran
E_{\Sigma}\otimes\F_{s}\otimes\F_{s}\) as explained above. Then
\begin{equation}\label{eq:ran-Omega}
  \ran \Omega_+ \cong \ran \Omega_+^{int} \otimes \F_s
\end{equation}
with respect to the factorization $\H\cong\H_i \otimes \F_s$. In
particular, the following statements are equivalent:
\begin{enumerate}
\item[i)] $\ran \Omega_+ \supset E_{\Sigma} (H_g) \H$.
\item[ii)] $\ran \Omega_+ \supset \Gamma (\chi_i) E_{\Sigma} (H_g) \H $.
\item[iii)] $\ran \Omega_+ \supset E_{\Sigma} (\Hmod) \H$.
\item[iv)] $\ran \Omega_+ \supset \Gamma (\chi_i) E_{\Sigma} (\Hmod) \H$.
\end{enumerate}
\end{lemma}
\begin{proof} Eq.~\eqref{eq:ran-Omega} follows from \eqref{eq:wo_fact} and
\eqref{eq:Omega_soft}. The equivalences i) $\Leftrightarrow$ ii)
and iii) $\Leftrightarrow$ iv) follow directly from
Eq.~\eqref{eq:ran-Omega}, while ii) $\Leftrightarrow$ iv) follows
from Eq.~\eqref{eq:int_ham}.
\end{proof}


\section{Propagation Estimates for Photons}\label{sec:propest}

The purpose of this Section is to prove a phase-space propagation
estimate (Proposition~\ref{prop:pe3}), which is used in the next
section to establish existence of the asymptotic observable $W$
and of the Deift-Simon wave operator $W_{+}$.

Henceforth we shall always work with the modified Hamiltonian
$\Hmod$, and we will use the shorthand notation
\begin{equation*}
H \equiv \Hmod = \Omega (p) + \dGamma (\omega) + g \, \phi (G_x) .
\end{equation*}
Moreover, for an operator $b$ acting on the one-boson space $\h$,
we define the Heisenberg derivative \[ db := [i \omega (k), b] +
\frac{\partial b}{\partial t}, \] while we define the Heisenberg
derivatives of an operator $A$ on $\H$ with respect to $H$ and
$H_0$, respectively, by
\begin{align*}
DA &:= i[H,A] + \frac{\partial A}{\partial t} \quad \text{and}
\\ D_{0} A &:= i[H_{0},A] +
\frac{\partial A}{\partial t}
\end{align*}
We observe that \[ D_0 (\dGamma (b)) = \dGamma (db). \]

The first propagation estimate is a maximal velocity propagation
estimate saying that photons cannot propagate into the region
$|y|>u t$, if $u=\max (1,\beta)$ (if $\beta >1$ there will always
be some photons, in the vicinity of the electron, propagating into
the region $|y| > t$).

\begin{prop}[Upper bound on the velocity of bosons]\label{prop:pe1}
Assume Hypotheses 0--3 are satisfied. Suppose $\beta$, $g$ and
\(\Sigma>\inf\sigma(H)\) are real numbers for which \(\| | \nabla
\Omega| E_{\Sigma}(H)\|\leq \beta\). Let \(f\in C_0^{\infty}(\R)\)
be real-valued with $\supp f \subset (-\infty , \Sigma)$, and
suppose $F \in \tf (\R)$, with $F\geq 0$ and $\supp F \subset
(-\infty, \beta]$. Then, for each pair of real numbers $\lambda,\,
\lambda'$ with $\max (1, \beta) < \lambda < \lambda'$, there
exists a constant $C_{\lambda,\lambda'}$ such that
\begin{equation*}
\int_1^{\infty} \frac{dt}{t} \, \sprod{\ph_t}{f\dGamma
(\chi_{[\lambda ,
  \lambda' ]}
  (|y|/t)) F(|x|/t) f\ph_t} \leq C_{\lambda , \lambda'} \| \ph \|^2
\end{equation*}
for all $\ph \in \H$. Here $f=f(H)$.
\end{prop}

\emph{Remark.} The lower bound, $1$, in the assumption
$\max(1,\beta) <\lambda$ is the upper bound on the photon
propagation speed $|\nabla\omega|$ in Hypothesis~3.

\begin{proof}
Choose $\eps >0$ so small that $3\eps<\lambda-\beta$ and
$\lambda-\eps >1$. Without loss of generality we may assume that
$F(s)=1$ for $s\leq \beta +\eps$, $F(s) =0$ for all $s\geq
\beta+2\eps$, and $F'\leq 0$. Choose $h\in \tf (\R;[0,1])$ with
$h=1$ on $[\lambda,\lambda']$ and $\supp(h)\subset
[\lambda-\eps,\lambda'+1]$. It is important that there are gaps
between $(-\infty,\beta]$ and $\supp(F')$, and between $\supp(F)$
and $\supp(h)$.

We define $\tilde{h} (s) =\int_0^s d\tau h^2 (\tau)$ and we use
the notation $h=h(|y|/t)$ and $\tilde{h}=\tilde{h} (|y|/t)$.
Consider the propagation observable
\begin{equation*}
\phi(t) = - f(H) \dGamma (\tilde{h}) F(|x|/t) f(H).
\end{equation*}
Since $\phi(t)$ is a bounded operator, uniformly in $t$, the
proposition follows if we show that
\begin{equation}\label{eq:pe1-claim}
  D\phi(t) \equiv \frac{\partial \phi(t)}{\partial t} + [iH , \phi
  (t)] \geq \frac{C}{t} \, f \dGamma (h^2) F(|x|/t) f + B(t),
\end{equation}
for some operator-valued function $B(t)$ with
\(\int_1^{\infty}|\sprod{\ph_t}{B(t)\ph_t}|dt\leq C\|\ph\|^2\). We
have that
\begin{eqnarray}
\frac{\partial \phi(t)}{\partial t} &=& \frac{1}{t} f \dGamma (h^2
\, |y|/t) F(|x|/t) f + \frac{1}{t} f \dGamma (\tilde{h})
F'(|x|/t) \, \frac{|x|}{t} \, f \nonumber \\
&\geq & \frac{\lambda-\eps}{t} f \dGamma (h^2) F f -
\frac{\beta+2\eps}{t} f\dGamma (\tilde{h}) |F'| f.
\label{eq:pe1-partial}
\end{eqnarray}
The second term on the right side gives a contribution to $B(t)$
by Proposition \ref{prop:pe-el}. In fact, since \(\supp(F')\subset
[\beta+\eps,\beta+2\eps]\subset[\beta+\eps,\lambda]\) and $F'\leq
0$ we have that
\begin{eqnarray*}
\lefteqn{\sprod{\ph_t}{f\dGamma (\tilde{h}) |F'| f\ph_t}\ \leq\
\|\chi_{[\beta+\eps,\lambda]}(|x|/t) f\ph_t\|\, \| \dGamma (\tilde{h}) F' f \ph_t \|} \\
&\leq & \|\chi_{[\beta+\eps ,\lambda]} (|x|/t) f\ph_t \| \,
\|\dGamma (\tilde{h}) (H+i)^{-1}\| \left(\| F' (H+i) f \ph_t \| + O(t^{-1})\|\ph\| \right) \\
&\leq & C \| \chi_{[\beta+\eps, \lambda]} (|x|/t) f\ph_t \|\, \|
\chi_{[\beta+\eps,\lambda]} ( |x|/ t) g (H) \ph_t \| +
O(t^{-1})\|\ph\|^2,
\end{eqnarray*}
where we used that $\|[H,F']f\|=O(t^{-1})$, by Lemma~\ref{lm:pdc1}
and Hypothesis 0, and put $g(s) := (s+i) f(s)$ and  $C=\|\dGamma
(\tilde{h}) (H+i)^{-1}\|$ in the last line. Thus, by the Schwarz
inequality and Proposition \ref{prop:pe-el}
\begin{equation}\label{eq:F'}
  \int_1^{\infty} \frac{dt}{t} \, \langle \ph_t , f \dGamma
  (\tilde{h}) |F'| f \ph_t \rangle \leq \const \, \| \ph \|^2,
\end{equation}
that is, the second term in \eqref{eq:pe1-partial} contributes to
$B(t)$ in \eqref{eq:pe1-claim}. To evaluate the commutator in
\eqref{eq:pe1-claim}, we use Lemma~\ref{lm:pdc1} and get
\begin{eqnarray}
- [iH , \phi(t)] &=& f [iH,\dGamma (\tilde{h})]Ff +
     f \dGamma (\tilde{h})[iH,F]f \nonumber \\
&=&  f [i\dGamma (\omega) , \dGamma (\tilde{h})] F f
   + f[ig\phi(G_x), \dGamma (\tilde{h})] Ff + f \dGamma (\tilde{h})
    [i\Omega (p),F] f \nonumber \\
&=& \label{eq:pe1-comm} \frac{1}{2t} f \dGamma\left(\nabla\omega
\cdot \frac{y}{|y|} h^2 + h^2
\frac{y}{|y|} \cdot \nabla \omega \right) F f + g f \phi(i\tilde{h}G_x) F f \\
& & + \frac{1}{2t} f \dGamma (\tilde{h}) \left(\nabla \Omega \cdot
\frac{x}{|x|} F' + F'  \frac{x}{|x|} \cdot \nabla \Omega \right) f
+ O(t^{-2}).\nonumber
\end{eqnarray}
The term that involves $F'$ is integrable w.r.~to $t$, by
Proposition \ref{prop:pe-el} and Hypothesis 0. This is seen in the
same way as the integrability of the second term of
\eqref{eq:pe1-partial}. Next we bound the second term of
\eqref{eq:pe1-comm}. By Lemma \ref{lm:srdecay} part ii),
\begin{equation}\label{eq:pe1-ww}
\| \phi (i\tilde{h} G_x) F (|x| /t)f\| \leq C \sup_{|x| \leq
(\beta+2\eps)  t} \|\chi (|y| \geq (\lambda -\eps) t) G_x \| \leq
C t^{-\mu}
\end{equation}
for some $\mu >1$, because $\supp(F)\subset
(-\infty,\beta+2\eps]$, $\supp(\tilde{h})\subset
[\lambda-\eps,\infty)$, and $\lambda - \eps> \beta+2\eps$.
Finally, in the first term of \eqref{eq:pe1-comm}, we commute one
factor of $h$ to the left and one to the right and conclude that
\begin{eqnarray*}
f [iH , \phi(t) ] f &= & - \frac{1}{2t} f \dGamma \Big(h (\nabla
\omega \cdot \frac{y}{|y|} +\frac{y}{|y|}\cdot \nabla
 \omega ) h\Big) F f + B(t)\\
 & \geq & -\frac{1}{t} f\dGamma (h^2) F f + B(t),
\end{eqnarray*}
where \(\int_1^{\infty}|\sprod{\ph_t}{B(t)\ph_t}|dt\leq
C\|\ph\|^2\). Together with \eqref{eq:pe1-partial}, \eqref{eq:F'},
and $\lambda-\eps> 1$ this proves Eq.~\eqref{eq:pe1-claim}.
\end{proof}


The following phase-space propagation estimate compares the group
velocity $\nabla \omega$ with the average velocity $y/t$ for
bosons that escape from the electron in the limit $t\to\infty$
(i.e., for bosons with asymptotic velocity greater than $\gamma$).
This result will be improved in Proposition \ref{prop:pe3}.

\begin{prop}\label{prop:pe2}
Assume Hypotheses 0--3 are satisfied. Suppose $\beta$, $g$ and
\(\Sigma>\inf\sigma(H)\) are real numbers for which \(\| | \nabla
\Omega| E_{\Sigma}(H)\| \leq \beta\). Let \(f\in
C_0^{\infty}(\R)\) be real-valued with $\supp f \subset (-\infty ,
\Sigma)$, and suppose $F \in \tf (\R)$, with $F \geq 0$ and $\supp
F \subset (-\infty, \beta]$. Then, for each pair of real numbers
$\lambda,\, \lambda'$ with $\beta < \lambda < \lambda'$, there
exists a constant $C_{\lambda,\lambda'}$ such that
\begin{equation*}
\int_1^{\infty} \frac{dt}{t} \, \sprod{\ph_t}{f\dGamma \left(
(\nabla \omega
    - \frac{y}{t}) \chi_{[\lambda,\lambda']} (|y|/t) (\nabla \omega
    - \frac{y}{t})\right) F(|x|/t) f\ph_t} \leq C_{\lambda,\lambda'} \| \ph \|^2 ,
\end{equation*}
for all $\ph \in \H$. Here $f=f(H)$.
\end{prop}

\begin{proof}
Choose $\eps >0$ so small that $3\eps<\lambda-\beta$. Without loss
of generality we may assume that $\lambda'>1$.  We may also assume
that $F(s)=1$ for $s\leq \beta +\eps$ and $F(s) =0$ for all $s\geq
\beta+2\eps$. Pick $R \in C_0^{\infty} (\R^3)$ with
\(\supp(R)\subset\{y:\lambda - \eps \leq |y| \leq \lambda' +1 \}\)
and
\begin{equation*}
  R''(y) \geq \chi_{[\lambda ,\lambda']}(|y|)-C \chi_{[\lambda',\lambda'+1]}(|y|).
\end{equation*}
It is easy to construct a function $R$ with these properties
explicitly. We work with the propagation observable
\begin{equation}\label{eq:pe2a}
    \phi(t) = f(H)\dGamma(b(t))F f(H)
\end{equation}
where
\begin{equation*}
b(t)=R(y/t)+ \frac{1}{2} [(\nabla\omega-y/t)\cdot (\nabla R)(y/t)+
(\nabla R)(y/t)\cdot (\nabla \omega-y/t)]
\end{equation*}
and $F$ denotes the operator of multiplication by $F(|x|/t)$. For
the reader who compares this proof with the proof of the related
Proposition~11 in \cite{FGS2} we remark that
\(b(t)=d(tR(y/t))+O(t^{-1})\), and that we could work with
$d(tR(y/t))$ here, too. The operator $\phi(t)$ is bounded
uniformly in $t\geq 1$, because \(b(t)\) is. Hence the proposition
follows if we show that
\begin{equation}\label{eq:pe2b}
\begin{split}
D\phi(t) \geq \; &\frac{C}{t} \sprod{\ph_t}{f\dGamma \left(
(\nabla \omega
    -\frac{y}{t}) \chi_{[\lambda,\lambda']} (|y|/t) (\nabla \omega
    -\frac{y}{t})\right) F f\ph_t} + B(t)
\end{split}
\end{equation}
for some operator-valued function $B(t)$ with
\(\int_1^{\infty}|\sprod{\ph_t}{B(t)\ph_t}|dt \leq C\|\ph\|^2\).
By the Leibniz rule for the Heisenberg derivative,
\begin{equation}\label{eq:pe2c}
   D\phi(t) = f\dGamma (d b(t)) F f + f \phi(ib(t)G_x) Ff + f\dGamma (b(t)) (DF)f.
\end{equation}
The second and the third term contribute to the integrable part
$B(t)$. For the second term this follows from Lemma
\ref{lm:srdecay}, since the distance between the support of $R$
and the support of $F$ is strictly positive. The integrability of
the third term follows from Proposition~\ref{prop:pe-el}, thanks
to the location of the support of $F'$, and from boundedness of
$\nabla \Omega$ w.r.to $H$ (Hypothesis 0); (see the proof of
Proposition~\ref{prop:pe1} for details). The first term in
\eqref{eq:pe2b} comes from the first term in \eqref{eq:pe2c}.
Using Lemma~\ref{lm:pdc1}, it is straightforward to show that
\begin{eqnarray*}
d b(t) &=& \frac{1}{t} (\nabla \omega -y/t)\cdot R''(|y|/t)\, (\nabla \omega -y/t)  + O(t^{-2})\\
&\geq & \frac{1}{t} (\nabla \omega -y/t)\cdot
\chi_{[\lambda,\lambda']} (|y|/t)
\, (\nabla\omega -y/t)\\
& & - \frac{C}{t} (\nabla \omega -y/t)\cdot
\chi_{[\lambda',\lambda'+1]}(|y|/t) \, (\nabla\omega -y/t) +
O(t^{-2})
\end{eqnarray*}
where
\begin{equation*}
  (\nabla \omega -y/t)\cdot \chi_{[\lambda,\lambda+1]}(|y|/t)\,
  (\nabla \omega -y/t) \leq C_{\eta}\chi_{[\lambda'-\eta,\lambda'+\eta+1]}(|y|/t) +O(t^{-1})
\end{equation*}
for some $\eta>0$ chosen so small that $\lambda'-\eta>\max (1,
\beta)$; (recall that $\lambda'>\max(1,\beta)$). Hence this term
contributes to $B(t)$, by Proposition~\ref{prop:pe1}, and
\eqref{eq:pe2b} is proven.
\end{proof}


Using Proposition~\ref{prop:pe2}, we can establish an improved
phase-space propagation estimate, which is the main result of this
section. Existence of an asymptotic observable, $W$, and of the
inverse wave operator, $W_+$, in Sections \ref{sec:observable} and
\ref{sec:inverse} will follow from this propagation estimate
alone; (see \cite{DG1} for a similar result). Some technical parts
in the proof of Proposition \ref{prop:pe3} are stated as Lemma
\ref{lm:pe3} below.

\begin{prop}\label{prop:pe3}
Assume Hypotheses 0--3 are satisfied. Suppose $\beta$, $g$ and
\(\Sigma>\inf\sigma(H)\) are real numbers for which \(\| | \nabla
\Omega| E_{\Sigma}(H)\| \leq \beta\). Let \(f\in
C_0^{\infty}(\R)\) be real-valued with $\supp f \subset (-\infty ,
\Sigma)$, and pick $F \in \tf (\R)$, with $F(s) \geq 0$ and $\supp
F \subset (-\infty, \beta]$. For each pair of real numbers
$\lambda,\lambda'$ with \(\max(1,\beta)<\lambda<\lambda'\) and
each $J=(J_1,J_2,J_3) \in \tf (\R^3 ; \R^3)$ with $\supp J_l
\subset \{ y \in \R^3 : \lambda<|y|<\lambda' \}$ there exists a
constant $C_{\lambda,\lambda'}$ such that
\begin{equation*}
\int_0^{\infty}\frac{dt}{t}\sprod{\ph_t}{f\dGamma \left(|J(y/t)
\cdot (\nabla\omega -y/t) + (\nabla \omega -y/t)\cdot
J(y/t)|\right) F(|x|/t) f\ph_t} \leq C_{\lambda,\lambda'}
\|\ph\|^2
\end{equation*}
for all $\ph\in \H$. Here \(f=f(H)\).
\end{prop}

\begin{proof}
Choose $\eps >0$ so small that $2\eps<\lambda-\beta$. Without loss
of generality we may assume that $F(s) =1$ for $s \leq \beta
+\eps$ and $F(s) =0$ for $s\geq\beta+2\eps$.

Let $A= (y/t -\nabla \omega)^2 + t^{-\delta}$, for some $\delta\in
(0,1]$, and set
\begin{equation*}
 b(t) = \tilde{J}(y/t)\cdot A^{1/2}\tilde{J}(y/t)
 = \sum_{i=1}^3 \tilde{J}_i(y/t) A^{1/2}\tilde{J}_i (y/t),
\end{equation*}
where $\tilde{J} \in \tf (\R^3 ; \R^3) $ is chosen such that
$\tilde{J}_i=1$ on the support of $J_i$ and with $\supp
\tilde{J}_i \subset \{ y \in \R^3 : \lambda < |y| < \lambda' \}$.
Note that the operator $b(t)$ is bounded uniformly in $t$, because
of the space cutoff $J$. We consider the propagation observable
\[\phi(t) = - f(H) \dGamma (b(t)) F(|x|/t) f(H). \] Because of the
boundedness of $b(t)$ and the energy cutoff $f$, the observable
$\phi (t)$ is bounded, uniformly in time. Thus, to prove the
proposition, it is enough to show that
\begin{equation}\label{eq:pe3-claim}
  D\phi(t) \geq \frac{C}{t} f(H) \dGamma \left(|J(y/t) \cdot (\nabla\omega
  -y/t) + \text{h.c}|\right) F(|x|/t) f(H) + B(t),
\end{equation}
for some operator-valued function $B(t)$ with $\int_1^{\infty} dt
|\sprod{\ph_t}{B(t)\ph_t}|\leq C\|\ph\|^2$. The Heisenberg
derivative of $\phi(t)$ is given by
\begin{equation}\label{eq:pe3-1}
\begin{split}
D\phi(t) &= - f \left( D\dGamma (b(t)) \right) F f - f \dGamma
(b(t))
(DF) f \\
&= -f \dGamma(db(t))F f - f \phi (i b(t)G_x) F f -f \dGamma(b(t))
(DF)f.
\end{split}
\end{equation}
The last term, involving $DF$, contributes to $B(t)$. This follows
from Proposition~\ref{prop:pe-el}, since, by Lemma~\ref{lm:pdc1},
\begin{equation*}
 DF = \frac{1}{t}\left(F' \frac{x}{|x|}
 \cdot \nabla\Omega - \frac{|x|}{t} F'\right) + O(t^{-2}),
\end{equation*}
where $F'$ is supported in the interval
$[\beta+\eps,\beta+2\eps]$, and $\nabla \Omega$ is bounded w.r.to
$H$, by Hypothesis 0 (see the proof of Proposition \ref{prop:pe1}
for more details). The term with the factor $\phi(ib(t)G_x)$ also
contributes to $B(t)$. This follows from Lemma \ref{lm:srdecay},
part ii), because the distance between the support of $F$ and the
support of $\tilde{J}$ is positive, and thus
\begin{equation*}
  \| \phi (ib(t) G_x) F(|x|/t)f \| \leq C t^{-\mu},
\end{equation*}
for some $\mu>1$. It remains to consider the contribution of the
first term on the r.h.s. of \eqref{eq:pe3-1}. To this end we use
that
\begin{equation}\label{eq:db}
\begin{split}
d b(t) &= \tilde{J}\cdot(d A^{1/2})\tilde{J} + (d \tilde{J})\cdot
A^{1/2} \tilde{J}
+ \tilde{J}\cdot A^{1/2} (d\tilde{J})  \\
&= \tilde{J}\cdot (d A^{1/2}) \tilde{J} + \sum_{i=1}^3 \left( (d
\tilde{J}_i) A^{1/2} \tilde{J}_i +  \tilde{J}_i A^{1/2} (d
\tilde{J}_i)\right).
\end{split}
\end{equation}
Applying Lemma \ref{lm:pe3} below, part ii) and part iii), we find
that
\begin{equation}\label{eq:dA1/2}
-\tilde{J} (y/t)\cdot (d A^{1/2}) \tilde{J} (y/t) \geq \frac{C}{t}
| J (y/t) \cdot (\nabla \omega - y/t) + (\nabla \omega - y/t)
\cdot J(y/t)| + O(t^{-1-\eta /2}),
\end{equation}
with $\eta=\min (\delta , 1-\delta/2)$. The other terms in
Eq.~\eqref{eq:db} turn out to contribute to $B(t)$ in
\eqref{eq:pe3-claim}, (a consequence of Proposition
\ref{prop:pe2}). To prove this, we start with the bound
\begin{equation}\label{eq:dJAJ}
\pm \left(d\tilde{J}_i A^{1/2} \tilde{J}_i + \tilde{J}_i A^{1/2}
d\tilde{J}_i \right) \leq \, t \, (d\tilde{J}_i)^2 + \frac{1}{t}
\tilde{J}_i A \tilde{J}_i .
\end{equation}
Observing that
\begin{equation*}
d\tilde{J}_i =\frac{1}{2t} \left( \nabla \tilde{J}_i \cdot (\nabla
\omega - y/t) + (\nabla
  \omega -y/t)\cdot \nabla \tilde{J}_i \right) + O(t^{-2})
\end{equation*}
we find that
\begin{equation*}
(d\tilde{J}_i)^2 \leq \frac{C}{t^2} \, (\nabla \omega -y/t)
\cdot\chi_{[\lambda,\lambda'] } (|y|/t) (\nabla \omega -y/t) +
O(t^{-3}).
\end{equation*}
To bound the second term on the r.h.s. of \eqref{eq:dJAJ}, we use
that
\begin{equation*}
\tilde{J}_i A \tilde{J}_i = \tilde{J}_i (\nabla \omega - y/t)^2
\tilde{J}_i + O(t^{-\delta}) = (\nabla \omega - y/t)\tilde{J}_i^2
(\nabla \omega - y/t)+O(t^{-\delta}) .
\end{equation*}
We then find that
\begin{equation*}
\pm \left(d\tilde{J}_i A^{1/2} \tilde{J}_i + \tilde{J}_i A^{1/2}
d\tilde{J}_i \right) \leq \frac{C+1}{t} (\nabla \omega -y/t)
\cdot\chi_{[\lambda,\lambda']} (|y|/t) (\nabla \omega -y/t) +
O(t^{-1-\delta}).
\end{equation*}
By \eqref{eq:db} and \eqref{eq:dA1/2} we thus conclude that
\begin{equation*}
\begin{split}
-f \dGamma (d b(t)) F(|x|/t) f \geq\; &\frac{C_1}{t} f \dGamma
\left( |J(y/t)\cdot (\nabla \omega -y/t) + (\nabla \omega
-y/t)\cdot J(y/t)|\right)
F f \\
 &-\frac{C_2}{t} f \dGamma \left((\nabla \omega -y/t) \chi_{[\lambda,\lambda']} (|y|/t)
 (\nabla \omega -y/t)\right) Ff + O(t^{-1-\eta}),
\end{split}
\end{equation*}
where the second term on the right hand side is integrable by
Proposition \ref{prop:pe2}. This, together with \eqref{eq:pe3-1},
proves Eq.~\eqref{eq:pe3-claim} and completes the proof of the
proposition.
\end{proof}

\begin{lemma}\label{lm:pe3}
Let $A=(y/t - \nabla \omega)^2 + t^{-\delta}$, $0< \delta \leq 1$,
and assume that $J \in \tf (\R^3 , \R^3)$ ($J$ has three
components $J_i$, $i=1,2,3$). Then
\begin{itemize}
\item[i)] $[ A^{1/2} , J (y/t) ] = O(t^{-1+\delta/2})$.
\item[ii)] $dA^{1/2} = -\frac{1}{t} A^{1/2} + O(t^{-1-\delta/2})$ .
\item[iii)] Suppose that $\tilde{J} \in \tf (\R^3 , \R^3)$ with
$\tilde{J}_i =1$ on the support of $J_i$, for $i=1,2,3$. Then
\[ | J(y/t)
\cdot (y/t - \nabla \omega) + (y/t - \nabla \omega)\cdot J(y/t)|
\leq C \tilde{J} A^{1/2} \tilde{J} + O(t^{-\eta/2}), \] where
$\eta = \min (\delta , 1-\delta/2)$.
\end{itemize}
\end{lemma}
This Lemma is taken from \cite{DG1}. For the sake of completeness
its proof is included in this paper.

\begin{proof}
i) Writing $A^{1/2}=A A^{-1/2}$ and using the representation \[
A^{-1/2} = \frac{1}{\pi} \int_0^{\infty} \frac{ds}{\sqrt{s}} \,
\frac{1}{s+A} \] one finds that
\begin{equation}\label{eq:A,J}
[A^{1/2} , J ] = \frac{1}{\pi} \int_0^{\infty} ds \sqrt{s}
\frac{1}{s+ A} [A,J ] \frac{1}{s+ A}.
\end{equation}
With the help of Lemma \ref{lm:pdc1} it is easy to see that $[A,
J]= O(t^{-1})$ and, by definition of $A$, $\| (s+A)^{-1} \| \leq
(s + t^{-\delta})^{-1}$. Hence \eqref{eq:A,J} implies that \[ \|
[A^{1/2} , J ] \| \leq \frac{C}{t} \int_0^{\infty} ds
\frac{\sqrt{s}}{(s +t^{-\delta})^2} = O(t^{-1+ \delta/2}).\]

ii) The main observation is that
\begin{equation}\label{eq:A1/2}
e^{it\omega(k)} A^{1/2} e^{-it\omega(k)} = \left(\frac{y^2}{t^2} +
    t^{-\delta}\right)^{1/2}.
\end{equation}
On the one hand, by definition of $dA^{1/2}$,
\begin{equation}\label{eq:dt_A}
\frac{d}{dt} \left(e^{it\omega(k)} A^{1/2} e^{-it\omega(k)}\right)
= e^{it\omega(k)} dA^{1/2} e^{-it\omega(k)},
\end{equation}
and, on the other hand, by \eqref{eq:A1/2},
\begin{equation*}
\frac{d}{dt}\left(e^{it\omega(k)} A^{1/2} e^{-it\omega(k)}\right)
=\frac{d}{dt} \left(  \frac{y^2}{t^2} + t^{-\delta}\right)^{1/2} =
-\frac{1}{t} \left (  \frac{y^2}{t^2} + t^{-\delta}\right)^{1/2}
+O(t^{-1-\delta/2}) .
\end{equation*}
Combining these two equations and using \eqref{eq:A1/2} again
proves the assertion.

iii) First we note that
\begin{equation*}
|J \cdot (\nabla \omega -y/t) + (\nabla \omega -y/t)\cdot J|^2
\leq \sum_{i,j}
 J_i (\partial_i \omega -y_i /t) (\partial_j \omega - y_j/t) J_j + O(t^{-1})
\end{equation*}
Using that $a_i^* a_j + a_j^* a_i \leq a_i^* a_i + a_j^* a_j$ it
follows that
\begin{equation}\label{eq:JAJ}
\begin{split}
|J \cdot (\nabla \omega -y/t) + (\nabla \omega -y/t)\cdot J|^2
&\leq C\sum_i
 J_i (\partial_i \omega - y_i /t)^2 J_i +O(t^{-1}) \\
&\leq C J A J + O(t^{-\delta}).
\end{split}
\end{equation}
Furthermore, by part i), and since $\tilde{J}^4 \geq J^2$ by our
choice of $\tilde{J}$,
\begin{eqnarray*}
\left( \tilde{J} A^{1/2} \tilde{J}\right)^2 &=& \sum_{i,j}
\tilde{J}_i A^{1/2} \tilde{J_i} \tilde{J}_j A^{1/2} \tilde{J}_j
\ =\ A^{1/2} \tilde{J}^4 A^{1/2} + O(t^{-1+\delta/2})\\
&\geq& A^{1/2} J^2 A^{1/2} +
  O(t^{-1+\delta/2})\ =\ J A J +O(t^{-1+\delta/2}).
\end{eqnarray*}
Combined with \eqref{eq:JAJ} this shows that
\begin{equation*}
 \left( \tilde{J} A^{1/2} \tilde{J}\right)^2 \geq C \, |J \cdot
(\nabla \omega -y/t) + (\nabla \omega -y/t)\cdot J|^2 +
O(t^{-\eta}),
\end{equation*}
where $\eta = \min (\delta , 1-\delta/2 )$. The assertion now
follows from the operator monotonicity of the square root.
\end{proof}



\section{The Asymptotic Observable}\label{sec:observable}

Let $\beta, g$ and $\Sigma$ be given real numbers for which
\(\||\nabla\Omega|E_{\Sigma}(H)\|\leq \beta\). Let $\gamma>\beta$
and pick \(\chi_{\gamma}\in \C^{\infty}(\R;[0,1])\) such that
$\chi_{\gamma}\equiv 1$ on $[\gamma,\infty)$ and
$\chi_{\gamma}\equiv 0$ on $(-\infty,\beta_3]$ for some
$\beta_3\in(\beta,\gamma)$ (see Figure~\ref{fig:beta}). Our goal,
in this section, is to establish existence of the \emph{asymptotic
observable}
\[  W= s-\lim_{t\to\infty} e^{iHt} f \dGamma (\chi_{\gamma} (|y| /t)) f e^{-iHt}
,\] where $f$ is a smooth energy cutoff supported in
$(-\infty,\Sigma)$. By construction of $W$, $\sprod{\psi}{W\psi}$
is the expectation value of the number of bosons present in
$f\psi$ that propagate into the region \(\{|y|\geq \gamma t\}\) as
$t\to\infty$. These bosons are asymptotically free, since the
energy cutoff and the assumption on $\nabla\Omega$ guarantee that
the electron stays confined to \(\{|x|\leq \beta t\}\) (cf.
Proposition~\ref{prop:pe-el}) and since $\beta<\gamma$. As a
consequence, the interaction strength between the electron and
those bosons counted by $W$ decays in $t$ at an integrable rate.
This is one of the two key ingredients for proving existence of
$W$ and of the Deift-Simon operator $W_{+}$. The other one is the
propagation estimate in Proposition~\ref{prop:pe3}.


\begin{theorem}[Existence of the asymptotic observable]\label{thm:W_exists}
Assume that Hypotheses 0 -- 3 are satisfied. Let $\beta,\, g$, and
$\Sigma$ be real numbers for which
\(\||\nabla\Omega|E_{\Sigma}(H)\|\leq \beta\). Suppose that \(f\in
C_0^{\infty}(\R)\) with \(\supp(f)\subset(-\infty,\Sigma)\). Let
$\beta$, $\gamma$, and $\chi_{\gamma}$ be as defined above, and
let $\chi_{\gamma,t}$ be the operator of multiplication with
$\chi_{\gamma}(|y|/t)$. Then
\begin{equation*}
W = s-\lim_{t\to \infty} e^{iHt}f\dGamma(\chi_{\gamma,t})f
e^{-iHt}
\end{equation*}
exists, $W=W^{*}$ and $W$ commutes with $H$. Here $f=f(H)$, (as
before).
\end{theorem}

\begin{proof}
Pick $F\in C_0^{\infty}(\R)$ with \(0\leq F\leq 1 \), $F(s)=1$ for
\(s\leq \beta_0\), and \(F(s)=0\) for $s \geq \beta_1$, where
\(\beta<\beta_0<\beta_1<\beta_3<\gamma\) (see Figure
\ref{fig:beta}, Sect.~\ref{sec:inverse}). We also use $F$ to
denote the operator of multiplication by $F(|x|/t)$. By
Proposition~\ref{prop:pe-el}~(ii) applied to $1-F$, and since
\(e^{iHt}f\dGamma(\chi_{\gamma,t})\) is bounded, it suffices to
prove the existence of
\begin{equation*}
  \lim_{t\to\infty} \ph(t),\quad \text{where}
  \quad \ph(t)=e^{iHt}f\dGamma(\chi_{\gamma,t}) F fe^{-iHt}\ph.
\end{equation*}
By a variant of Cook's argument this limit will exist if there
exists a constant $C$ such that
\begin{equation*}
\int_1^{\infty}|\sprod{\psi}{\ph'(t)}|\, dt \leq C\|\psi\|
\end{equation*}
for all $\psi\in\H$. We have
\begin{equation}\label{eq:W123}
\begin{split}
\frac{d}{dt}\sprod{\psi}{\ph(t)} =& \sprod{\psi_t}{f D
\Big[\dGamma(\chi_{\gamma,t})F \Big]f\ph_t}\\
 =& \sprod{\psi_t}{f \dGamma(d\chi_{\gamma,t})F f\ph_t}
 + g \sprod{\psi_t}{ f \phi(i\chi_{\gamma,t} G_x)F f\ph_t}
 + \sprod{\psi_t}{ f \dGamma(\chi_{\gamma ,t})(DF) f\ph_t},
\end{split}
\end{equation}
and we shall prove integrability of these three terms, beginning
with the third one.

Since $\supp(F')\subset [\beta_0,\beta_1]$ and by
Lemma~\ref{lm:pdc1},
\begin{eqnarray*}
  DF &=&\frac{1}{t} \, F'\left(\frac{x}{|x|}\cdot\nabla\Omega - \frac{|x|}{t}\right) +
O(t^{-2})\\
 &=& \frac{1}{t} \chi_{[\beta_0,\beta_1]}(|x|/t)
     \left(\frac{x}{|x|}\cdot\nabla\Omega - \frac{|x|}{t}\right) F' + O(t^{-2})
\end{eqnarray*}
and hence, using that, by Hypothesis 0, $|\nabla \Omega|$ is
bounded w.r.t. $H$,
\begin{eqnarray}
\lefteqn{|\sprod{\psi_t}{f\dGamma(\chi_{\gamma,t})(DF) f\ph_t}| } \nonumber\\
  &\leq & \frac{1}{t} \|\chi_{[\beta_0,\beta_1]}f\psi\|\,
  \|\dGamma(\chi_{\gamma,t})F' f\ph_t\| + O(t^{-2}) \|\psi\|
\|\ph\|.\label{eq:W1}
\end{eqnarray}
On the right hand side the operator $F'f$ can be replaced by
\((H+i)^{-1}F'g(H)\), $g(s)=(s+i)f(s)$, at the expense of another
term of order $t^{-2}$ originating from   \(t^{-1}[H,F'] =
O(t^{-2})\). The integrability of \eqref{eq:W1} then follows from
Proposition~\ref{prop:pe-el}.

The second term on the r.h.s. of \eqref{eq:W123} is integrable
because \(|x|/t \leq \beta_1 \) on $\supp(F)$, while $|y|/t\geq
\beta_3$ on $\supp(\chi_{\gamma,t})$, and hence, by
Lemma~\ref{lm:srdecay},
\begin{eqnarray*}
|\sprod{\psi_t}{ f \phi(i\chi_{\gamma,t}G_x)F f\ph_t}| &\leq &
C\sup_{|x|/t\leq \beta_1} \|\chi_{\gamma,t}G_{x}\| \|\psi\| \|\ph\|\\
& \leq & \const \, \, t^{-\mu} \|\psi\|\, \|\ph\|,
\end{eqnarray*}
with $\mu>1$. This is integrable in $t$.

To bound the first term on the r.h.s. of \eqref{eq:W123}, we note
that
\begin{eqnarray*}
  d\chi_{\gamma,t} &=& \frac{1}{2}[(\nabla\omega-y/t)\cdot
  \nabla\chi_{\gamma,t} +
h.c.] + O(t^{-2})\\
  &=:& \frac{1}{t}P_t + O(t^{-2}),
\end{eqnarray*}
where  $1/t$ has been factored out from $\nabla\chi_{\gamma,t}=
(1/t)\chi_{\gamma}' (|y|/t) \, y/|y|$. It follows that
\begin{eqnarray*}
| \sprod{\psi_t}{f \dGamma(d\chi_{\gamma,t})F f\ph_t}| &\leq &
\frac{1}{t}\big|\sprod{\psi_t}{f F^{1/2} \dGamma(P_t)
F^{1/2}f\ph_t} \big| +
O(t^{-2}) \|\psi\| \|\ph\| \\
& \leq & \frac{1}{t} \sprod{\psi_t}{f F^{1/2}\dGamma(|P_t|)
F^{1/2} f\psi_t}^{1/2} \sprod{\ph_t}{f F^{1/2}\dGamma(|P_t|)
F^{1/2} f\ph_t}^{1/2}\\ && + O(t^{-2}) \|\psi\|
\|\ph\|. \\
\end{eqnarray*}
Since $F^{1/2}$ commutes with $\dGamma(|P_t|)$, this is integrable
thanks to Proposition~\ref{prop:pe3}.

To prove that $W$ commutes with $H$ we show that
$e^{-iHs}W=We^{-iHs}$ for all $s\in \R$. By definition of $W$
\begin{equation*}
[e^{-iHs}W e^{iHs}-W]\ph = \lim_{t\to\infty}
e^{iHt}f[\dGamma(\chi_{\gamma,\tau})]_{\tau=t}^{\tau=t+s}f
e^{-iHt}\ph.
\end{equation*}
This limit vanishes because
\(\partial_{\tau}\chi_{\gamma,\tau}=O(\tau^{-1})\) and hence
\(\|\dGamma(\chi_{\gamma,\tau})|_t^{t+s}(N+1)^{-1/2}\|\leq Cs/t\).
\end{proof}


\subsection{Positivity of $W$}\label{sec:Wpos}

The upper bound $\beta$ on the electron speed (cf.
Proposition~\ref{prop:pe-el}) could usually be chosen arbitrarily,
so far. Only in our proof of the existence of the wave operator we
required $\beta< 1$. To prove positivity of $W$, we must require
that $\beta< 1/3$.

Recall that $\sprod{\psi}{W\psi}$ is the number of bosons in
$f\psi$ with asymptotic speed $\gamma$ or higher, while the energy
cutoff $f$ in $W$ ensures that the speed of the electron does not
exceed $\beta$. By the positive commutator estimate, Theorem
\ref{thm:pos_comm2}, in a state orthogonal to $\Hgs$ with energy
in the support of $f$, the photons have a speed, relative to the
electron, of at least $1-\beta$. Their speed relative to the
origin is thus bounded below by $1-2\beta$. By assuming
$\gamma\leq 1-2\beta$ we we can ensure that these bosons are
counted by $W$. (Their number is positive by our smallness
assumption on $g$.) Since $\beta<\gamma$ is required for the
existence of $W$, we need to assume that $\beta<1/3$.


\begin{theorem}\label{thm:W_pos}
Assume Hypotheses 0 -- 3 are satisfied. Given $\beta<1/3$, pick
$\Sigma<O_{\beta}$ and suppose that $g_{\Sigma}>0$ is so small
that \(\sup_{|g|\leq g_{\Sigma}}\| |\nabla \Omega| E_{\Sigma}
(H_{g}) \| \leq \beta \) (cf. Hypothesis 2 and
Lemma~\ref{lm:grad-Omega2}). Pick \(\gamma\in (\beta,1-2\beta)\),
and let $W$ be defined as in Theorem~\ref{thm:W_exists}. Choosing
$g_{\Sigma}$ even smaller if necessary, there exists a constant
$C>0$ such that
\begin{equation*}
W\restricted{\Pgs^{\perp} \Gamma (\chi_i) \H} \geq C f(H)^2 .
\end{equation*}
for $|g|\leq g_{\Sigma}$. In particular, if $f=1$ on an interval
$\Delta \subset (-\infty , \Sigma)$, then
\begin{equation*}
W\restricted{E_{\Delta} (H) \Pgs^{\perp} \Gamma (\chi_i) \H} \geq
C > 0 .
\end{equation*}
\end{theorem}

\emph{Remark.} Our proof shows that $g_{\Sigma}=O(1-3\beta)$, as
$(1-3\beta)\to 0$, is sufficient if $\gamma>\beta$ is chosen close
to $\beta$.

\begin{proof}
Let $\mathcal{D} = D(\dGamma ( a )) \cap \ran \Pgs^{\perp} \Gamma
(\chi_i)$, where $a=1/2 (\nabla \omega \cdot (y-x) + (y-x)\cdot
\nabla \omega)$. Since $\mathcal{D}$ is dense in $ \ran
\Pgs^{\perp}\Gamma (\chi_i)$ (see Lemma \ref{lm:dicht} in Appendix
\ref{sec:inv}), and since $W$ is bounded, it suffices to prove
that there is a constant $C>0$ such that
\begin{equation}\label{eq:beha1}
  \sprod{\ph}{W\ph} \equiv
  \lim_{t \to \infty} \sprod{\ph_t}{f \dGamma (\chi_{\gamma,t}) f \ph_t}
  \geq C \| f \ph \|^2
\end{equation}
for all $\ph \in \mathcal{D}$. In the following $\ph \in
\mathcal{D}$ is fixed. The proof of \eqref{eq:beha1} is based on
estimates of \(\sprod{\ph_{t}}{f \dGamma (a/t) f \ph_{t}}\) from
above and from below. The upper bound relates \(\sprod{\ph_{t}}{f
\dGamma (a/t) f \ph_{t}}\) to $\sprod{\ph}{W\ph}$ and the lower
bound uses the positive commutator estimate,
Theorem~\ref{thm:pos_comm2}. We begin with the estimate from
above.

\emph{Step 1.} Let $\eps>0$. There exists a finite constant $C$
such that
\begin{eqnarray*}
\sprod{\ph_t}{f \dGamma(a/t) f\ph_t} &\leq& C
\sprod{f\ph_t}{\dGamma (\chi_{\gamma,t})f \ph_t}^{1/2} \|f\ph\|\\
& & + (\gamma+\beta+\eps) \sprod{\ph_t}{f N f \ph_t} + o(1),
\qquad t\to\infty.
\end{eqnarray*}

To see this, suppose \(F\in C^{\infty}(\R;[0,1])\),
\(\supp(F)\subset (-\infty,\beta+\eps]\) and $F(s)=1$ for $s\leq
\beta$. Then
\begin{eqnarray*}
\chi_{\gamma} (|y|/t) &\geq& \chi (|y|/t \geq \gamma) \\
&\geq & \chi (|x|/t \leq \beta + \eps)\, \chi (|x-y|/t \geq \gamma+\beta +\eps ) \\
&\geq &  F (|x|/t)\, \chi (|x-y|/t \geq \gamma+\beta +\eps).
\end{eqnarray*}
It follows that
\begin{equation}\label{eq:step30}
\begin{split}
  \sprod{\ph_t}{f \dGamma (\chi_{\gamma} (|y|/t)) f \ph_t} &\geq
  \sprod{\ph_t}{f F (|x|/t) \dGamma (\chi (|x-y|/t \geq \gamma+
  \beta + \eps)) f \ph_t}\\ &= \sprod{\ph_t}{f \dGamma (\chi
  (|x-y|/t \geq \gamma+ \beta +\eps)) f \ph_t} + o(1),
\end{split}
\end{equation}
where we used Proposition~\ref{prop:pe-el} to get rid of the
factor $F(|x|/t)$. Next we estimate the right side from below by
showing that
\begin{eqnarray}
\sprod{\ph_t}{f \dGamma(a/t) f\ph_t} &\leq& C
\sprod{f\ph_t}{\dGamma (\chi(|x-y|/t \geq \gamma+\beta+\eps))f
\ph_t}^{1/2}\|f\ph\|
\nonumber \\
& & + (\gamma+\beta +\eps) \sprod{\ph_t}{f N f \ph_t} + O(t^{-1}),
\qquad t\to\infty , \label{eq:step30b}
\end{eqnarray}
for some $\sigma$-dependent but finite constant $C$. Combined with
\eqref{eq:step30} this will prove Step 1.

From now on $\lambda:=\gamma+\beta+\eps$, $\chi\equiv \chi
(|x-y|/t \geq \lambda)$ and $\bar{\chi} \equiv 1 - \chi$, for
short. Using the identity
\(1=\Gamma(\bar{\chi})+(1-\Gamma(\bar{\chi}))\) we split each
photon wave function into parts in- and outside of the sphere
$|x-y|/t=\lambda$. We find the bound
\begin{multline}\label{eq:step31}
\begin{aligned}
\sprod{\ph_t}{f \dGamma (a/t) f \ph_t} =&
 1/2 \sprod{\ph_t}{f\dGamma (a/t) \Gamma(\bar{\chi}) f \ph_t} + h.c\\
&+ 1/2 \sprod{\ph_t}{f \dGamma (a/t)(1-\Gamma (\bar{\chi})) f \ph_t} + h.c.\\
\end{aligned}\\
\leq  \sprod{\ph_t}{f \dGamma \left(\bar{\chi} , 1/2 ((a/t) \,
\bar{\chi} + \bar{\chi}\, (a/t))\right) f \ph_t} + \| \dGamma
(a/t) f \ph_t \| \| (1 - \Gamma (\bar{\chi})) f \ph_t \|.
\end{multline}
To estimate the first term on the right hand side, note that
\(\dGamma(\bar{\chi},b)\leq \dGamma(b)\leq \|b\|N\) for every
symmetric one-photon operator $b$. Since
\begin{eqnarray*}
\| (a/t) \, \bar{\chi} \| &\leq& 1/t \, \| \nabla \omega (k) \cdot
(y-x)\, \bar{\chi}\| + 1/2t \| \Delta \omega (k) \, \bar{\chi} \|
\\ &\leq& \lambda + O(t^{-1}),
\end{eqnarray*}
one arrives at
\begin{equation}\label{eq:step31b}
\sprod{\ph_t}{f \dGamma [\bar{\chi}, 1/2 ((a/t) \, \bar{\chi} +
\bar{\chi} (a/t))] f \ph_t} \leq \lambda \sprod{\ph_t}{f Nf \ph_t}
+ O(t^{-1}).
\end{equation}
The first factor in the second term of \eqref{eq:step31} is
estimated by
\begin{equation}\label{eq:step32}
\| \dGamma (a/t) f \ph_t \| \leq C(\|f\ph\| + 1/t \|\dGamma(a)\ph
\|),
\end{equation}
by Lemma \ref{lm:inv1} (use $f=gf$, for a suitable $g\in
C_0^{\infty}(\R)$ to see this). This is finite, since $\ph \in
D(\dGamma (a))$ by assumption. For the second factor in the second
term of \eqref{eq:step31} we use that
\begin{equation}\label{eq:step33}
\| (1 - \Gamma (\bar{\chi})) f \ph_t \|^2 = \sprod{\ph_t}{f(1 -
\Gamma(\bar{\chi})) f \ph_t} \leq \sprod{\ph_t}{f \dGamma (\chi) f \ph_t} \\
\end{equation}
since $\bar{\chi}$ and hence $(1-\Gamma (\bar{\chi}))$ is a
projection. The bound \((1-\Gamma(\bar{\chi}))\leq \dGamma
(\chi)\) is easily verified on each $n$-boson sector separately.

After inserting \eqref{eq:step31b}, \eqref{eq:step32} and
\eqref{eq:step33} into \eqref{eq:step31} one arrives at
\eqref{eq:step30b}, which proves Step 1.
\\

\emph{Step 2.} For each $\delta > 0$, there is a sequence $t_n\to
\infty$ such that
\begin{equation}\label{eq:step1}
\sprod{\ph_{t_n}}{f \dGamma (a/t_{n}) f \ph_{t_n}} \geq
\frac{1}{1+ \delta} (1- \beta) \sprod{\ph_{t_n}}{f N f \ph_{t_n}}
- C_{M} g \| f \ph \|^2 + o(1)
\end{equation}
as $n\to\infty$.

By the positive commutator estimate, Theorem \ref{thm:pos_comm2},
\begin{equation*} \sprod{\ph_t}{f \dGamma (a) f \ph_t} \geq
\sprod{\ph}{f \dGamma (a) f \ph} + (1 - \beta) \int_0^t ds
\sprod{\ph_s}{f N f \ph_s} - C_{M} g t \| f \ph \|^2,
\end{equation*}
and, after dividing both sides by $t$,
\begin{equation*}
\sprod{\ph_t}{f \dGamma ( a/t ) f \ph_t} \geq (1 - \beta) \,
\frac{1}{t} \, \int_0^t ds \sprod{\ph_s}{f N f \ph_s} - C_{M} g
\|f \ph \|^2 + O(t^{-1}),
\end{equation*}
as $t\to\infty$. This inequality proves Step 2 thanks to the
following general fact: for every bounded, continuous function
$h(t)\geq 0$ and for each $\delta >0$, there exists a sequence
$t_n\to \infty$ such that
\begin{equation*}
m(t):= \frac{1}{t} \int_0^t ds h(s) \geq \frac{1}{1 + \delta} h(t)
\end{equation*}
for all $t\in \{ t_n\}_{n\in \N}$. In fact, the opposite
assumption that \( h(t) \geq (1+ \delta) m(t)\), for all $t>T_0$
and some $T_0 \in \R$, would imply that \[ \frac{d}{dt} \log m(t)
= \frac{m'(t)}{m(t)} \geq \frac{\delta}{t}
\] for all $t >T_0$. This is impossible since $m(t)$ is bounded.\\

Combining Steps 1 and 2 we get
\begin{eqnarray}\label{eq:step1+2}
C\|f\ph\| \sprod{\ph_{t_n}}{f \dGamma (\chi_{\gamma,t}) f
\ph_{t_n}}^{1/2} &\geq &
\left\{\frac{1}{1+\delta}(1-\beta)-(\gamma+\beta+\eps)\right\}\sprod{\ph_{t_n}}{f
Nf \ph_{t_n}}\\
& & - C_M g \|f\ph\|^2 + o(1),\qquad n\to\infty.\nonumber
\end{eqnarray}
Using $(1+\delta)^{-1}\geq 1-\delta$ and the assumption on
$\gamma$, one finds that $\{\ldots\}\geq
(1-2\beta-\gamma-\eps-\delta)\geq (1-2\beta-\gamma)/2>0$ for
$\eps$ and $\delta$ small enough. To bound the second factor on
the r.h.s of \eqref{eq:step1+2}, we use that $N\geq 1-P_{\Omega}$
and that \(fP_{\Omega}f \geq f \Pgs f - D_{\Sigma}|g|^{1/2} f^2\),
by the remark after Theorem~\ref{cor:gs-only} (here we use that
$\supp f \subset (-\infty , \Sigma)$). Since $\Pgs\ph=0$ by
assumption on $\ph$, we conclude that
\begin{equation*}
 \sprod{\ph_{t_n}}{f \dGamma (\chi_{\gamma,t}) f \ph_{t_n}}
 \geq \frac{1}{C}\left\{\frac{1}{2}(1-2\beta-\gamma)(1-D_{\Sigma}|g|^{1/2})
 -C_{M} |g| \right\}^2 \|f\ph \|^2 + o(1),
\end{equation*}
as $n\to\infty$. For $|g|$ small enough this proves
Eq.~\eqref{eq:beha1}, because
\(\lim_{n\to\infty}\sprod{\ph_{t_n}}{f \dGamma (\chi_{\gamma}) f
\ph_{t_n}}=\sprod{\ph}{W\ph}\) by Theorem~\ref{thm:W_exists}, and
the proof is complete.
\end{proof}


\section{The Inverse of the Wave Operator}\label{sec:inverse}

The purpose of this section is the construction of an operator
\(W_{+}:\H\to \HxF\) inverting the extended wave operator
$\tilde{\Omega}_+$ with respect to the asymptotic observable $W$;
that is \( W = \tilde{\Omega}_+ W_+ \). To this end one needs to
show that the dynamics of bosons that escape from the electron
ballistically - if there are any - is well approximated by the
free-boson dynamics. We shall prove this with the help of
Proposition~\ref{prop:pe3}, which was established for exactly this
purpose.

\begin{figure}
\begin{center}
\epsfig{file=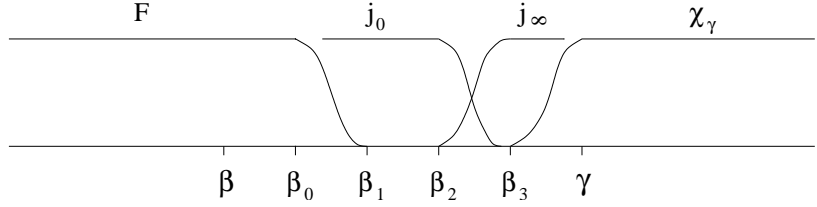,scale=.75}
\end{center}
\caption{Typical choice of the function $\chi_{\gamma}$, of the
electron space cutoff $F$ and of the partition in the photon space
$j_0, j_{\infty}$.\label{fig:beta}}
\end{figure}

Many elements in the construction of $W_{+}$ are familiar from the
construction of $W$. We recall from Section~\ref{sec:observable}
that $\beta,g$, and $\Sigma$ are real numbers with
\(\||\nabla\Omega|E_{\Sigma}(H)\|\leq \beta\) and that
$\gamma>\beta$. Then
\[  W_{+} := s-\lim_{t\to\infty} e^{i\Hex t} \tilde f
\uGamma(j_t)\dGamma(\chi_{\gamma,t}) f e^{-iHt} , \] where
$\tilde{f}=f(\Hex)$ and $f=f(H)$ are smooth energy cutoffs
supported in $(-\infty,\Sigma)$. As in
Section~\ref{sec:observable}, $\chi_{\gamma,t}$ is the operator of
multiplication with $\chi_{\gamma}(|y|/t)$ where $\chi_{\gamma}\in
C^{\infty}(\R;[0,1])$, $\chi_{\gamma}\equiv 1$ on
$[\gamma,\infty)$ and
\(\supp(\chi_{\gamma})\subset[\beta_3,\infty)\) for some
$\beta_3>\beta$. The purpose of \(\uGamma(j_t):\F\to \F\otimes\F\)
is to split each boson state into two parts, the second part being
mapped to the second Fock-space of prospective asymptotically
freely moving bosons. We introduce $\beta_1$ and $\beta_2$ such
\[  \beta < \beta_1 < \beta_2 < \beta_3 < \gamma  \]
and define $j_t:\h = L^2 (\R^3,dk) \to \h \oplus \h$ as follows:
let \(j_th=(j_{0,t}h,j_{\infty,t}h)\), where
$j_{\sharp,t}(y)=j_{\sharp}(|y|/t)$, $j_{\sharp}\in
C^{\infty}(\R;[0,1])$, $j_0+j_{\infty}\equiv 1$, $j_0\equiv 1$ on
$(-\infty,\beta_2]$, $\supp(j_0)\subset(-\infty,\beta_3]$ while
$j_{\infty}\equiv 1$ on $[\beta_3,\infty)$ and
$\supp(j_{\infty})\subset[\beta_2,\infty)$ (see
Figure~\ref{fig:beta}, below).

As in the last section, we work with the modified Hamiltonian
$\Hmod =\Omega (p) + \dGamma (\omega) + g \phi (G_x)$ and with the
extended modified Hamiltonian $\Hmodex = \Hmod \otimes 1 + 1
\otimes \dGamma (\omega)$, and we use the notation $H \equiv
\Hmod$, $\Hex \equiv \Hmodex$. Moreover, as in Section
\ref{sec:observable}, we use the notation  $DA$ and $D_0 A$ to
denote Heisenberg derivatives of operators $A$ on $\H$. If $B$ is
an operator on the extended Hilbert space $\HxF$, and if $C$ maps
$\H$ to $\HxF$ we set
\begin{align*}
DB &:= i[\Hex,B] + \frac{\partial B}{\partial t}\\ \tilde{D} C &:=
i\big(\Hex C-C H\big) +\frac{\partial C}{\partial t}.
\end{align*}
The derivatives $D_0$, and $\tilde{D}_0$ are defined in a similar
way, using $H_0$ and $\Hex_0$ instead of $H$ and $\Hex$. The
Heisenberg derivative of an operator $a$ on $L^2 (\R^3)$ is
denoted by $da=[i \omega (k) , a]+ \partial a / \partial t$.
Finally, the Heisenberg derivative $db$ of an operator $b$ mapping
the one-boson space $\h$ to $\h \oplus \h$ is defined by
\begin{equation*}
db = i\left(\begin{array}{cc} \omega & 0\\ 0 &\omega
\end{array}\right)b - b\,i\omega + \frac{\partial b}{\partial t}
=: \left(\begin{array}{l} db_0\\ db_{\infty}\end{array}\right).
\end{equation*}

\begin{theorem}[Existence of $W_{+}$]\label{thm:W+exists}
Assume Hypotheses 0 -- 3 are satisfied. Let $\beta, g$ and
$\Sigma$ be real numbers for which \(\| |\nabla \Omega| E_{\Sigma}
(H)\|\leq \beta\). Suppose that \(f\in C_0^{\infty}(\R)\) with
\(\supp(f)\subset (-\infty,\Sigma)\), and that \(\beta, \gamma\)
and $\chi_{\gamma}$ are defined as described above. Then
\begin{itemize}
\item[(i)] The limit
\[  W_{+} = s-\lim_{t\to\infty} e^{i\Hex t} \tilde f
\uGamma(j_t)\dGamma(\chi_{\gamma,t}) f e^{-iHt} \] exists, and
\(e^{-i\Hex s}W_{+} = W_{+}e^{-iHs}\), for all $s\in \R$.

\item[(ii)] \( ( 1\otimes \chi(N=0)) W_{+} = 0\).

\item[(iii)] \(W=\tilde \Omega_{+} W_{+}\).
\end{itemize}
\end{theorem}

\begin{proof}
Statement (ii) follows from \( (1\otimes \chi(N=0)) \uGamma(j_t) =
\uGamma(j_{0,t},0)\) and \(j_{0,t}\chi_{\gamma,t}=0\).

(i) Pick \(F\in C_0^{\infty}(\R)\) with $F(s)=1$ for $s\leq
\beta_0$ and $F(s)=0$ for $s\geq \beta_1$, where
$\beta_0\in(\beta,\beta_1)$. We also use $F$ to denote the
operator of multiplication with $F(|x|/t)$. By
Proposition~\ref{prop:pe-el}, it suffices to prove the existence
of
\begin{equation*}
\lim_{t\to \infty} \ph(t), \quad \text{where} \quad \ph(t) =
e^{i\Hex t} \tilde f \uGamma(j_t)\dGamma(\chi_{\gamma,t})F f
e^{-iHt}\ph
\end{equation*}
for all $\ph\in\H$. Using Cook's argument one is led to show that
\begin{equation*}
\int_1^{\infty} |\sprod{\psi}{\ph'(t)}| dt \leq C\|\psi\|
\end{equation*}
for all $\psi\in \H$. We have
\begin{equation}\label{eq:W+123}
\begin{split}
\frac{d}{dt}\sprod{\psi}{\ph(t)} = \; & \sprod{\psi_t}{\tilde
f\tilde
D\Big[\uGamma(j_t)\dGamma(\chi_{\gamma,t})F\Big]f\ph_t}\\
 =\; & \sprod{\psi_t}{\tilde f \udGamma(j_t,dj_t)\dGamma(\chi_{\gamma,t})F f\ph_t}\\
 & + \sprod{\psi_t}{\tilde f \uGamma(j_t)\dGamma(d\chi_{\gamma,t})F f\ph_t}\\
 & + g \sprod{\psi_t}{\tilde f \Big[\big(i\phi(G_x)\otimes
1\big)\uGamma(j_t)-\uGamma(j_t)i\phi(G_x)\Big]\dGamma(\chi_{\gamma,t})Ff\ph_t}\\
 & + g \sprod{\psi_t}{\tilde f \uGamma(j_t)\phi(i\chi_{\gamma,t}G_x)F f\ph_t}\\
 & + \sprod{\psi_t}{\tilde f \uGamma(j_t)\dGamma(\chi_{\gamma,t})(DF) f\ph_t}.\\
\end{split}
\end{equation}
We now prove integrability of all these terms, beginning with the
last one. Since
\begin{equation*}
DF = \frac{1}{t}\chi_{[\beta_0,\beta_1]}(|x|/t) \left(F'
\frac{x}{|x|}\cdot \nabla\Omega - \frac{|x|}{t}
F'\right)\chi_{[\beta_0,\beta_1]}(|x|/t) + O(t^{-2})
\end{equation*}
the last term on the r.h.s. of \eqref{eq:W+123} is integrable, by
Proposition~\ref{prop:pe-el} and the remarks thereafter and
because $|\nabla \Omega|$ is bounded w.r.to $H$, by Hypothesis 0.
See the proof of Proposition~\ref{prop:pe1} for details.

The second but last term on the r.h.s. of \eqref{eq:W+123} decays
like $t^{-\mu}$ with $\mu>1$, because \(|x|/t\leq \beta_1\) on
$\supp(F)$, \(|y|/t\geq \beta_3\) on $\supp(\chi_{\gamma,t})$ and
hence \(|x-y|\geq
 t (\beta_3 - \beta_1)\) on \(\supp(\chi_{\gamma,t}G_x F)\);
(see the proof of Proposition~\ref{prop:pe1} for details). Similar
remarks prove the integrability of the third term, because
\begin{equation*}
  [\phi(G_x)\otimes 1 ]\uGamma(j_t) - \uGamma(j_t)\phi(G_x) =
  \big[\phi((1-j_{0,t})G_x)\otimes 1 - 1\otimes
\phi(j_{\infty}G_x)\big]\uGamma(j_t) ,
\end{equation*}
where $1-j_{0,t}$ and $j_{\infty,t}$ are supported in \(|y|/t \geq
\beta_2\), while \(|x|/t \leq \beta_1\) on $\supp(F)$, hence
$|x-y|\geq t (\beta_2 - \beta_1)$.

The integrability of the first and second term on the r.h.s. of
\eqref{eq:W+123} will follow from the improved propagation
estimate in Proposition~\ref{prop:pe3}. For the second term we use
that \(\uGamma(j_t)\dGamma(d\chi_{\gamma,t}) = \udGamma(j_t,j_t
d\chi_{\gamma,t})\) where
\begin{eqnarray*}
  j_t d\chi_{\gamma,t} &=& \frac{1}{2}\big[(\nabla\omega-y/t)\cdot\nabla
\chi_{\gamma,t}j_t +
  j_t \nabla\chi_{\gamma,t}\cdot(\nabla\omega-y/t)\big] +O(t^{-2})\\
  &=:& \frac{1}{t} P_{t} + O(t^{-2}),
\end{eqnarray*}
where one power of $1/t$ has been factored out from $\nabla
\chi_{\gamma,t} = (1/t) \chi_{\beta ,t}^{\prime} (|y| /t) y /|y|$.
The error term $O(t^{-2})$ is integrable. By
Lemma~\ref{lm:udGamma} and since \(P_{0,t}=0\), $P_{0,t}$ being
the first component of $P_{t}=(P_{0,t},P_{\infty,t})$,
\begin{eqnarray*}
  \lefteqn{\big|\sprod{\psi_t}{\tilde f \udGamma(j_t,P_{t}) F f
\ph_t}\big|}\\
&\leq& \sprod{\tilde f\psi_t}{\big[1\otimes
\dGamma(|P_{\infty,t}|)\big]F\tilde f\psi_t}^{1/2}
\sprod{f\ph_t}{\dGamma(|P_{\infty,t}|)Ff\ph_t}^{1/2}.
\end{eqnarray*}
This is integrable by Proposition~\ref{prop:pe3} and the remarks
thereafter.

Finally, we estimate the first term on the r.h.s. of
\eqref{eq:W+123}. Let \(K_t=1/2((\nabla\omega-y/t)\cdot\nabla j_t
+ h.c.)\) and let the operator $\underline{\chi}_{\gamma}$ be
defined by
\(\underline{\chi}_{\gamma}(h_1,h_2)=(0,\chi_{\gamma,t}h_2)\) on
\(L^2(\R^3) \oplus L^2(\R^3)\). Then \(dj_t = K_t + O(t^{-2})\),
$j_t\chi_{\gamma,t} = \underline{\chi}_{\gamma}j_t$ and
\(K_t\chi_{\gamma,t} = \underline{\chi}_{\gamma} K_t +
O(t^{-2})\). Therefore
\begin{equation*}
\udGamma (j_t,dj_t) \dGamma(\chi_{\gamma,t}) = \big[1\otimes
\dGamma(\chi_{\gamma,t})\big]\dGamma(j_t,K_t) + O(t^{-2})N^2.
\end{equation*}
We write
\begin{equation}\label{eq:W+2}
   \big[1\otimes \dGamma(\chi_{\gamma,t})\big]\udGamma(j_t,K_t) =
\udGamma(j_t,\underline{\chi}_{\gamma}K_t) +UR_{t}
\end{equation}
where $R_t$ is defined by this equation and $U$ is as in
$\udGamma=U\dGamma$. The term
\(\udGamma(j_t,\underline{\chi}_{\gamma}K_t)\) is treated very
much like \(\udGamma(j_t,j_t d\chi_{\gamma,t})\) above, and it
leads to an integrable contribution thanks to the choice of
$\supp(\nabla j)$ and Proposition~\ref{prop:pe1}. On $\otimes_s^n
L^2(\R^3)$ the operator $R_t$ is given by
\begin{equation*}
  \sum_{l=1}^n \sum_{k=1,\, k\neq l}^n j_t\otimes \ldots
\underbrace{(\underline{\chi}_{\gamma}j_t)}_{kth}\otimes \ldots
\underbrace{K_t}_{lth}\ldots\otimes j_t.
\end{equation*}
From the defining equation \eqref{eq:W+2} for $R_t$ and from
Lemma~\ref{lm:udGamma} it is plausible that
\begin{eqnarray}\label{eq:W+3}
  \lefteqn{|\sprod{\psi_t}{\tilde f UR_t F f \psi_t}|} \nonumber \\
    &\leq & \sprod{\psi_t}{\tilde f \big[1\otimes \dGamma(|K_{\infty,t}|)\big]
N_{\infty}^2 F \tilde f \psi_t}^{1/2}
           \sprod{\ph_t}{f \dGamma(|K_{\infty,t}|) F f \ph_t}^{1/2} \\
      & & +  \sprod{\psi_t}{\tilde f \big[\dGamma(|K_{0,t}|)\otimes N^2\big] F
\tilde f \psi_t}^{1/2}
           \sprod{\ph_t}{f \dGamma(|K_{0,t}|) F f \ph_t}^{1/2}.\nonumber
\end{eqnarray}
To prove this, we return to the proofs of Lemma~\ref{lm:dGamma}
and Lemma~\ref{lm:udGamma} with $K_t=r_2^{*} r_1$, and
$r_2^{*}r_2=|K_{\sharp,t}|=r_1^{*}r_1$. The number operators in
\eqref{eq:W+3} prevent us from applying
Proposition~\ref{prop:pe3}. We choose \(g\in C_{0}^{\infty}(\R)\)
with \(\supp(g)\subset (-\infty,\Sigma)\) and $gf=f$. Then
\begin{equation*}
N_{\infty}\tilde f \psi_t = g(\Hex) e^{-i\Hex t}(N_{\infty}\tilde
f)\psi
\end{equation*}
where $N_{\infty}\tilde f$ is a bounded operator. Now the
integrability of \eqref{eq:W+3} follows from \(\supp(\nabla
j)\subset \{\beta_2 \leq |y| \leq \beta_3 \}\) and
Proposition~\ref{prop:pe3}.

The second assertion in (i) is proved in the same way as the
corresponding statement for $W$. By definition of $W_{+}$
\begin{equation*}
  \left[ e^{-i\Hex s} W_{+} e^{iHs} - W_{+}\right] \ph =
  \lim_{t\to\infty} e^{i\Hex t}\tilde f
  \big[\uGamma(j_t)\dGamma(\chi_{\gamma,t})\big]_t^{t+s} f e^{-iHt}\ph.
\end{equation*}
Since \(\partial_t j_{t} = O(t^{-1})\) and \(\partial_t
\chi_{\gamma,t}= O(t^{-1})\) we conclude that
\begin{equation*}
  \frac{d}{dt} \uGamma(j_t)\dGamma(\chi_{\gamma,t})f =
  \big[\udGamma(j_t,\partial_tj_t) \dGamma(\chi_{\gamma,t}) +
\uGamma(j_t)\dGamma(\partial_t \chi_{\gamma,t})\big] f = O(t^{-1})
\end{equation*}
and hence
$\|\big[\uGamma(j_t)\dGamma(\chi_{\gamma,t})\big]_t^{t+s}
f\|=O(t^{-1})$.

It remains to prove (iii). Recall from Eq.~\eqref{eq:Igamma} that
$I\Gamma(j_t)=1$, because $j_0+j_{\infty}=1$. Furthermore
\begin{equation}\label{eq:W+4}
  I\tilde f \uGamma(j_t) F = fF + o(1), \qquad (t\to\infty).
\end{equation}
as can be shown using Lemma~\ref{lm:comestnew1} in
Appendix~\ref{sec:comm_est} (see the proof of Lemma~16 in
\cite{FGS2} for details). Let \(g\in C_0^{\infty}(\R)\) with
$gf=f$, and let \(\tilde g=g(\Hex)\). By definition of $W$,
Proposition~\ref{prop:pe-el}, and by \eqref{eq:W+4},
\begin{eqnarray*}
W\ph &=& e^{iHt}fF\dGamma(\chi_{\gamma,t})f e^{-iHt}\ph + o(1)\\
     &=& e^{iHt} I\tilde g \Big(e^{-i\Hex t}e^{i\Hex t}\Big)\tilde f
     \uGamma(j_t) F
\dGamma(\chi_{\gamma,t})f e^{-iHt}\ph
     + o(1)\\
     &=& e^{iHt} I\tilde g e^{-i\Hex t} W_{+}\ph +o(1),
\end{eqnarray*}
where the last step uses that $I\tilde g$ is a bounded operator.
Since \(\tilde g W_{+}= W_{+}\) the assertion follows.
\end{proof}


\section{Putting It All Together: Asymptotic Completeness}\label{sec:AC}

As explained in the introduction, we prove asymptotic completeness
by induction in the energy measured in units of $\sigma/2$,
$\sigma$ being the infrared cutoff. The first step is the
following essentially trivial lemma. The idea is that AC on
$E_{\eta}(H)$, as characterized by Eq.~\eqref{eq:decay}, implies the
same property for $Ie^{-i\Hex t}$ on \(E_{\eta} (H)\otimes\F\),
the photons from $\F$ merely contributing to the asymptotically
free radiation.

\begin{lemma}\label{lm:induction}
Assume that Hypotheses 0 -- 3 are satisfied. Suppose $g$ and
$\Sigma>\inf \sigma (H)$ are real numbers for which $\| | \nabla
\Omega (p) | E_{\Sigma} (H) \| < 1$. Let the wave operators
$\tilde{\Omega}_+$ and $\Omega_+$ be defined as in Lemma
\ref{lm:tilde-Omega-mod} and in Theorem \ref{thm:Omega_+}, respectively.
Suppose $\ran \Omega_+ \supset
E_{\eta} (H) \H$, for some $\eta < \Sigma$. Then, for every $\ph
\in \ran E_{\Sigma} (\Hex)$, there exists a $\psi \in \ran
E_{\Sigma} (\Hex)$ such that
\[\tilde{\Omega}_{+}(E_{\eta}(H)\otimes 1)\ph = \Omega_{+}\psi.\]
If $\Delta \subset ( -\infty , \Sigma)$ and \(\ph\in E_{\Delta}(\Hex)\HxF\) then
\(\psi\in E_{\Delta}(\Hex)\HxF\).
\end{lemma}

\begin{proof}
By Lemma \ref{lm:dps} (Appendix~\ref{app:dps}), every given $\ph
\in \ran E_{\Sigma} (\Hex)$ can be approximated by a sequence of
vectors $\ph_n \in E_{\Sigma} (\Hex)$ which are finite linear
combinations of vectors of the form
\begin{equation}\label{eq:ind1}
\gamma = \alpha \otimes a^* (h_1) \dots a^*(h_n) \Omega, \hspace{3em}
\lambda + \sum_{i=1}^n M_i <\Sigma ,
\end{equation}
for some $\lambda$, where \(\alpha = E_{\lambda}(H)\alpha\) and
\(M_i=\sup\{|k|:h_i(k)\neq 0\}\). Let $\gamma\in\HxF$ be of the
form \eqref{eq:ind1}. Then
\begin{equation}\label{eq:ind2}
\begin{split}
e^{iH t} I e^{-i\Hex t} (E_{\eta}(H) \otimes 1) \, \gamma
&= e^{iH t} a^{*} (h_{1,t}) \dots a^{*}(h_{n,t}) \,
e^{-iH t} \,  E_{\eta}(H)\alpha \\ &= a_+^* (h_1) \dots a_+^* (h_n)
E_{\eta}(H)\alpha + o(1),
\end{split}
\end{equation}
as $t \to \infty$. By assumption, $E_{\eta}(H)\alpha = \Omega_+ \beta$, for some
$\beta\in \HxF$, and we may assume that $\beta=E_{\eta}(\Hex)\beta$, thanks to
the intertwining relation for
$\Omega_{+}$. From \eqref{eq:ind2} it follows that
\begin{equation*}
\begin{split}
\tilde{\Omega}_+ (E_{\eta}(H)\otimes 1)\gamma &=  a_+^* (h_1) \dots a_+^* (h_n)
\Omega_+ \beta \\
&= \Omega_+ (1 \otimes a^* (h_1) \dots a^* (h_n)) \beta,
\end{split}
\end{equation*}
where, in the second equation, we have used
Lemma~\ref{lm:wo_prop}. Hence, to each vector $\ph_n$ as in
Eq.~\eqref{eq:ind1}, there corresponds a vector $\psi_n\in
E_{\mu}(H)\HxF$ such that \(\tilde{\Omega}_{+}(E_{\eta}(H)\otimes
1)\ph_n = \Omega_{+}\psi_n.\) The left side converges to
\(\tilde{\Omega}_{+}(E_{\eta}(H)\otimes 1)\ph\), as $n\to \infty$,
and hence the right side converges as well. Since $\Omega_{+}$ is
isometric on $\Hgs \otimes\F$, it follows that $(\Pgs \otimes
1)\psi_n$ is Cauchy and hence has a limit $\psi\in
E_{\mu}(H)\HxF$. Thus \( \tilde{\Omega}_{+}(E_{\eta}(H)\otimes
1)\ph = \Omega_{+}\psi\) which proves the lemma.
\end{proof}

\begin{theorem}\label{thm:AC}
Assume Hypotheses 0 -- 3 are satisfied. Suppose that
$\Sigma>\inf\sigma (H)$ and $g_0 >0$ are so small that
$\||\nabla\Omega| E_{\Sigma} (H) \| < 1/3$, for all $g < g_0$.
Then, if $g < g_0$ is sufficiently small (compared to $(1-3 \||
\nabla \Omega | E_{\Sigma} (H) \| )$)
\[ \Ran \Omega_+ \supset E_{(-\infty , \Sigma)} (H) \H . \]
\end{theorem}

\begin{proof} The proof is by induction in energy steps of size $m=\sigma/2$.
We show that
\begin{equation}\label{eq:to-prove}
\Ran(\Omega_{+})
\supset E_{(-\infty,\Sigma-km)}(H) \H ,
\end{equation}
for $k=0$, by
proving it for all \(k\in \{0,1,2,\ldots\}\).  Since $H$ is
bounded below, \eqref{eq:to-prove} is certainly correct for $k$
large enough.  Assuming that \eqref{eq:to-prove} holds for $k=n+1$,
we now prove it for $k=n$. Since $\Ran \Omega_{+}$ is closed, by
Theorem~\ref{thm:Omega_+}, it suffices to prove that \(\Ran
\Omega_{+}\supset E_{\Delta}(H)\H\) for all compact intervals
\(\Delta\subset (-\infty,\Sigma-nm)\), which is equivalent to
\begin{equation*}
  \Ran{\Omega_{+}} \supset \Pgs^{\perp} \Gamma (\chi_i) E_{\Delta}(H)\H,
\end{equation*}
by Lemma~\ref{lm:ran-Omega} and because \(\Ran \Omega_{+}\supset
\Hgs\). Choose \(f\in C_0^{\infty}(\R;\R)\) with $f=1$ on $\Delta$
and \(\supp(f)\subset(-\infty,\Sigma)\), and define $W$ in terms of
$f$ as in Theorem~\ref{thm:W_exists}. By Theorem~\ref{thm:W_pos},
the operator \(\Gamma(\chi_i)\Pgs^{\perp} W \Pgs^{\perp} \Gamma
(\chi_i)\) is positive on \(\Pgs^{\perp} \Gamma (\chi_i)
E_{\Delta}(H)\H\), and hence onto, if $g<g_0$ is small enough.
Given $\psi$ in this space we can thus find \(\ph = \Pgs^{\perp}
\Gamma (\chi_i)\ph\) such that
\begin{equation*}
  \Pgs^{\perp} \Gamma (\chi_i) W\ph =\psi.
\end{equation*}
By Theorem~\ref{thm:W+exists}, \(W\ph=\tilde \Omega_{+}W_{+}\ph\)
and \(W_{+}\ph = E_{\Sigma-mn}(\Hex)W_{+}\ph\). Furthermore, by
part (ii) of Theorem~\ref{thm:W+exists}, $W_{+}\ph$ has at least
one boson in the outer Fock space, and thus an energy of at most
\(\Sigma-(n+1)m\) in the inner one. That is,
\begin{equation*}
  W_{+}\ph = [E_{\Sigma-(n+1)m}(H)\otimes 1]W_{+}\ph ,
\end{equation*}
and we can now use the induction hypothesis \(\Ran \Omega_{+}
\supset E_{\Sigma-(n+1)m}(H)\H\). Using Lemma~\ref{lm:induction},
it follows that \(\tilde \Omega_+ W_{+}\ph = \Omega_{+}\gamma\)
for some \(\gamma\in E_{\Delta}(\Hex)\H\). We conclude that
\begin{eqnarray*}
  \psi &=& \Gamma(\chi_i) \Pgs^{\perp} \Omega_{+} \gamma\\
       &=& \Gamma(\chi_i) \Omega_{+} (1\otimes P_{\Omega}^{\perp})\gamma\\
       &=& \Omega_{+} (\Gamma(\chi_i)\otimes \Gamma(\chi_i) P_{\Omega}^{\perp})\gamma,
\end{eqnarray*}
where $P_{\Omega}^{\perp}$ is the projection onto the orthogonal
complement of the vacuum. This proves the theorem.
\end{proof}



\section{Outlook}
\label{sec:out}

It is clear that the infrared cutoff $\sigma >0$ has played an
unpleasantly crucial role in our proof of AC for Compton
scattering. We do not know how to remove this cutoff in several
key estimates used in our proof; see Sect. \ref{sec:inverse}.

However, the construction of a suitable M\o ller wave operator in
the limit $\sigma \to 0$ has been accomplished by Pizzo \cite{P},
using results of \cite{Froe1} and of \cite{Chen}.

In the presence of an infrared cutoff we are also able to
construct M\o ller wave operators for the scattering theory of
$N\geq 2$ conserved electrons interacting with scalar bosons or
photons. The proof follows arguments used in Haag--Ruelle
scattering theory; see \cite{Jost} and refs. given there. However,
because the models studied here are neither Galilei--, nor Lorentz
covariant, in particular, because the dispersion law $E_g(P)$ of
dressed one--electron states does not reflect any symmetries other
than Euclidian motions and hence the center of mass motion of
bound clusters does not factor out, there are no methods known to
us enabling one to attack the problem of proving AC for the
scattering of many electrons.

By combining the methods developed in this paper with those in
\cite{FGS2} and with elements of Mourre theory for Schr{\"o}dinger
operators, we expect to be able to extend the results of this
paper to a model, where the electron moves under the influence of
a screened electrostatic force generated by some static nuclei. We
thus expect to be able to describe scattering processes
corresponding to ionization of an atom and electron capture by a
nucleus (Bremsstrahlung).

\appendix

\section{Functional Calculus}\label{sec:HScalc}
The Helffer-Sj{\"o}strand Functional Calculus is a useful tool in
the computation of commutators of functions of self adjoint
operators. Suppose that $f\in C_0^{\infty}(\R;\C)$ and that $A$ is
a self adjoint operator. A convenient representation for $f(A)$,
which is often used in this paper, is then given by
\begin{equation*}
f (A) = -\frac{1}{\pi} \int
dxdy\,\frac{\partial\tilde{f}}{\partial\bar{z}} (z)\, (z -A)^{-1},
\hspace{2em}z=x+iy,
\end{equation*}
which holds for any extension $\tilde{f}\in C_0^{\infty}(\R^2;\C)$
of $f$ with $|\partial_{\bar{z}}\tilde{f}|\leq C|y|$,
\begin{equation}\label{eq:alman}
\tilde{f} (z) = f(z), \makebox[5em]{and}
\frac{\partial\tilde{f}}{\partial\bar{z}} (z)
=\frac{1}{2}\left(\frac{\partial f}{\partial x}+ i\frac{\partial
f}{\partial y} \right)(z)= 0, \makebox[5em]{for all}z \in \R.
\end{equation}
Such a function $\tilde{f}$ is called an {\em almost analytic
extension} of $f$. A simple example is given by \(\tilde{f} (z) =
(f(x) + i y f^{\prime} (x)) \, \chi (z)\) where $\chi \in
C_0^{\infty} (\R^2)$ and $\chi = 1$ on some complex neighborhood
of $\supp f$. Sometimes we need faster decay of
$|\partial_{\bar{z}}\tilde{f}|$, as $|y|\to 0$; namely
$|\partial_{\bar{z}}\tilde{f}|\leq C |y|^n$. We then work with the
almost analytic extension
\begin{equation*}
\tilde{f}(z) = \left(\sum_{k=0}^n
f^{(k)}(x)\frac{(iy)^k}{k!}\right)\chi(z) ,
\end{equation*}
with $\chi$ as above. We call this an almost analytic extension
{\em of order $n$}. For more details and extensions of this
functional calculus the reader is referred to \cite{HunSig} or
\cite{Dav}.

To estimate commutators involving $\Omega (p)= \sqrt{p^2 + M^2}$
we will use the following lemma.

\begin{lemma}\label{lemma:Omega-comm}
Let $B$ be an operator on $\mathcal{H}$. Then
\begin{equation}\label{eq:commut_omega}
[ \Omega (p) , B ] = \frac{1}{\pi} \int_{M^2}^{\infty} dy \,
\frac{\sqrt{y-M^2}}{y+p^2} \, [ p^2 , B ]\,\frac{1}{y+p^2}.
\end{equation}
\end{lemma}

\section{Pseudo Differential Calculus}\label{app:pdc}

In order to compute commutators of functions of the
momentum-coordinates with functions of the position-coordinates
the following lemma is very useful.
\begin{lemma}\label{lm:pdc1}
Suppose $f\in \mathcal{S}(\R^d)$, $g\in C^{n}(\R^{d})$ and
$\sup_{|\alpha|=n}\|\partial^{\alpha}g\|_{\infty}<\infty$. Let
$p=-i\nabla$. Then
\begin{align*}
i[g(p),f(x)]&= i\sum_{1\leq|\alpha|\leq
n-1}\frac{(-i)^{|\alpha|}}{\alpha!}
(\partial^{\alpha}f)(x)(\partial^{\alpha}g)(p) + R_{1,n}\\
&= (-i)\sum_{1\leq|\alpha|\leq n-1}\frac{i^{|\alpha|}}{\alpha!}
(\partial^{\alpha}g)(p)(\partial^{\alpha}f)(x) + R_{2,n}
\end{align*}
where \[\Vert R_{j,n}\Vert\leq C_n
\sup_{|\alpha|=n}\|\partial^{\alpha}g\|_{\infty} \int
dk\,|k|^n|\hat{f}(k)|.\] In particular, and most importantly, if
$n=2$ then
\begin{align*}
i[g(p),f(\eps x)] &= \eps \nabla g(p)\cdot\nabla f(\eps x) + O(\eps^2)\\
   &= \eps \nabla f(\eps x)\cdot\nabla g(p) + O(\eps^2),
\end{align*}
as $\eps \to 0$.
\end{lemma}
For the proof of this lemma see \cite{FGS2}.

\section{Representation of States in $\chi (\Hex \leq c) \HxF$}
\label{app:dps}

The representation of states in \(\Ran\chi (\Hex \leq c)\) proved
in this section is used in Section \ref{sec:waveop} to prove the
existence of the wave operator and in Lemma~\ref{lm:induction}.
See \cite{FGS2} for the proofs.

\begin{lemma}\label{lm:slater}
Suppose $\omega(k)=|k|$ or that $\omega$ satisfies Hypothesis 3,
and let $c>0$. Then the space of linear combinations of vectors of
the form \(a^*(h_1)\ldots a^*(h_n)\Omega\) with $h_i\in
L^2(\R^{d})$ and \(\sum_{i=1}^n\sup\{\omega(k):k\in
\supp(h_i)\}\leq c\) is dense in $\chi(\dGamma(\omega)\leq c)\F$.
\end{lemma}

\begin{lemma}\label{lm:dps} 
Suppose that $\omega$ satisfies Hypothesis 3, set $H=\Hmod$ and
$\Hex =\Hmodex$, where $\Hmod$ and $\Hmodex$ are the Hamiltonians
on $\H$ and on $\HxF$ introduced in Section \ref{sec:modham}. Let
$c>0$. Then the set of all linear combinations of vectors of the
form
\begin{equation}\label{eq:dps1}
\ph\otimes a^*(h_1)\ldots a^*(h_n)\Omega,
\hspace{2em}\lambda+\sum_{i=1}^N M_i \leq c
\end{equation}
where \(\ph=\chi(H\leq \lambda)\ph\) for some $\lambda \leq c$,
$n\in\N$ and \(M_i=\sup\{\omega(k):h_i(k)\neq 0\}\), is dense in
$\chi(\Hex\leq c)\HxF$.
\end{lemma}

\section{Spectral Results}

In the first subsection  of this appendix we prove the existence
of ground state vectors for $H_g (P)$, which are used in Section
\ref{sec:DES} to construct the dressed electron states (DES). In
the second subsection we prove a version of the Virial Theorem for
the modified Hamiltonian $\Hmod (P)$ introduced in Section
\ref{sec:modham}, which together with the positive commutator
discussed in Section \ref{sec:poscomm} allows us to prove the
absence of eigenvalues of $H_g (P)$ above its ground state energy.

\subsection{Existence of DES}\label{app:DES}

Our proof that \( E_g (P) = \inf \sigma (H_g (P)) \) is an
eigenvalue of the Hamiltonian $H_g (P)$ for $\sigma>0$ relies on
the Lipshitz property
\begin{equation}\label{eq:Lipschitz}
\inf_{|k|\geq\eps} \big\{E_g(P-k)+|k|-E_g(P)\big\} >0
\end{equation}
valid whenever $\eps>0$, $\Omega(P)<O_{\beta=1}$, and $|g|$ is
small enough. To prove Eq.~\eqref{eq:Lipschitz}, we argue by way
of perturbation theory and we use that
\begin{equation}\label{eq:gs1}
(1 - \alpha) E_0 (P) - \frac{g^2}{\alpha} \, \int
\frac{|\kappa(k)|^2}{|k|}\, dk \leq E_g (P) \leq \Omega(P)
\end{equation}
for all $P\in \R^3, g\in \R$ and \(\alpha\in (0,1]\). The upper
bound in \eqref{eq:gs1} follows from
\(\sprod{\Omega}{\phi(\kappa_{\sigma})\Omega}=0\) (Rayleigh--Ritz
principle) and the lower bound from \( H_g(P) \geq (1-\alpha)
H_0(P) + \alpha \dGamma(|k|) + g \phi(\kappa_{\sigma})\) and from
Lemma~\ref{lm:estim}. Note that the lower bound is independent of
the IR cutoff $\sigma$, because, by Hypothesis 1, $\kappa_{\sigma}
(k) = \kappa (k) \chi (|k| /\sigma)$, and $0 \leq \chi \leq 1$.

\begin{lemma}\label{lm:gap}
Assume Hypotheses 0 -- 2 and define \(B:=
\sup_P\|\partial^2\Omega(P)\|<\infty\) and \(C:=
\int|\kappa(k)|^2/|k|\, dk <\infty\) . If $\beta<1$,
$\Omega(P)\leq O_{\beta}$, and
\[|g| < g_{\beta}:=\min\left(1, \frac{(1-\beta)^{3/2}}{3(BC)^{1/2}},
\frac{(1-\beta)^2}{3B(C+O_{\beta})}\right) \] then, for all
$\eps>0$, Eq.~\eqref{eq:Lipschitz} holds true.
\end{lemma}

\begin{proof}
For shortness we write $P_f$ and $H_f$ instead of $\dGamma (k)$
and $\dGamma (|k|)$ in the following. Let $P\in\R^3$ with
\(\Omega(P)\leq O_{\beta}\) be fixed. Given \(\delta>0\) and
\(k\in \R^3\) pick $\psi_{\delta}\in D(H_g(P-k))$ with
\(\|\psi_{\delta}\|=1\) and
\begin{equation}\label{eq:gap1}
  \sprod{\psi_{\delta}}{H_g(P-k)\psi_{\delta}} \leq
E_{g}(P-k)+\delta
\end{equation}
Since \(\sprod{\psi_{\delta}}{H_g(P)\psi_{\delta}}\geq E_g(P)\),
it follows that
\begin{equation}\label{eq:gap2}
\begin{aligned}
  E_g(P-k) - E_g(P) &\geq
  \sprod{\psi_{\delta}}{[H_{g}(P-k)-H_{g}(P)]\psi_{\delta}} - \delta\\
  &=\sprod{\psi_{\delta}}{[\Omega(P-k-P_f)-\Omega(P-P_f)]\psi_{\delta}}
  - \delta.
\end{aligned}
\end{equation}
From the formula
\begin{eqnarray*}
\lefteqn{\Omega(P-k-q) - \Omega(P-q)}\\
 &= &  \Omega(P-k) - \Omega(P) + \int_0^1 dt\int_0^1 ds
 \sum_{i,j}(\partial_i\partial_j\Omega)(P-sk-tq)k_i q_j,
\end{eqnarray*}
the assumptions and \eqref{eq:hyp2low}, we obtain the estimate
\begin{equation}\label{eq:gap4}
   \Omega(P-k-q) - \Omega(P-q) \geq -\beta|k|-B|k| |q|
\end{equation}
valid for all $k,q\in \R^3$. Since $|P_f|\leq H_f$,
Eq.~\eqref{eq:gap4} leads to the operator bound
\begin{equation}\label{eq:gap5}
   \Omega(P-k-P_f) - \Omega(P-P_f) \geq -\beta|k|-B|k| H_f.
\end{equation}
In conjunction with \eqref{eq:gap2} this proves that
\begin{equation}\label{eq:gap6}
 E_{g}(P-k) - E_{g}(P) \geq
 -\beta|k|-B|k|\sprod{\psi_{\delta}}{H_f\psi_{\delta}}-\delta
\end{equation}
and hence we need a bound on
$\sprod{\psi_{\delta}}{H_f\psi_{\delta}}$ from above.

From the bound \eqref{eq:gap1} characterizing $\psi_{\delta}$ we
see that
\begin{align*} \Omega(P-k)+\delta &\geq E_{g}(P-k)+\delta\geq
\sprod{\psi_{\delta}}{H_{g}(P-k)\psi_{\delta}}\\
&= \sprod{\psi_{\delta}}{[\Omega(P-k-P_f)+H_f+
g\phi(\kappa_{\sigma})]\psi_{\delta}}
\end{align*}
which we estimate from below using the operator bounds
\begin{eqnarray*}
\Omega(P-k-P_f) &\geq & \Omega(P-k) - (\beta+B|k|)H_f\\
g\phi &\geq & -\alpha H_f -\frac{g^2C}{\alpha},
\end{eqnarray*}
obtained from \eqref{eq:gap4} with $q$ and $k$ interchanged, and
Lemma~\ref{lm:estim}, respectively . We conclude that
\begin{equation}\label{eq:gap7}
\delta \geq (1-\beta
-B|k|-\alpha)\sprod{\psi_{\delta}}{H_f\psi_{\delta}}-\frac{g^2}{\alpha}C.
\end{equation}
Inserting this bound on $\sprod{\psi_{\delta}}{H_f\psi_{\delta}}$
in \eqref{eq:gap6} and letting $\delta\to 0$ leads to
\begin{eqnarray*}
   E_{g}(P-k)+|k| - E_{g}(P) &\geq &
   \left(1-\beta-\frac{g^2BC/\alpha}{1-\beta-B|k|-\alpha}\right)|k|\\
   &\geq & \left(1-\beta-g^2\frac{9 BC}{(1-\beta)^2}\right)\eps
\end{eqnarray*}
for \(\alpha=(1-\beta)/3\) and \(\eps\leq |k|\leq
(1-\beta)/(3B)\). This is positive under our assumption on $g$. It
remains to estimate the left hand side from below when $|k|\geq
(1-\beta)/(3B)$.

To this end we note that for $g=0$
\begin{equation}\label{eq:gap9}
  E_{0}(P-k)+|k| - E_{0}(P) \geq (1-\beta)|k|
\end{equation}
while, by \eqref{eq:gs1} with $\alpha=|g|$,
\begin{eqnarray}\label{eq:gap10}
E_g(P-k) &\geq & (1-|g|)E_0(P-k)-C|g|\\
E_g(P) &\leq & \Omega(P)=E_0(P).\label{eq:gap11}
\end{eqnarray}
Eq.~\eqref{eq:gap9} follows from
\(E_{0}(P-k)=\inf_{q}(\Omega(P-k-q)+|q|)\geq
\Omega(P)-\beta|k+q|+|q|\geq \Omega(P)-\beta|k|\geq
E_0(P)-\beta|k|\). By \eqref{eq:gap9}, \eqref{eq:gap10},
\eqref{eq:gap11}, and \(E_0(P)=\Omega(P)\leq O_{\beta}\),
\begin{eqnarray*}
\lefteqn{E_{g}(p-k)+|k|-E_{g}(P)} \\
&\geq & (1-|g|)(E_{0}(P-k) -
E_{0}(P))- C|g| +|k|-|g|E_0(P)\\
&\geq & (1-\beta)^2/3B - |g|(C+O_{\beta}) > 0
\end{eqnarray*}
where $|k|\geq (1-\beta)/(3B)$ and \(|g|<
(1-\beta)^2/(3B(C+O_{\beta}))\) was used in the last line.
\end{proof}

To prove that \( E_g (P) = \inf \sigma (H_g (P)) \) is an
eigenvalue of the Hamiltonian $H_g (P)$ we first show the
corresponding result for the modified Hamiltonian
\[ \Hmod (P) = \Omega (P - \dGamma ( k)) + \dGamma (\omega
) + g \phi (\kappa_{\sigma}) \] introduced in Section
\ref{sec:modham}.

\begin{lemma}\label{lm:gsmod}
Assume Hypotheses 0, 1, and 3. Let $E_{\text{mod}}(P) := \inf
\sigma (\Hmod (P))$, and
\(\Delta(P):=\inf_{k}(E_{\text{mod}}(P-k)+\omega(k)-
E_{\text{mod}}(P)) \). Then
\[ \inf\sigma_{\rm ess}(\Hmod(P)) \geq E_{\text{mod}} (P) + \Delta(P).\]
In particular, if $\Delta(P)>0$ then $E_{\text{mod}} (P)$ is an
eigenvalue of $\Hmod(P)$ .
\end{lemma}

\noindent \emph{Remark.} The assumption that $\Delta(P)>0$ will be
derived from Hypothesis 3 in the proof of Theorem~\ref{thm:gs}
below.

\begin{proof}
Let \(\lambda\in \sigma_{ess}(\Hmod(P))\). Then there exists a
sequence \((\ph_n)_{n\in\N}\subset D(\Hmod(P))\), $\|\ph_n\|=1$,
such that \(\|(\Hmod(P)-\lambda)\ph_n\|\to 0\) and $\ph_n\weak 0$
(weakly) as $n\to\infty$. Hence
\[ \lambda =\lim_{n\to\infty}\sprod{\ph_n}{\Hmod(P)\ph_n}.\]
To estimate \(\sprod{\ph_n}{\Hmod(P)\ph_n}\) from below, we need
to localize the photons. Pick \(j_0,j_{\infty}\in
C^{\infty}(\R^3)\) with \(j_0^2+j_{\infty}^2=1\), $j_0(y)=1$ for
$|y|\leq 1$ and $j_0(y)=0$ for $|y|\geq 2$. Given $R>0$ set
\(j_{\sharp,R}(y)=j_{\sharp}(y/R)\) where $\sharp=0$ or $\infty$.
Let \(j_R:\h\to\h\oplus\h\) be defined by
\(h\mapsto(j_{0,R}h,j_{\infty,R}h)\) and let $j_{x,R}$ be defined
in a similar way with $j_{\sharp}(y)$ replaced by
$j_{\sharp}(y-x)$. By Lemma~\ref{lm:comest1}
\begin{multline}\label{eq:des2}
\text{esssup}_P\|[\Hmod(P)-
\uGamma(j_R)^{*}\tilde{H}_{\text{mod}}(P)\uGamma(j_R)](N+1)^{-1}\| \\
=\|[H-\uGamma(j_{x,R})^{*}\tilde{H}\uGamma(j_{x,R})](N+1)^{-1}\| =
O(R^{-1})\qquad\text{as}\ R\to\infty,
\end{multline}
where
\begin{equation*}
 \tilde{H}_{\text{mod}}(P) = \Omega(P-\dGamma(k)\otimes 1-
1\otimes \dGamma(k)) + \dGamma(\omega)\otimes 1 +
1\otimes\dGamma(\omega) + g\phi(\kappa_{\sigma})\otimes 1.
\end{equation*}
In \eqref{eq:des2} we may replace "$\text{essup}_P$" by "$\sup_P$"
because \(\|[\Hmod(P)-
\uGamma(j_R)^{*}\tilde{H}_{\text{mod}}(P)\uGamma(j_R)](N+1)^{-1}\|\)
is continuous as a function of $P$. Using that
\(\sum_{i=1}^N\omega(k_i)\geq \omega(\sum_{i=1}^N k_i)\), by
Hypothesis 3, and the definition of $\Delta(P)$, we arrive at the
lower bound
\[ \tilde{H}_{\text{mod}}(P) \geq \Emod(P)+\Delta(P) - \Delta(P)E_{\{0\}}(N_{\infty}), \]
which, in conjunction with \eqref{eq:des2} and
\(\uGamma(j_R)^{*}E_{\{0\}}(N_{\infty})\uGamma(j_R)=\Gamma(j_{0,R}^2),\)
shows that
\begin{equation*}
\begin{aligned}
\sprod{\ph_n}{\Hmod(P)\ph_n} &=
\sprod{\ph_n}{\uGamma(j_R)^{*}\tilde{H}_{\text{mod}}(P)\uGamma(j_R)\ph_n}
+ O(R^{-1})\\ & \geq \Emod(P)+\Delta(P) -
\sprod{\ph_n}{\Gamma(j_{0,R}^2)\ph_n}\Delta(P) + O(R^{-1})
\end{aligned}
\end{equation*}
where $O(R^{-1})$ is independent of $n$. Now let $n\to\infty$ and
observe that \(\Gamma(j_{0,R}^2)(\Hmod(P)+i)^{-1}\) is compact to
get
\[ \lambda \geq \Emod(P)+\Delta(P) + O(R^{-1})\qquad\text{for all}\ R>0.\]
Letting $R\to\infty$ this proves the theorem.
\end{proof}

\begin{theorem}\label{thm:gs}
Assume Hypotheses 0 -- 3. Suppose $\beta<1$ and $|g|<g_{\beta}$,
with $g_{\beta}$ defined by Lemma~\ref{lm:gap}.
\begin{itemize}
\item[i)] If $\Omega(P)\leq O_{\beta}$ then \( E_g (P) = E_{\text{mod}} (P) \) and $E_g(P)$ is an
eigenvalue of $H_g (P)$.
\item[ii)] Suppose $\Omega(P)\leq O_{\beta}$. If $\psi_P \in \F$ is a ground state of $H_g (P)$ or
of $\Hmod(P)$, then it belongs to $\ran \Gamma (\chi_i)$. In
particular, by i), $\psi_P$ is ground state of $H_g(P)$ if and
only if it is a ground state of $\Hmod(P)$.
\item[iii)] The mapping $P\mapsto E_g (P)$ is twice continuously
differentiable on \(\{P\in \R^3| \Omega(P)\leq O_{\beta}\}\).
\end{itemize}
\end{theorem}

\begin{proof}
Recall from the proof of Theorem~\ref{cor:gs-only}, that \(\F\cong
\oplus_{n\geq 0}\F_{s,n}\) where each subspace $\F_{s,n}$ is
invariant under $H_g(P)$ and that on \(\F_{s,n} = L_s^2
(B_{\sigma}(0)^{\times n} , dk_1 \dots dk_n ; \F_i)\) the operator
$H_g(P)$ is given by \[ (H_g (P) \psi ) (k_1, \dots ,k_n) = H_P
(k_1 , \dots ,k_n) \psi (k_1 , \dots , k_n) , \] where
\begin{equation}\label{eq:gs-key}
\begin{aligned}
H_P(k_1,\ldots,k_n) &= H_g (P-k_1\ldots-k_n)+|k_1|+\ldots+|k_n| \\
&> H_g(P) \qquad \text{if}\ (k_1,\ldots,k_n)\neq(0,\ldots,0)
\end{aligned}
\end{equation}
as an operator inequality on $\F_i$. In the last equation we used
that \(\Omega(P-k)+|k|>\Omega(P)\) by assumption and Hypothesis 2.
\begin{itemize}
\item[i)] Inequality \eqref{eq:gs-key} proves that
\begin{equation*}
\inf\sigma(H_g(P)\restricted\F_{s,n})\geq
\inf\sigma(H_g(P)\restricted\F_i) =
\inf\sigma(\Hmod(P)\restricted\F_i)\geq \Emod(P)
\end{equation*}
for each $n\in \N$. This shows that \(E_g(P)\geq \Emod(P)\) and
hence that \(E_g(P)= \Emod(P)\). We next verify that $\Delta(P)>0$
in Lemma~\ref{lm:gsmod}. In fact,
\(\inf_{|k|\geq\sigma/4}(\Emod(P-k)+\omega(k)-\Emod(P))\geq
\inf_{|k|\geq\sigma/4}(E_{g}(P-k)+|k|-E_{g}(P))>0\) by
Lemma~\ref{lm:gap} while, for $|k|\leq \sigma/4$, by
\eqref{eq:gs-key}, \(\Emod(P-k)+\omega(k)-\Emod(P)\geq
\sigma/2-|k|\geq \sigma/4\). Hence, by Lemma~\ref{lm:gsmod},
$\Emod(P)$ is an eigenvalue of $\Hmod(P)$, and that $E_g(P)$ is an
eigenvalue will now follow from ii) because \(\Hmod(P)=H_g(P)\) on
\(\ran \Gamma(\chi_i)\).
\item[ii)] By \eqref{eq:gs-key}, \(H_P(k_1,\ldots,k_n)>E_g(P)\) if
\((k_1,\ldots,k_n)\neq(0,\ldots,0)\). This shows that any
hypothetical eigenvector of $H_g(P)$ with eigenvalue $E_g(P)$
belongs to $\ran\Gamma(\chi_i)$. The corresponding result for
\(\Hmod(P)\) follows from an inequality similar to
\eqref{eq:gs-key} for $\Hmod(P)$.
\item[iii)] This statement follows by analytic perturbation theory, because
$E_g (P) = E_{\text{mod}} (P)$, and because $E_{\text{mod}} (P)$
is an isolated eigenvalue of $\Hmod (P)$.
\end{itemize}
\end{proof}

\begin{lemma}\label{lm:unique}
Assume Hypotheses 0--2 are satisfied. Suppose that $\Omega (P)
\leq O_{\beta}$ for some $\beta <1$ (see Hypothesis 2 for the
definition of $O_{\beta}$) and that $E_g (P)= \inf \sigma (H_g
(P))$ is an eigenvalue of $H_g (P)$. Then $E_g (P)$ is a simple
eigenvalue.
\end{lemma}

\begin{proof}
If $g=0$ (or if $\kappa_{\sigma} (k) = 0$ a.e.) the lemma is true,
under our assumptions, because the only ground state of $H_{g=0}
(P)$ is the vacuum. In fact, in this case $H_g (P)$ commutes with
$N$ and the absence of ground state vectors in the $n$ particle
sector, for any $n>0$, can easily be proven using the equation
\begin{equation*}
\Omega (P-k_1 -\dots k_n) \geq \Omega (P) - \beta |k_1| - \dots -
\beta |k_n|
\end{equation*}
with $\beta <1$ (see the remark after Hypothesis 2). Thus, without
loss of generality we can assume that $g \neq 0$ and that the set
$\{ k \in \R^3: \kappa_{\sigma} (k) \neq 0 \}$ has positive
measure. We consider here the case $g >0$ . The proof for $g<0$ is
then similar. Suppose that $\psi = \{ f^{(n)} (k_1 , \dots, k_n)
\}_{n=0}^{\infty} \in \F$ is an eigenvector of $H_g (P)$
corresponding to the eigenvalue $E_g (P)$. Then we have
\begin{equation*}
\begin{split}
\sprod{ \psi}{ H_g(P) \psi}= \; &\sum_{n=0}^{\infty} \int dk_1
\dots dk_n |f^{(n)} (k_1 , \dots , k_n)|^2  \, \left\{ \Omega (P
-\sum_{i=1}^{n} k_i) + \sum_{i=1}^{n} |k_i| \right\} \\ &+ 2 g
\Ree \sum_{n=0}^{\infty} \sqrt{n+1} \int dk_1 \dots dk_n  \,
\overline{f^{(n)} (k_1 , \dots k_n)}\, \int dk \, \kappa_{\sigma}
(k) f^{(n+1)} (k , k_1 ,\dots ,k_n).
\end{split}
\end{equation*}
Now define
\begin{equation*}
g^{(n)} (k_1 , \dots k_n) = (-1)^{n} | f^{(n)} (k_1 , \dots k_n)|
\end{equation*}
and set $\tilde{\psi} = \{ g^{(n)} (k_1 , \dots, k_n)
\}_{n=0}^{\infty}$. Then $\| \tilde{\psi} \| = \| \psi \|$ and
since $\kappa_{\sigma}\geq 0$ we have
\begin{equation*}
\begin{split}
\sprod{ \tilde{\psi}}{ H_g(P) \tilde{\psi}}= \;
&\sum_{n=0}^{\infty} \int dk_1 \dots dk_n |f^{(n)} (k_1 , \dots ,
k_n)|^2  \, \left\{ \Omega (P -\sum_{i=1}^{n} k_i) +
\sum_{i=1}^{n} |k_i| \right\} \\ &- 2 g \Ree \sum_{n=0}^{\infty}
\sqrt{n+1} \int dk_1 \dots dk_n  \, |f^{(n)} (k_1 , \dots k_n)| \\
&\times \int dk \, \kappa_{\sigma} (k) |f^{(n+1)} (k , k_1 , \dots
,k_n)|
\\ \leq \; & \sprod{ \psi}{ H_g(P) \psi},
\end{split}
\end{equation*}
where the equality holds if and only if there is some real
$\theta$ with
\begin{equation}\label{eq:unique1}
g^{(n)} (k_1, \dots , k_n)= e^{i\theta} f^{(n)} (k_1 , \dots,
k_n), \quad \text{for all} \quad n \geq 0 .
\end{equation}
Since $\psi$ is a ground state vector for $H_g (P)$,
Eq.~\eqref{eq:unique1} has to be satisfied.

Now suppose that $\psi_1 = \{ f^{(n)}_1 (k_1 , \dots, k_n)
\}_{n=0}^{\infty}$ and $\psi_2 = \{ f^{(n)}_2 (k_1 , \dots, k_n)
\}_{n=0}^{\infty}$ are two orthonormal ground state vectors of
$H_g (P)$. Then, by \eqref{eq:unique1},
\begin{align*}
f^{(n)}_1 (k_1 , \dots, k_n) &= e^{i \theta_1} \, (-1)^{n}
|f_1^{(n)} (k_1 ,
\dots , k_n) | \quad \text{and} \\
f^{(n)}_2 (k_1 , \dots, k_n) &= e^{i \theta_2} \, (-1)^{n}
|f_2^{(n)} (k_1 , \dots , k_n) | ,
\end{align*}
for some constants $\theta_1, \theta_2$ and thus
\begin{equation}\label{eq:unique_last}
\begin{split}
0 &= \sprod{\psi_1}{\psi_2} = \sum_{n=0}^{\infty} \int dk_1 \dots
dk_n
\overline{f_1^{(n)} (k_1 , \dots , k_n)} f_2^{(n)} (k_1 , \dots , k_n) \\
&=e^{i(\theta_2 -\theta_1)} \, \sum_{n=0}^{\infty} \int dk_1 \dots
dk_n |f_1^{(n)} (k_1 , \dots , k_n)| \, |f_2^{(n)} (k_1 , \dots ,
k_n)| .
\end{split}
\end{equation}
This implies, in particular, that $f^{(0)}_1 \cdot f^{(0)}_2 =0$.
We claim that this is not possible. In fact, let $\psi = \{
f^{(n)} (k_1 , \dots , k_n) \}_{n=0}^{\infty}$ be an eigenvector
of $H_g (P)$, and suppose that $f^{(n)} = 0$ for all $n < n_0$ for
some $n_0 >0$, and that $f^{(n_0)}\neq 0$, that is, $f^{(n_0)}(k_1
, \dots k_{n_0}) \neq 0$ on a set $G$ of positive measure. Since
$f^{(n_0)} (k_1 , \dots k_n) = 0$ unless $k_i \in \supp
\kappa_{\sigma}$, for all $i=1, \dots n_0$ (this can be proved in
the same way as the absence of soft bosons in the ground state,
see Theorem \ref{thm:gs}), the set $G$ must (essentially) belong
to $(\supp \kappa_{\sigma})^{\times n_0}$. Using that
$\kappa_{\sigma} (k) \geq 0$ and that $f^{(n_0)} (k_1, \dots,
k_{n_0})= (-1)^{n_0} e^{i\theta} |f^{(n_0)}(k_1, \dots ,k_{n_0})|$
it follows that
\begin{equation*}
\begin{split}
(H_g (P) \psi)^{(n_0 -1)} (k_1 , \dots k_{n_0 -1}) &= ( g
a(\kappa_{\sigma}) \psi)^{(n_0 -1)} (k_1 , \dots k_{n_0 - 1}) \\
&= g \sqrt{n_0} \int dk \, \kappa_{\sigma} (k) f^{(n_0)} (k, k_1,
\dots , k_{n_0 -1}) \neq 0,
\end{split}
\end{equation*}
which is in contradiction with $(H_g(P)\psi)^{(n_0-1)} = E
f^{(n_0-1)}=0$. Hence $n_0=0$ and $f^{(0)} \neq 0$. Thus
Eq.~\eqref{eq:unique_last} cannot be true.
\end{proof}

The following Lemma is needed to apply Theorem~\ref{thm:gs} in
cases where an upper bound on $E_{g}(P)$, rather than $\Omega(P)$,
is given.

\begin{lemma}\label{lm:Omega-E}
Suppose $\beta\leq 1$ and $\Sigma<O_{\beta}$. If \(|g|\leq
(O_{\beta}-\Sigma)/(O_{\beta}+C)\) and $E_{g}(P)\leq \Sigma$, then
$\Omega(P)\leq O_{\beta}$.
\end{lemma}

\begin{proof}
Recall from \eqref{eq:gs1} that
\[ E_{g}(P) \geq (1-|g|)E_0(P) -C|g|\]
for all $P\in \R^3$ and all $g$. Hence $E_{g}(P)\leq \Sigma$ and
\(|g|\leq (O_{\beta}-\Sigma)/(O_{\beta}+C)<1\) imply that
\begin{eqnarray*}
  E_0(P) \leq \frac{\Sigma+C|g|}{1-|g|} \leq O_{\beta}
\end{eqnarray*}
It remains to prove that \(E_0(P)\leq O_{\beta}\) implies that
$\Omega(P)\leq O_{\beta}$ for $\beta\leq 1$. This is fairly
obvious from \(E_0(P)=\inf_{k}(\Omega(P-k)+|k|)\) and a sketch of
$E_0(P)$ for a typical $\Omega$. We nevertheless give an
analytical proof. Since $\Omega(P)\leq O_{\beta=1}$ implies that
$E_0(P)= \Omega(P)$ it suffices to consider the case $\beta=1$.
Let $A:=\{P: \Omega(P)\leq O_{\beta=1}\}\neq \emptyset$. We derive
a contradictions from the two assumptions $P\not\in A$ and
$E_0(P)\leq O_{\beta=1}$. Let \(d:=\mathrm{dist}(P,A)>0\), let $k$
be any vector with $P-k\in A$ and choose a point $P'$ on the
intersection of $\partial A$ and the line segment from $P-k$ to
$P$. Then $\Omega(P')=O_{\beta=1}$ and hence
\begin{eqnarray*}
\Omega(P-k) &\geq & \Omega(P') - |P'-(P-k)|\\
& = & O_{\beta=1} -(|k|-|P-P'|)\\
& \geq & E_0(P) -|k| + d.
\end{eqnarray*}
Using again that $E_0(P)\leq O_{\beta=1}$ and the above inequality
we get
\begin{eqnarray*}
E_0(P) &= & \min_{k}(\Omega(P-k)+|k|)\\
& = & \min_{k: (P-k)\in A}(\Omega(P-k)+|k|) \geq  E_0(P)+ d,
\end{eqnarray*}
a contradiction.
\end{proof}

\subsection{Virial Theorem for the modified Hamiltonian}
\label{app:modivir}

Let $A_{\text{mod}}=\dGamma(a)$ where \( a=1/2(\nabla\omega\cdot
y+ y\cdot\nabla\omega) \) and define the commutator
$[i\Hmod(P),A_{\text{mod}}]$ by the quadratic form
\begin{equation*}
\sprod{\ph}{[i\Hmod(P),A_{\text{mod}}]\ph} :=
\sprod{\ph}{\dGamma(|\nabla\omega|^2)\ph}
-\sprod{\nabla\Omega(P-\dGamma(k))\ph}{\dGamma(\nabla\omega)\ph}
-\sprod{\ph}{\phi(ia\kappa_{\sigma})\ph}
\end{equation*}
for $\ph\in D(\Hmod(P))$.

\begin{lemma}[Virial theorem]\label{lm:modivir}
Let Hypothesis 0 be satisfied. If $\ph$ is an eigenvector of
$\Hmod(P)$ then
\[ \sprod{\ph}{[i\Hmod(P),A_{\text{mod}}]\ph} = 0. \]
\end{lemma}

\begin{proof}
We adapt the strategy used to prove Lemma~3 in \cite{FGS2} to the
present situation. Let $\eps>0$ and define \(y_{\eps}=y/(1+\eps
y^2)\), \(a_{\eps}=1/2(\nabla\omega\cdot y_{\eps}+
y_{\eps}\cdot\nabla\omega)\) and \(A_{\eps}=\dGamma(a_{\eps})\).
The subspace \({\mathcal D}=\{\ph\in \F_0:\ph_n\in
\tf(\R^{3n},dk_1\ldots dk_n )\}\) is a core of
$\Omega(P-P_f)+\dGamma(\omega)$, and hence it is also a core of
$\Hmod(P)$. On ${\mathcal D}$
\begin{equation}\label{eq:approx-vir}
i\sprod{\Hmod(P)\ph}{A_{\eps}\ph} -
i\sprod{A_{\eps}\ph}{\Hmod(P)\ph} =
\sprod{\ph}{\big\{[i\Omega(P-P_f),A_{\eps}]+\dGamma(i[\omega,a_{\eps}])-
\phi(a_{\eps}\kappa_{\sigma})\big\}\ph}
\end{equation}
where
\begin{equation*}
2i[\omega,a_{\eps}] = |\nabla\omega|^2\frac{1}{1+\eps y^2} -
(\nabla\omega\cdot y)\frac{\eps}{1+\eps y^2}(y\cdot
\nabla\omega+\nabla\omega\cdot y)\frac{1}{1+\eps y^2} +
\text{h.c.}
\end{equation*}
and, on $\otimes_s^n L^2(\R^3,dk)$,
\begin{eqnarray*}
2i[\Omega(P-P_f),A_{\eps}] & =& -\sum_{i=1}^n
\nabla\omega(k_i)\cdot
\nabla\Omega(P-P_f)\frac{1}{1+\eps y_i^2}\\
& & +\nabla\omega(k_i)\cdot y_i \frac{\eps}{1+\eps
y_i^2}(y_i\cdot\nabla\Omega(P-P_f)+\nabla\Omega(P-P_f)\cdot
y_i)\frac{1}{1+\eps y_i^2} \\ && +\text{h.c.}
\end{eqnarray*}
Since ${\mathcal D}$ is a core of $\Hmod(P)$, since $A_{\eps}$ is
bounded w.r.to $\Hmod(P)$ and the quadratic for on the right side
of \eqref{eq:approx-vir} is form bounded with respect to
$\Hmod(P)^2$, this equation carries over to all $\ph\in
D(\Hmod(P))$. If $\ph$ is an eigenvector of \(\Hmod(P)\) then the
left side vanishes because $A_{\eps}$ is symmetric, and thus it
remains to show that the right side converges to
\([i\Hmod(P),A_{\mathrm{mod}}]\) as $\eps\to 0$. This is done by
repeated application of Lebesque's dominated convergence theorem,
see \cite{FGS2} for more details.
\end{proof}

\section{Number--Energy Estimates.}\label{sec:num_en}

In this section we consider the modified Hamiltonian \[ \Hmod =
\Omega(p) + \dGamma (\omega) + g \phi (G_x) \] introduced in
Section \ref{sec:modham}, where the dispersion relation $\omega$
satisfies Hypothesis 3. We use the notation $H \equiv \Hmod$.
Thanks to the lower bound $\omega(k)\geq \sigma/2>0$, one has the
operator inequality
\begin{equation}\label{eq:N<H}
N\leq a H + b,
\end{equation}
for some constants $a$ and $b$. The purpose of this section is to
prove that also higher powers of $N$ are bounded with respect to
the same powers of $H$. This easily follows from \eqref{eq:N<H} if
the commutator \([N,H]\) is zero, that is, for vanishing
interaction. Otherwise it follow from the boundedness of
\(ad_N^k(H)(H+i)^{-1}\) for all $k$.

\begin{lemma}\label{lm:num_en_est}
Assume the Hypotheses 0, 1 and 3 are satisfied and suppose $m \in
\N \cup \{ 0 \}$.
\begin{enumerate}
\item[i)] Then uniformly in $z$, for $z$ in a compact subset of $\C$,
\begin{equation*}
\|(N+1)^{-m} (z-H)^{-1} (N+1)^{m+1}\| = O(|\Ima z|^{-m}).
\end{equation*}
\item[ii)] $(N+1)^m (H+i)^{-m}$ is a bounded operator. In particular $(N+1)^m
\chi (H)$ is bounded, for all $m \in \N$, if $\chi \in \tf (\R)$.
\end{enumerate}
\end{lemma}

\begin{proof}
This lemma follows from Lemma~31 i) and ii) in \cite{FGS2}, where
it is proved for a class of Hamiltonians which is larger than the
one we consider here. Note that Hypothesis 3 in this paper implies
Hypothesis (H1) in \cite{FGS2}, and that Hypothesis (H1) in
\cite{FGS2} is sufficient to prove parts i) and ii) of Lemma~31 in
\cite{FGS2}.
\end{proof}

\section{Commutator Estimates}\label{sec:comm_est}

In this section we consider the modified Hamiltonians $\Hmod =
\Omega (p) + \dGamma (\omega) + g \phi (G_x)$ and $\Hmodex=
\Hmod\otimes 1 + 1 \otimes \dGamma (\omega)$ introduced in Section
\ref{sec:modham}. We use the notation $H = \Hmod$ and $\Hex =
\Hmodex$.

Let \(j_0,\ j_{\infty}\in C^{\infty}(\R^d)\) be real-valued with
\(j_0^2+j_{\infty}^2\leq 1\), \(j_0(y)=1\) for $|y|\leq 1$ and
\(j_0(y)=0\) for $|y|\geq 2$. Given $R>0$ set \(j_{\#,R}
=j_{\#}((x-y) / R)\) and let $j_{R,x} = (j_{0,R} ; j_{\infty ,R})$
($j_{R,x}$ is an operator from $L^2 (\R^3) \otimes \h$ to $L^2
(\R^3) \otimes (\h\oplus\h)$).

\begin{lemma}\label{lm:comest1}
Assume Hypotheses 0,1 and 3 are satisfied. Suppose $m \in \N \cup
\{0 \}$, and $j_{R,x}$ is as above. Suppose also that $\chi ,
\chi^{\prime} \in \tf (\R)$. Then, for $R \to \infty$,
\begin{enumerate}
\item[i)] \((N_0 + N_{\infty} +1)^m
\left(\uGamma(j_{R,x})H-\Hex\uGamma(j_{R,x})\right) \chi^{\prime}
= O(R^{-1})\),
\item[ii)] \((N_0 + N_{\infty} +1)^m \left(\chi (\Hex) \uGamma (j_{R,x}) -
\uGamma (j_{R,x}) \chi (H)\right) \chi^{\prime}(H) = O(R^{-1})\).
\end{enumerate}
\end{lemma}
\emph{Remark:} this Lemma also holds if we replace the modified
Hamiltonian $H \equiv \Hmod$ with the original Hamiltonian $H_g$
and if we restrict the equality to states with no soft bosons,
that is to states in the range of the orthogonal projection
$\Gamma (\chi_i)$.
\begin{proof}
\begin{enumerate}
\item[i)] From the intertwining relations \eqref{eq:ugamma-phi}, and
\eqref{eq:ugamma-o} we have that
\begin{align*}
\uGamma(j_{R,x})H-\Hex\uGamma(j_{R,x})= &\udGamma(j_{R,x},
[j_{R,x} , \omega (k) + \Omega (p)])\\ &+
[\phi((j_{0,R}-1)G_x)\otimes 1 +
1\otimes\phi(j_{\infty,R}G_x)]\uGamma(j_{R,x}).
\end{align*}
By Lemma \ref{lm:pdc1}, and because of Hypothesis 0 (which
guarantees that $\nabla \Omega$ is bounded with respect to $H$),
we have
\begin{equation*}
(N_0 + N_{\infty} +1)^m  \udGamma(j_{R,x},[ j_{R,x} , \omega (k) +
\Omega (p)]) \chi^{\prime} (H) = O(R^{-1}).
\end{equation*}
To see that the other two terms lead to contributions of order
$O(R^{-1})$ write
\begin{equation*}
\begin{split}
(N_0 + &N_{\infty} +1)^m [\phi((j_{0,R}-1)G)\otimes 1 +
1\otimes\phi(j_{\infty,R}G)] \\ = \; &[\phi((j_{0,R}-1)G)\otimes 1
+ 1\otimes\phi(j_{\infty,R}G)] (N_0 + N_{\infty} + 1)^{m} \\ &+
\sum_{l=1}^{m} \binom{m}{l} (-i)^{l} [\phi(i^l (
j_{0,R}-1)G)\otimes 1 + 1\otimes\phi(i^l j_{\infty,R}G)] (N_0 +
N_{\infty}+1)^{m-l},
\end{split}
\end{equation*}
and then use $(N_0 + N_{\infty} +1)^{m-l} \uGamma (j_{R,x})
=\uGamma (j_{R,x}) (N+1)^{m-l}$, the fact that $(N+1)^{m-l}
\chi^{\prime} (H)$ is bounded (see Lemma \ref{lm:num_en_est}) and
Lemma \ref{lm:srdecay}.
\item[ii)] Let $\tilde{\chi}$ be an almost analytic
extension of $\chi$ of order $m$, as defined in
Appendix~\ref{sec:HScalc}. Then we have
\begin{equation*}
\begin{split}
(N_0 + N_{\infty} +1)^m &(\chi (\Hex) \uGamma (j_{R,x}) - \uGamma
(j_{R,x}) \chi (H)) \chi^{\prime}(H) \\ = \; -\frac{1}{\pi} \int
&dx dy \, \partial_{\bar{z}} \tilde{\chi} (N_0 + N_{\infty} +1)^m
(z-\Hex)^{-1} (\Hex \uGamma (j_{R,x}) - \uGamma (j_{R,x}) H ) \\
&\times \chi^{\prime}(H) (z-H)^{-1}.
\end{split}
\end{equation*}
Then the statement follows by i) because
\begin{equation}\label{eq:N_H_N}
(N_0 + N_{\infty} + 1)^m (z-\Hex)^{-1} ( N_0 + N_{\infty} +
1)^{-m+1} = O( |\Ima z|^{-m}).
\end{equation}
\end{enumerate}
\end{proof}

Now suppose that $j_0, j_{\infty} \in C^{\infty} (\R^3)$, with
$j_0^2 +j_{\infty}^2 \leq 1$, $j_0 \in \tf (\R^3)$ and with $j_0
(y) = 1$ for $|y|<\lambda_0$, for some $\lambda_0 >0$. Set
$j_{\sharp ,R} = j_{\sharp} (y/R)$ and $j_R = (j_{0,R} ,
j_{\infty,R})$ (note that here the operator $j_R$ does not depend
on the electron position $x$). Suppose moreover that $F \in \tf
(\R)$ with $F(s) = 0$ for $s > \lambda_1$, for some $\lambda_1 <
\lambda_0$.

\begin{lemma}\label{lm:comestnew1}
Assume that Hypotheses 0, 1 and 3 are satisfied. Suppose that $m
\in \N$ and that $j_R$ and $F$ are defined as above and that $f,f'
\in \tf (\R)$. Then, if $R \to \infty$,
\begin{enumerate}
\item[i)] \( (N_0 + N_{\infty} +1)^m \left(\uGamma (j_R) H - \Hex \uGamma (j_R)
  \right) F(|x| /R) (N+1)^{-m-1} = O(R^{-1}) \),
\item[ii)] \( (N_0 + N_{\infty} +1)^m \left( f(\Hex) \uGamma (j_R) - \uGamma
  (j_R) f(H) \right) F(|x| /R) f' (H) = O(R^{-1}). \)
\end{enumerate}
\end{lemma}

The proof of the last lemma is very similar to the proof of Lemma
\ref{lm:comest1}. The only difference is that now, in order to
bound the commutator with the interaction $\phi (G_x)$ we use the
space cutoff $F(|x|/t)$ and part ii) of Lemma \ref{lm:srdecay}.

\section{Invariance of Domains}
\label{sec:inv}

In this section the invariance of the domain of $\dGamma(\nabla
\omega \cdot (y -x) + (y-x) \cdot \nabla \omega)$ with respect to
the action of $f(H)$ for smooth functions $f$ is proven. Here $H$
denotes the modified Hamiltonian $\Hmod = \Omega (p) + \dGamma
(\omega) + g\phi (G_x)$ introduced in Section \ref{sec:modham}.
Moreover we prove in Lemma \ref{lm:inv1} that the norm of $\dGamma
(a) f(H) e^{-iHt} \ph$ can only grow linearly in $t$ if $\ph \in
D(\dGamma (a))$. All these results are only used in Section
\ref{sec:Wpos} to prove the positivity of the asymptotic
observable $W$.

In the following we use the notation $a = 1/2 \left( \nabla \omega
\cdot (y -x) + (y-x) \cdot \nabla \omega \right)$.

\begin{lemma}\label{lm:inv1}
Assume Hypotheses 0,1 and 3 are satisfied and let $f \in \tf
(\R)$. Then \\$f(H) D(\dGamma (a)) \subset D(\dGamma (a))$ and
\begin{equation*}
\| \dGamma (a) e^{-iHt} f(H) \ph \| \leq C ( \| \dGamma (a) \ph \|
+ t \, \| \ph \| ),
\end{equation*}
for all $t \geq 0$ and for all $\ph \in D(\dGamma (a))$.
\end{lemma}
\begin{proof}
First we note, that
\begin{equation*}
\begin{split}
e^{iHt} \dGamma (a) e^{-iHt} &f (H) - \dGamma (a) f (H) =
\int_{0}^{t} ds \,
e^{iHs} \, [ iH , \dGamma (a) ] \, f (H) e^{-iHs} \\
&= \int_{0}^{t} ds \, e^{iHs} \, \left( \dGamma (\nabla \omega
\cdot (\nabla \omega -\nabla \Omega)) - \phi ( i a G_x )  \right)
f(H) e^{-iHs} .
\end{split}
\end{equation*}
Since the operator in the integral on the r.h.s. of the last
equation is bounded (because of the energy cutoff $f(H)$ and
because, by Hypothesis 0, $\nabla \Omega$ is bounded w.r.t. $H$)
it follows that
\begin{equation}\label{eq:inv10}
\| \dGamma (a) e^{-iHt} f(H) \ph \| \leq C \left( \| \dGamma (a)
f(H) \ph \| + t \| \ph \| \right).
\end{equation}
Now we have
\begin{equation}\label{eq:inv11}
\dGamma (a) f(H) \ph = f(H) \dGamma (a) \ph + \left[ \dGamma (a) ,
f (H) \right] \ph .
\end{equation}
To compute the commutator in the last equation we choose an almost
analytic extension $\tilde{f}$ of $f$, and we expand $f (H)$ in an
Helffer-Sj\"ostrand integral (see Appendix \ref{sec:HScalc}).
\begin{equation*}
\begin{split}
\left[ \dGamma (a) , f (H) \right] = \; &\frac{-1}{\pi} \int dxdy
\partial_{\bar{z}} \tilde{f} \, (z -H)^{-1} [\dGamma (a) , H ]
(z-H)^{-1}\\
= \; & \frac{-i}{\pi} \int dxdy \partial_{\bar{z}} \tilde{f} \,
(z-H)^{-1}
\dGamma (\nabla \omega \cdot (\nabla \omega -\nabla \Omega)) (z-H)^{-1} \\
&+ \frac{i}{\pi} \int dxdy \partial_{\bar{z}} \tilde{f} \,
(z-H)^{-1} \phi ( i a G_x ) (z-H)^{-1}.
\end{split}
\end{equation*}
Both integral on the r.h.s. of the last equation are bounded
(because, by Hypothesis 0, $\nabla \Omega$ is bounded w.r.t. $H$).
This together with \eqref{eq:inv11} and \eqref{eq:inv10} completes
the proof of the lemma.
\end{proof}

In the following lemma we prove the invariance of the domain of
$\dGamma (a+1)$ with respect to the action of operators like
$\Gamma (\chi (k))$, where $\chi$ is a smooth function. This
result is used below, in the proof of Lemma \ref{lm:dicht}.

\begin{lemma}\label{lm:inv4}
Assume Hypothesis 3 is satisfied. Suppose moreover that $\ph \in
D( \dGamma (a + 1))$ and that $\chi \in \tf (\R^3)$ with $\chi (k)
\leq 1$ for all $k \in \R^3$. Then
\[ \| \dGamma (a) \Gamma (\chi (k)) \ph \| \leq C \| \dGamma (a + 1 )\ph
\| \]
\end{lemma}
\begin{proof}
For $\ph \in D(\dGamma (a))$ we have
\begin{equation*}
\dGamma (a) \Gamma (\chi (k)) \ph = \Gamma (\chi (k))\dGamma
(a)\ph + \dGamma ( \chi (k) , [a , \chi (k)]) \ph.
\end{equation*}
The lemma follows because
\begin{equation*}
[ a , \chi (k)] = i \nabla \omega (k) \cdot \nabla \chi (k)
\end{equation*}
is a bounded operator (and thus the operator $ \dGamma ( \chi (k)
, [a , \chi (k)])$ can be estimated by the number-operator $N$).
\end{proof}

Next, using Lemma \ref{lm:inv4}, we prove that vectors in the
domain of $\dGamma (a+1)$ are dense in the range of $\Gamma
(\chi_i)$, the orthogonal projection onto the subspace of vectors
without soft bosons. This is used in the proof of Theorem
\ref{thm:W_pos}, where the positivity of the asymptotic observable
$W$ is proven.

\begin{lemma}\label{lm:dicht}
Suppose Hypothesis 3 is satisfied and that $\chi_i$ is the
characteristic function of the set $\{ k \in \R^3 : |k| \geq
\sigma \}$. Let $\mathcal{D} := D(\dGamma (a+1))$ and $\H_i = \ran
\Gamma (\chi_i)$. Then the linear space $\H_i \cap \mathcal{D}$ is
a dense subspace of $\H_i$.
\end{lemma}
\begin{proof} First, we note that $\H_i \cap D(N)$ is dense in $\H_i$.
This is clear, since $[N, \Gamma (\chi_i)]=0$. The lemma follows
if we show that $\H_i \cap \mathcal{D}$ is dense in $\H_i \cap
D(N)$. To this end choose an arbitrary $\ph \in \H_i \cap D(N)$.
Then, since $\mathcal{D}$ is dense in $\H$, we find a sequence
$\ph_n \in \mathcal{D}$ with $\ph_n \to \ph$, as $n \to \infty$.
Moreover we find functions $f_n \in \mathcal{C}^{\infty} (\R^3)$
with $f_n (k) = 0$, if $|k| < \sigma$, and with $f_n \to \chi_i$,
as $n \to \infty$, pointwise. Then we define $\psi_n := \Gamma
(f_n) \ph_n$. On the one hand, by Lemma~\ref{lm:inv4}, $\psi_n \in
\H_i \cap \mathcal{D}$ for all $n \in \N$. On the other hand
\begin{equation*}
\begin{split}
\| \psi_n - \ph \| &= \| \Gamma(f_n) \ph_n - \ph \| \leq \| \Gamma
(f_n) (\ph_n
-\ph)\| + \| (\Gamma (f_n) - \Gamma (\chi_i)) \ph \| \\
&\leq \const \| \ph_n - \ph \| +   \| (\Gamma (f_n) - \Gamma
(\chi_i)) \ph \| \to 0
\end{split}
\end{equation*}
for $n \to \infty$. In the last step we used that, by assumption,
$\ph \in \H_i \cap D(N)$.
\end{proof}


\bibliographystyle{alpha}

\end{document}